\documentclass[floatfix,showpacs,superscriptaddress,prl]{revtex4-2}
\usepackage{amsbsy,amssymb,amsmath,bm}
\usepackage{float}
\usepackage{epsf}
\usepackage{graphicx,color}
\usepackage[caption = false]{subfig}
\usepackage{amsfonts}
\usepackage{amssymb}
\usepackage{amsmath}
\usepackage[T1]{fontenc}
\usepackage{xspace}
\usepackage{epsfig}
\usepackage{ctable}
\usepackage{enumitem}
\newcommand{\rrangle}{>\kern-1.2ex~>\xspace}

\definecolor{red}{rgb}{1,0,0}           
\definecolor{green}{rgb}{0,1,0}
\definecolor{blue}{rgb}{0,0,1}
\definecolor{darkblue}{rgb}{0,0,0.5}
\definecolor{lightblue}{rgb}{.5,.5,1}
\definecolor{lightgray}{gray}{.87}          
\definecolor{Dark}{gray}{.20}
\definecolor{pink}{rgb}{.95,0.82,0.92}  
\definecolor{yellow}{rgb}{1,1,0}
\definecolor{lightyellow}{rgb}{1,1,.5}
\definecolor{purple}{rgb}{0.7,0,0.85}
\definecolor{darkgreen}{rgb}{0,0.5,0}
\definecolor{orange}{rgb}{0.8,0.2,0.2}

\def \be {\bea}
\def \ee {\eea}
\def \bea {\begin{eqnarray}}
\def \eea {\end{eqnarray}}
\def \bse {\begin{subequations}}
\def \ese {\end{subequations}}
\def \bde {\begin{description}}
\def \ede {\end{description}}
\def \bee {\begin{enumerate}}
\def \eee {\end{enumerate}}
\def \nn {\nonumber}

\def \nt {\tilde{\nu}}

\def \rt {\tilde{r}}

\def \tt {\tilde{t}}

\def \sb {\bar{s}}
\def \tb {\bar{t}}
\def \ub {\bar{u}}
\def \nb {\bar{\nu}}

\def \a {\alpha}
\def \b {\beta}

\def \dag {\dagger}
\def \g {\gamma}

\def \n {\nu}

\def \tA {\tilde{A}}
\def \tB {\tilde{B}}

\begin{document}

\title{ Quasi-Hermitian extended SSH models}

\author{Iao-Fai Io\footnote{e-mail address: jackyphys@gmail.com}}
\affiliation{\textit{Physics Department, National Taiwan University, Taipei 10617, Taiwan}}
\author{Cheng-Yuan Huang\footnote{e-mail address: s963037@gmail.com}}
\affiliation{\textit{Physics Department, National Taiwan University, Taipei 10617, Taiwan}}
\author{Jhih-Shih You\footnote{e-mail address: jhihshihyou@ntnu.edu.tw}}
\affiliation{\textit{Physics Department, National Taiwan Normal University, Taipei 11677, Taiwan}}
\author{Hao-Chun Chang\footnote{e-mail address: haochun71@yahoo.com.tw}}
\affiliation{\textit{Electrophysics Department, National Yang Ming Chiao Tung University, Hsinchu 30050, Taiwan}}
\author{Hsien-chung Kao\footnote{e-mail address: hckao@phy.ntnu.edu.tw}}
\affiliation{\textit{Physics Department, National Taiwan Normal University, Taipei 11677, Taiwan}}

\date{\today }

\begin{abstract}
We consider the quasi-Hermitian limit of a non-Hermitian extended Su–Schrieffer–Heeger model, in which the hopping amplitudes obey a specific relation so that the system may be mapped to a corresponding Hermitian one and its energy spectrum is completely real.  Analogous to the Hermitian case, one may use the modified winding number $\nb$ to determine the total number of edge states on the boundaries so that a modified bulk-boundary correspondence may be achieved. Due to the skin effect in non-Hermitian systems, the spectral winding numbers $\n_E^L$ and $\n_E^R$ must be used to classify such systems further. It dictates how the edge states would be distributed over the left and right boundaries. We then naively extend the criteria to the cases that the quasi-Hermitian condition is violated.  For all the cases that we consider no inconsistency has been found.

\end{abstract}

\maketitle

\subsection{I. Introduction}

Topological materials have been at the center of attention in the community of condensed matter physicists since their discovery~\cite{Review1, Review2}.  Bulk-boundary correspondence (BBC) is arguably the most distinct among all the intriguing properties of topological insulators and superconductors. It is well known that they may be classified into a ``periodic table'' according to the symmetries and dimensionalities of the system~\cite{Periodic table1, Periodic table2, Periodic table3}. Topological invariants associated with the Bloch bands of a system may be used to tell whether the system is in the topological phase.  Moreover, it dictates the number of edge states that appear on the system's boundaries.

The Su–Schrieffer–Heeger (SSH) model is the simplest topological insulator~\cite{SSH}. Originally, it was proposed to model a dimerized polyacetylene chain. It has also been realized using various systems.  See Ref.~\cite{SSH-QD} and references therein for details. The one-dimension (1D) winding number $\n$ is the topological invariant that may be used to predict the number of edge states on the boundaries of the system. It is known that the 1D winding number is closely related to the Zak phase, $\g$~\cite{Zak}, which is basically similar to the Berry phase~\cite{Berry}. It has been measured directly using a dimerized optical lattice~\cite{Zak phase-OL}.

Because of its simplicity, the SSH model has been one of the favorite prototypes to study topological insulators. First by adding an on-site energy term in the SSH model, one may extend it to the Rice-Mele model~\cite{Rice-Mele}. One may then use it to relate the SSH model to a Chern insulator by considering a charge-pumping process in such a system~\cite{Thouless, polarization1, polarization2}. One may also add more ``atoms'' in a unit cell.  As a result, there could be more than one occupied band in the system~\cite{Multi-band1, Multi-band2, Multi-band3} and holonomies or equivalently, Wilson-loop, may be used to describe the topology of such a system~\cite{Holonomy, W-loop}.  This is effectively the non-Abelian generalization of the Berry phase~\cite{Zak, Berry}.  Another interesting generalization is to introduce third-nearest-neighbor hopping amplitudes in the system. Such systems would admit topological phases with higher winding numbers, and they are referred to as the extended SSH models. It is known that there is an ambiguity in defining the Zak phase due to its dependence on the real-space origin and unit cell~\cite{Kudin, Rhim}.  Although a proposal that resolves such an ambiguity in the SSH model has been put forward, this kind of difficulty remains for extended SSH models, since the {\it intercellular} Zak phase from a band may be as large as $2\pi$~\cite{Kudin, Rhim}.  Consequently, extended SSH models could serve as an arena to study such ambiguity.

Although the Hamiltonian of a closed system is typically Hermitian so that unitarity is preserved, real physical systems are usually coupled to their environment. Therefore, dissipative processes generally exist and the effective Hamiltonian would become non-Hermitian (NH). In fact, an effective NH Hamiltonian approach provides a simple and intuitive way to describe open systems and thus has been broadly used in electrical circuits, optical and mechanical systems, etc. See Ref.~\cite{NH-Review1, NH-Review2} and references therein. In the past few years, there has been a growing interest in NH topological systems motivated by experimental observation of topological states in dissipative systems~\cite{NH-Exp1, NH-Exp2, NH-Exp3, NH-Exp4, NH-Exp5, NH-Exp6, NH-Exp7, NH-Exp8, NH-Exp9, NH-Exp10}. Many surprises appear when one extends a Hermitian topological system to its NH counterpart~\cite{NH-Review1, NH-Review2, NH-Review3, NH-Review4}. First of all, the energy spectrum for a system with periodic boundary conditions (PBCs) is generally very different from that with open boundary conditions (OBCs)~\cite{Hatano-Nelson}.  Since the energy spectrum of an NH system is generally complex, a new topological invariant, the spectral winding number, has been introduced~\cite{Spectral winding number1, Spectral winding number2}. It has been incorporated to derive an NH counterpart of the periodic table of topological insulators~\cite{NH Periodic table}.  Secondly, there is also the skin effect and hence even the wave functions of ``bulk'' states have an exponential damping or growing factor~\cite{Skin-effect1, Skin-effect2, Skin-effect3, Skin-effect4}. This in turn will lead to the breakdown of the conventional BBC~\cite{NH-Review1, BBC, Skin-effect1, Skin-effect3}.  In an effort to generalize the BBC, the generalized Brillouin zone (BZ) has been introduced~\cite{GBZ1, Skin-effect4, GBZ3, GBZ4, GBZ5}.  It has been shown that using the generalized BZ, one can obtain the correct winding number for NH topological systems and reestablish the BBC.  In Ref.~\cite{Skin-effect1}, the authors propose the idea of biorthogonal BBC, which is then further generalized~\cite{Biorthogonal1, Biorthogonal2}. They study the NH SSH model and its extension to demonstrate that the biorthogonal polarization may be used to determine the total number of edge states on the boundaries of the topological systems~\cite{Skin-effect1}.  Although generalized BZ and biorthogonal polarization are useful to restore the BBC in NH systems~\cite{Biorthogonal-polarization}, it is still worthwhile to seek a classification based only on the suitably modified winding number, $\nb$, and spectral winding numbers, $\n_E^L$ and $\n_E^R$, along the same lines as in Ref.~\cite{Spectral winding number1, Spectral winding number2}. For this purpose, we study in detail the quasi-Hermitian (QH) limit of the NH extended SSH models with next-to-nearest neighbor hopping amplitudes. In such a limit, the system may be mapped to a Hermitian one. As a result, the energy spectrum is real and we can use a simple modified winding number to classify the topological phases of the system.

The rest of the paper is organized in the following way. In Sec. II, we first use the NH SSH model to carry out a detailed analysis of its secular and characteristic equations.  Based on the study, we conclude the NH SSH model may be mapped to a Hermitian one. In other words, it is pseudo-Hermitian (PH)~\cite{PH}. Moreover, one can use the modified winding number, $\nb$, and spectral winding number $\n_E$ to reestablish the BBC. In particular, $\nb$ may be used to determine the total number of edge states on the boundaries, while $\n_E$ dictates how the edge states would distribute over the left and right boundaries. The results obtained are confirmed using numerical calculation and are consistent with those found in the literature~\cite{Skin-effect1, Skin-effect2, Skin-effect3, Skin-effect4, Biorthogonal1, Biorthogonal2, NH-SSH1}.  Next, we try to establish the so-called quasi-Hermitian condition (QHC), under which this may be generalized to the NH extended SSH models. When QHC is satisfied, we again show explicitly in terms of the system's secular and characteristic equations that it may be mapped to a Hermitian one. After some analysis, we show that one may still use $\nb$, $\n_E^L$ and $\n_E^R$ to reestablish the BBC. The results are also verified using numerical calculation. In Sec. IV,  we carry out a study of general NH extended SSH models. Again we analytically find the characteristic equation and numerically calculate the energy spectrum of such systems. Here, we naively extend the criteria that we obtained in the quasi-Hermitian (QH) limit and check whether they still hold. Finally, we make conclusions and discuss possible extensions in Sec. V.

\subsection{II. NH SSH model}
Let's begin with a finite chain of NH SSH models with OBCs:
\bea
&\;& \hskip -3.1cm H_{\rm SSH}=\sum_{j=1}^{N-1}\left\{ \left(t_0^L  A_j^\dag  +t_1^R A_{j+1}^\dag\right) B_j  + \left(t_0^R  A_j  +t_1^L A_{j+1}\right) B_j^\dag  \right\}  + t_0^L  A_N^\dag B_N  + t_0^R  A_N B_N^\dag.
\eea
There are $N$ unit cells in the system, and $j$ is used to denote them.  $t_0^{L,R}, t_1^{L,R}$ are the intra-cell and inter-cell left and right hopping amplitudes, respectively.  It is straightforward to obtain the equation of motion (EOM) satisfied by the energy eigenstates:
\bea \label{SSH EOM 1}
&\;& \hskip -3.1cm E A_j - \left(t_0^L B_j +t_1^R B_{j-1} \right) =0, \quad \mbox{\rm for }  j = 2, \ldots,N; \cr
&\;& \hskip -3.1cm E B_j - \left(t_0^R  A_j +t_1^L A_{j+1}\right) =0, \quad \mbox{\rm for }  j = 1, \ldots, N-1.
\eea
The boundary conditions (BCs) are specified by
\bea \label{BC 1}
&\;& \hskip -3.1cm E A_1 - t_0^L B_1 = 0; \cr
&\;& \hskip -3.1cm E B_N - t_0^R  A_N = 0,
\eea
which may be simplified to
\bea \label{Simplified BC 1}
&\;& \hskip -3.1cm  B_0  = 0; \cr
&\;& \hskip -3.1cm  A_{N+1}   = 0.
\eea
To find the solutions to the system, we let
\bea
&\;& \hskip -3.1cm A_j =\a s^j, B_j=\b s^j,
\eea
and it would lead to
\bea \label{EOM SSH}
&\;& \hskip -3.1cm E \a- \left(t_0^L +t_1^R s^{-1} \right)\b=0, \cr
&\;& \hskip -3.1cm E \b -\left(t_0^R  +t_1^L s \right)\a =0.
\eea
Non-trivial solutions for $\a$ and $\b$ exist only if the determinant formed by the coefficients of $\a$ and $\b$ is vanishing, i.e.
\bea \label{Seculae eq 1}
&\;& \hskip -3.1cm E^2 - \left(t_0^L +t_1^R s^{-1} \right)\left(t_0^R  +t_1^L s \right)=0,
\eea
which we call the secular equation.  Since the above equation is quadratic in $s$, the most general solutions of $A_j$ and $B_j$ to eq.~(\ref{EOM SSH}) are given by
\bea
&\;& \hskip -3.1cm A_j =\a_1 s_1^j +\a_2 s_2^j , B_j=\b_1 s_1^j +\b_2 s_2^j.
\eea
Here, $s_1$ and $s_2$ are the two roots of the secular equation in eq.~(\ref{Seculae eq 1}). Obviously, the product and sum of the two roots satisfy the following relations:
\bea \label{Prod and sum}
&\;& \hskip -3.1cm  s_1 s_2 = t_0^R t_1^R/(t_0^L t_1^L), \nn \cr
&\;& \hskip -3.1cm   s_1 + s_2 = \left(E^2- t_0^L t_0^R- t_1^L t_1^R\right)/(t_0^L t_1^L).
\eea
Substituting the above expressions back to the BCs given in eq.~(\ref{Simplified BC 1}) and then expressing $\a_1, \a_2$ in terms of $\b_1, \b_2$, we see the BCs may be simplified to
\bea
&\;& \hskip -3.1cm  t_1^R  \b_1 + t_1^R \b_2  =0; \cr
&\;& \hskip -3.1cm  t_1^L s_1^{N}  \left(t_0^Ls_1 +t_1^R  \right) \b_1  + t_1^L s_2^{N} \left(t_0^L s_2+t_1^R  \right)\b_2 =0.
\eea
Again, non-trivial solutions for $\b_1$ and $\b_2$ exist only if the determinant consisting of the coefficients of $\b_1$ and $\b_2$ in the above equations is vanishing:
\bea
&\;& \hskip -3.1cm t_1^R t_1^L \left( s_1^N-  s_2^N  \right) + t_0^R t_0^L  \left(s_1^{N+1}-  s_2^{N+1} \right)  = 0,
\eea
which we call the characteristic equation of $s$ hereafter. Define the skin effect factor
\bea\label{skin effect}
r=\sqrt{t_0^R t_1^R/(t_0^L t_1^L)},
\eea
so that $s_1 = r \sb, s_2 = r \sb^{-1}$, and the above equation may be cast into the following form:
\bea  \label{Characteristic eq 1}
&\;& \hskip -3.1cm   \tb_0 U_{N}(\ub) + \tb_1 U_{N-1}(\ub) = 0.
\eea
Here, $\tb_0 = \sqrt{t_0^R t_0^L }$ and $\tb_1 = \sqrt{t_1^R t_1^L }$. $\ub = \left(\sb+\sb^{-1}\right)/2$ and $U_N(\ub)$ is the Chebyshev polynomials of the second kind of degree $N$, which has also been mentioned in Ref.~\cite{Chebyshev}. Note that the above equation is equivalent to that in the Hermitian case when we identify the variables properly. From Eq.~(\ref{Prod and sum}), we see the energy of the system is always real, with its spectrum given by
\bea
E=\pm\left(\tb_0^2 + \tb_1^2  + 2 \tb_0 \tb_1 \ub \right)^{1/2}.
\eea

From our experience with the Hermitian SSH model, it is straightforward to come to the conclusion that the system would be in the topological phase if  $\tb_1/\tb_0>1$. In this case, there exist two solutions of $\sb$, that give rise to two almost zero energy edge states. They are approximately $\sb_0$ and $\sb_0^{-1}$ with $\sb_0 = - \tb_0/\tb_1$.  Since now it is the ratio $\tb_0/\tb_1$ that determines whether the system is in the topological phase, it is now the modified winding number, $\nb$, that plays the role of topological invariant. We expect that $\nb$ is determined by whether $h(p) = \tb_0 + \tb_1 e^{ip}$ wrap around the origin when $p$ ranges over the BZ.  Furthermore, it is known that there is the so-called ``skin effect'' in the NH SSH model, which may be described by the parameter $r$ defined in eq.~(\ref{skin effect}). The most striking feature here is that even the ``bulk'' states would exhibit exponential growing or damping behavior in general. For $r>1$ and $r<1$, all the ``bulk'' states would be ``tilted'' to the right and left boundaries, respectively. Because of this effect, the exponential factors of the edge states become $r \sb_0, \approx -\left(t_0^R/t_1^L \right)$ and $r \sb_0^{-1}\approx -\left(t_1^R/t_0^L \right)$. In the topological phase, if both $t_0^R/t_1^L$ and $t_1^R/t_0^L$ are smaller (larger) than 1, then both two edge states would appear on the left (right) boundary, in stark contrast with the Hermitian case. We would also like to mention that even if $t_1^R/t_0^L <1$ the edge state associated with  $s=-t_1^R/t_0^L$ would not exist in a right semi-infinite chain, different from the one associated with  $s=-t_0^R/t_1^L$.

\begin{figure}[hbt!]
\centering
\subfloat[]{\includegraphics[width=0.50\textwidth]{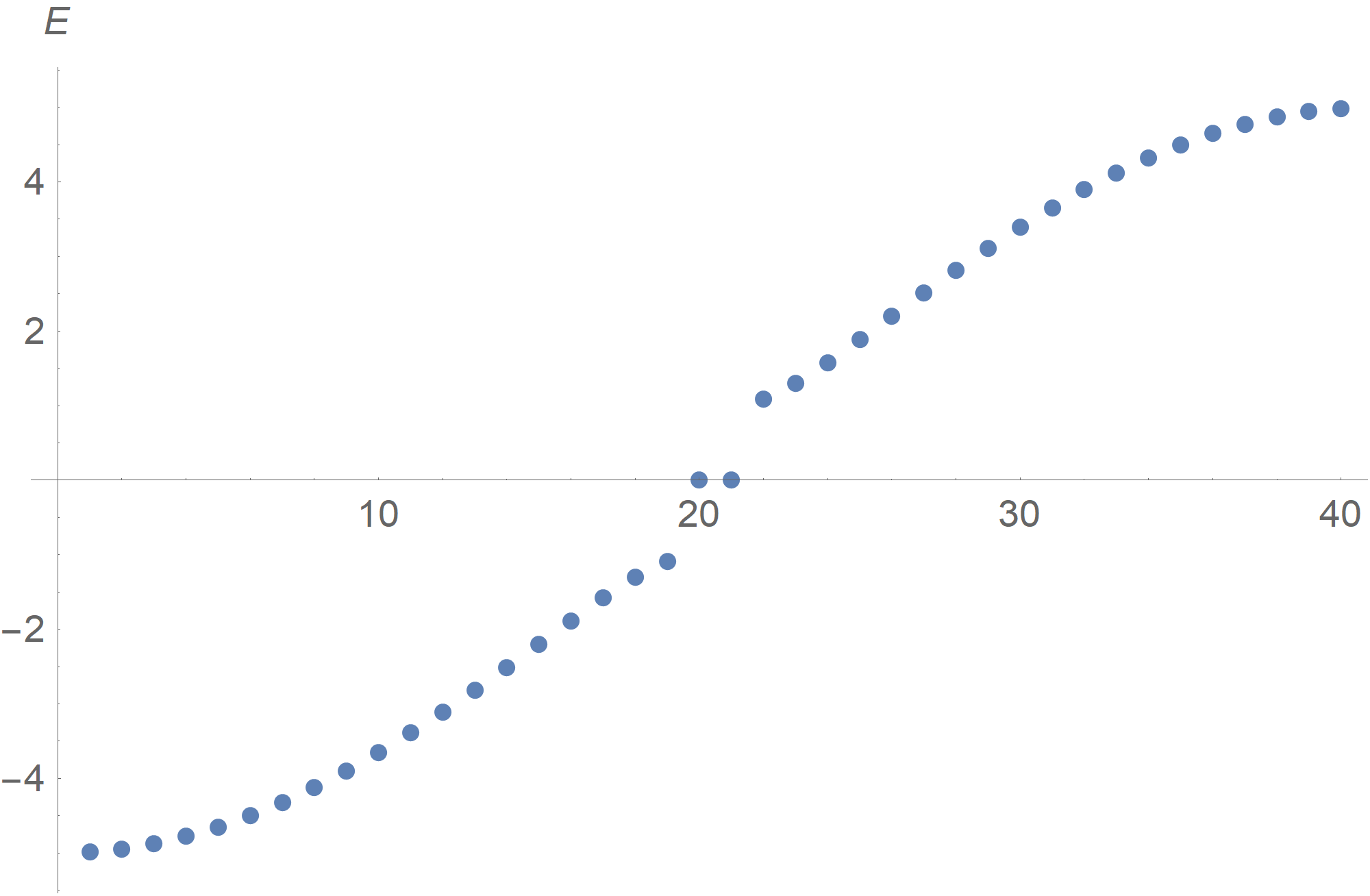}}\\
\subfloat[]{\includegraphics[width=0.30\textwidth]{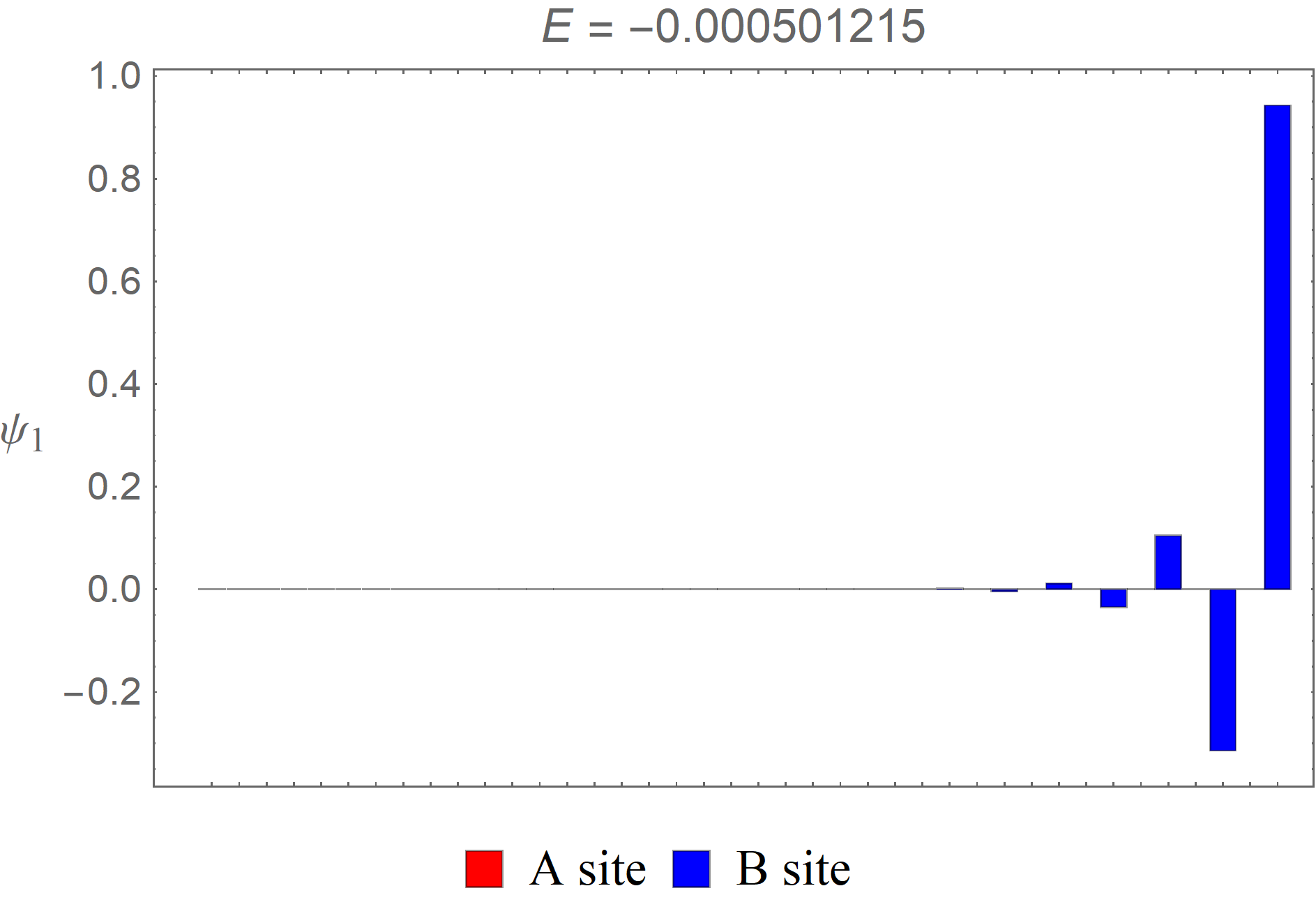}}\hskip 0.5cm
\subfloat[]{\includegraphics[width=0.30\textwidth]{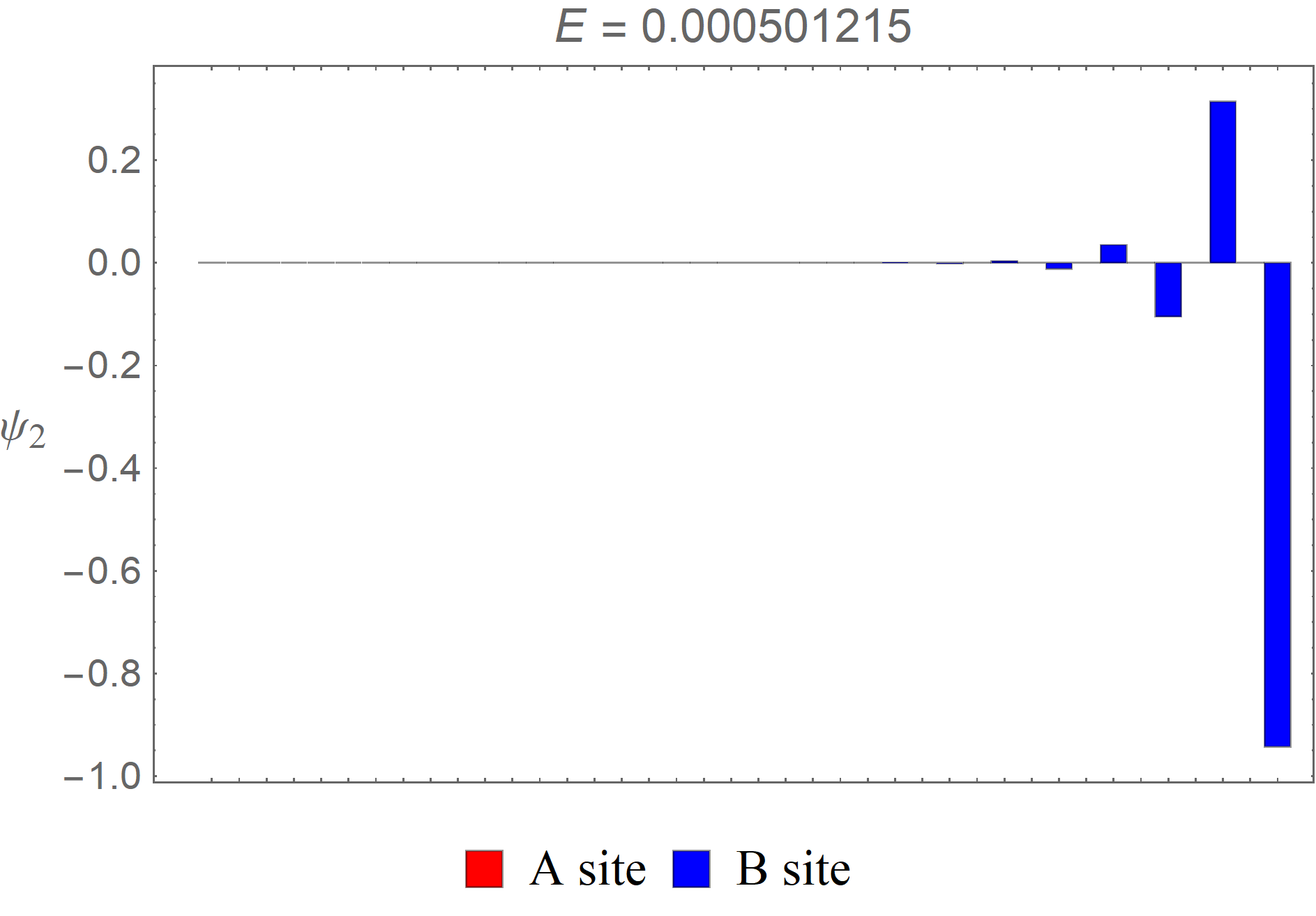}}\\
\subfloat[]{\includegraphics[width=0.30\textwidth]{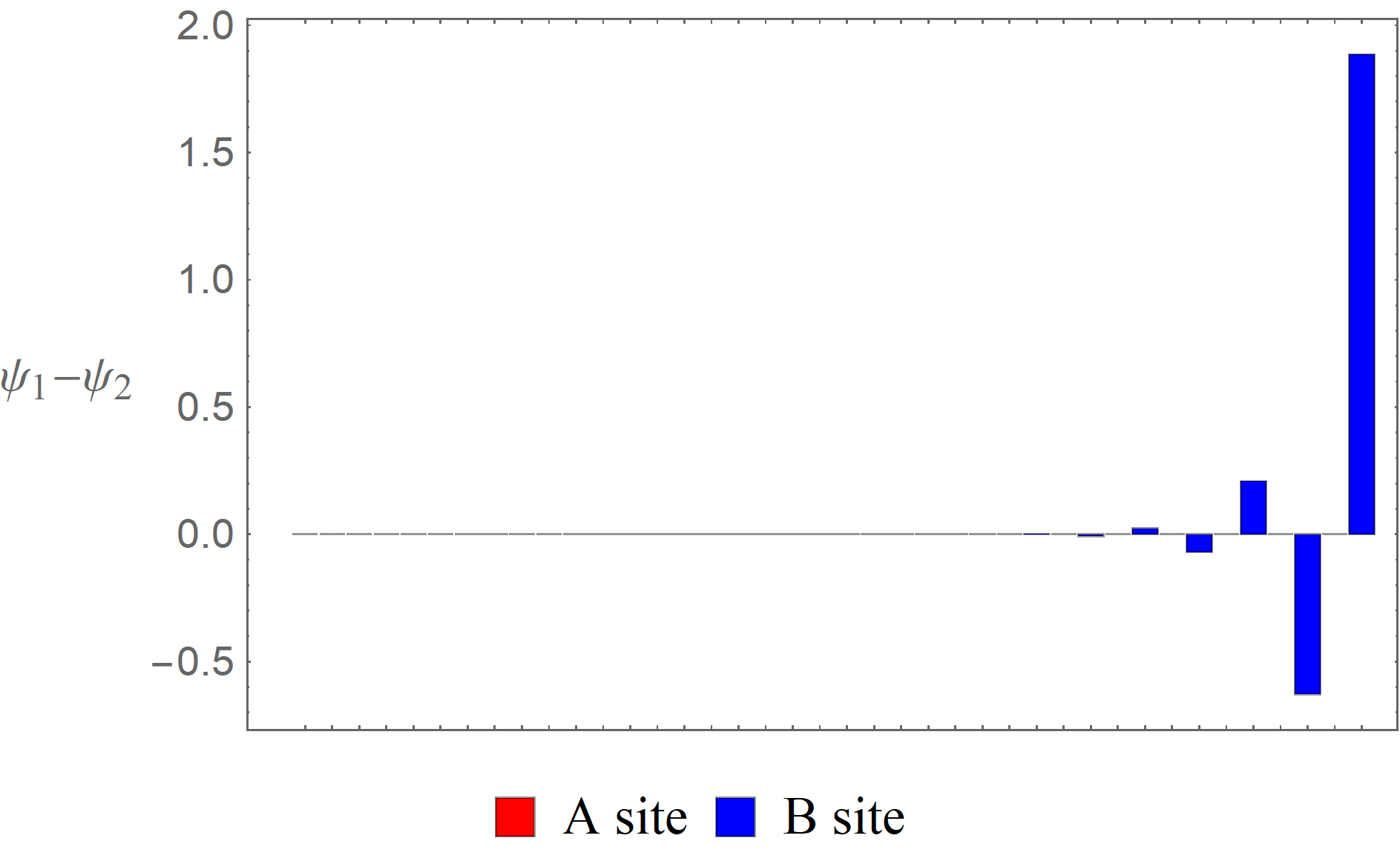}}\hskip 0.5cm
\subfloat[]{\includegraphics[width=0.30\textwidth]{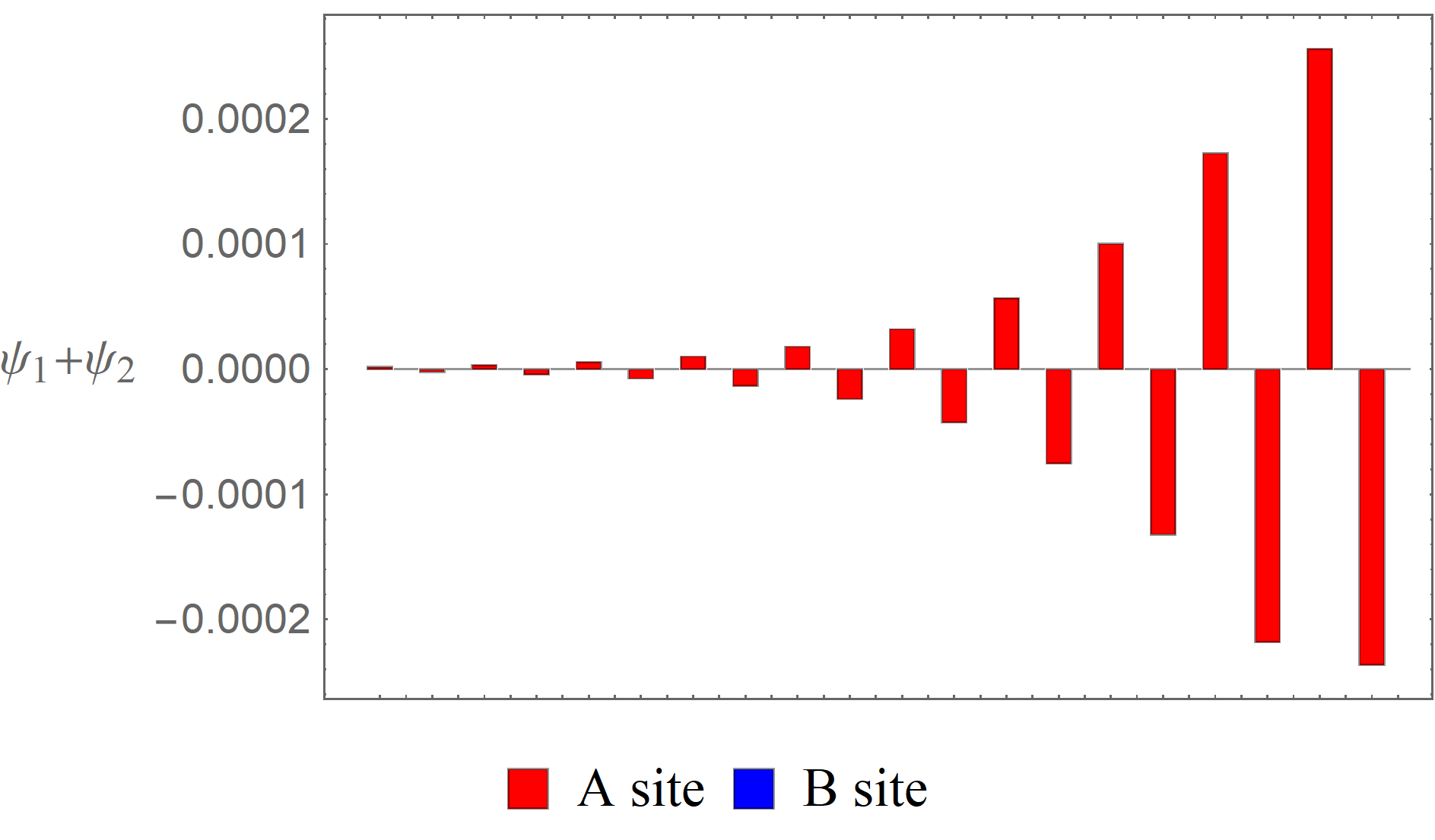}}
\caption{(a)The energy spectrum in the topological phase of the NH SSH model, where $\left( t_0^L, t_0^R, t_1^L, t_1^R \right) = (1, 4, 3, 3)$. The energies of the two edge states $\psi_1, \psi_2$ are both approximately zero.  (b,), (c) The wave functions of the two edge states in the system. Note that they are both right edge states. (d), (e) The wave functions of the difference and sum of $\psi_1$ and $\psi_2$.  Note that although $\psi_1 + \psi_2$ is also a right edge state, it is non-vanishing only on the A sites in contrast to $\psi_1 - \psi_2$.}  \label{fig1nHSSH}
\end{figure}

These results may be confirmed numerically, and the spectrum and wave functions of the edge states are shown in Fig.~\ref{fig1nHSSH}.  Here, we choose $N=20$, $\left( t_0^L, t_0^R, t_1^L, t_1^R \right) = (1, 4, 3, 3)$ so that $(\tb_0, \tb_1) = (2, 3)$, $r=2$ and $\nb = 1$.  At first glance, it seems that both edge states, $\psi_1$ and $\psi_2$, are non-vanishing only on the B sites and they form a chiral conjugate pair.  This is of course a reflection of the skin effect, but it is in fact misleading and arises from the limitation in numerical accuracy.  This may be seen by taking the difference and sum of the wave functions of the two edge states, $\psi_1$ and $\psi_2$.  The resulting wave functions correspond to two right edge states. The former one is non-vanishing on B sites only and the latter one is on A sites only.  This is a signature that they are derived from the Hermitian SSH model.

As shown above, although $\nb$ may tell us the total number of edge states in the system, the edge states now may no longer distribute evenly on the left or right boundaries in contrast to the Hermitian case. To further classify the topological phase, we must consider the case with PBCs to see how to determine which boundaries the edge states would show up.  Under such BCs, we have
\bea
&\;& \hskip -3.1cm s_k = \exp\left( i2\pi k/N \right), \mbox{\rm  and } E = \pm\left(t_0^L t_0^R +  t_1^L t_1^R  + t_0^L t_1^L s_k +t_0^R t_1^R s_k^{-1} \right)^{1/2},
\eea
with $k=0,1,\ldots, N-1$.  It is obvious that the energy is generally complex and the spectrum is very different from that of the OBC case. In the continuum limit, we have
\bea
&\;& \hskip -3.1cm  \det H(p) = E^2(p) = \left(t_0^R  +t_1^L  e^{ip} \right) \left(t_0^L +t_1^R e^{-ip} \right),
\eea
where $p \in (-\pi, \pi] $.  Since $ \det H(p)$ is a product of two factors, the spectral winding number $\n_E$ for its  trajectory on the complex plane is given by
\bea
&\;& \hskip -3.1cm  \n_E = \n_E^R + \n_E^L.
\eea
Here $\n_E^L$ is the spectral winding number corresponding to the factor $\left(t_0^R  +t_1^L  e^{ip} \right)$, and $\n_E^R$ corresponding to $\left(t_0^L +t_1^R e^{-ip} \right)$. It is $\n_E $ that will determine how the edge states would be distributed over the two boundaries.
\bee[label=\arabic*)]

\item $\nb =1$:

\bee[label=\roman*)]

\item $\n_E =1~(t_0^L > t_1^R, t_0^R < t_1^L)$. Both edge states appear on the left boundary.

\item $\n_E =0~(t_0^L < t_1^R, t_0^R < t_1^L )$. One edge state appears on the left boundary and the other on the right.

\item $\n_E =-1~(t_0^L < t_1^R, t_0^R > t_1^L)$. Both edge states appear on the right boundary.

\eee

\item $\nb =0$: $\n_E =1,0,-1$. There is no edge state.

\eee
The modified winding number $\nb$ may also be obtained using the so-called modified PBCs~\cite{MPBC}.  Of course, it is equivalent to using a generalized BZ  so that $\nb = \n_E\left|_{C_s} \right.$, where $C_s$ is a circle about the origin with radius equal to the skin effect factor $r$~\cite{Skin-effect4, GBZ4, GBZ5}.  As we shall see, this criterion may also be generalized to the NH extended SSH models.

\subsection{III. Quasi-Hermitian Extended SSH models}

Let's now turn to the NH extended SSH models, again with OBCs.  To be specific, we first consider the type 1 case, and the Hamiltonian for a finite chain is given by
\bea
&\;& \hskip -3.1cm H_{\rm ext1}=\sum_{j=1}^{N-2}\left\{ \left(t_0^L  A_j^\dag  + t_1^R A_{j+1}^\dag + t_2^R A_{j+2}^\dag \right) B_j  + \left(t_0^R  A_j  +t_1^L A_{j+1} + t_2^L A_{j+2} \right) B_j^\dag  \right\}  \nn \\
&\;& \hskip -1.1cm + \left(t_0^L  A_{N-1}^\dag  + t_1^R A_{N}^\dag\right) B_{N-1} + \left(t_0^R  A_{N-1}  + t_1^L A_{N}\right) B_{N-1}^\dag + t_0^L  A_N^\dag B_N  + t_0^R  A_N B_N^\dag.
\eea
Again, $j$ denotes the unit cell and there are $N$ unit cells in total. In addition to $t_0^{L, R}$ and $t_1^{L, R}$, now we also have $t_2^{L, R}$, the third-nearest-neighbor left and right hopping amplitudes.  It is straightforward to obtain the EOM:
\bea \label{SSH EOM 1A}
&\;& \hskip -3.1cm E A_j - \left(t_0^L B_j + t_1^R B_{j-1} + t_2^R B_{j-2} \right) =0, \quad \mbox{\rm for }  j = 3, \ldots,N; \cr
&\;& \hskip -3.1cm E B_j - \left(t_0^R  A_j +t_1^L A_{j+1} + t_2^L A_{j+2} \right) =0, \quad \mbox{\rm for }  j = 1, \ldots, N-2,
\eea
and BCs after some algebra
\bea \label{Simplified BC 1A}
&\;& \hskip -3.1cm B_{-1} = B_0 =0; \cr
&\;& \hskip -3.1cm  A_{N+1} = A_{N+2} =0.
\eea

By substituting $A_j =\a s^j, B_j=\b s^j$ into the EOM, we obtain
\bea \label{Simplified SSH EOM 1A}
&\;& \hskip -3.1cm E \a- \left(t_0^L + t_1^R s^{-1} + t_2^R s^{-2} \right)\b=0, \cr
&\;& \hskip -3.1cm E \b -\left(t_0^R + t_1^L s + t_2^L s^2 \right)\a =0,
\eea
and hence the secular equation:
\bea \label{Secular eq1A}
&\;& \hskip -3.1cm E^2 - \left(t_0^L + t_1^R s^{-1} + t_2^R s^{-2} \right)\left(t_0^R  +t_1^L s + t_2^L s^2 \right)=0.
\eea
Since the above equation is quartic in $s$, the most general solutions of $A_j$ and $B_j$ would be given by
\bea
&\;& \hskip -3.1cm A_j = \sum_{I=1}^4 \a_I s_I^j , B_j =  \sum_{I=1}^4 \b_I s_I^j.
\eea
Here, $s_1, s_1, s_3$ and  $s_4$ are the four roots of the secular equation, and they satisfy the following relations:
\bea \label{Prod and sum 1A}
&\;& \hskip -3.1cm  s_1 s_2 s_3 s_4 = t_0^R t_2^R/(t_0^L t_2^L), \cr
&\;& \hskip -3.1cm  s_1 s_2 s_3 + s_1 s_2 s_4 + s_1 s_3 s_4 + s_2 s_3 s_4= -\left(t_0^R t_1^R + t_1^L t_2^R\right)/(t_0^L t_2^L), \cr
&\;& \hskip -3.1cm  s_1 s_2 + s_1 s_3 +  s_1 s_4 + s_2 s_3 + s_2 s_4 + s_3 s_4 = \left(t_0^L t_0^R + t_1^L t_1^R + t_2^L t_2^R - E^2\right)/(t_0^L t_2^L), \cr
&\;& \hskip -3.1cm  s_1 + s_2 + s_3 + s_4= -\left(t_0^L t_1^L + t_1^R t_2^L\right)/(t_0^L t_2^L).
\eea
By using the most general forms of $A_j$ and $B_j$, the BCs given in eqs.~(\ref{Simplified BC 1A}) becomes
\bea \label{Simplified BC 2A}
&\;& \hskip -3.1cm  \b_1 s_1^{-1} + \b_2 s_2^{-1} + \b_3 s_3^{-1} + \b_4 s_4^{-1} = 0, \cr
&\;& \hskip -3.1cm  \b_1 + \b_2 + \b_3 + \b_4 = 0, \cr
&\;& \hskip -3.1cm  \a_1 s_1^{N+1} + \a_2 s_2^{N+1} + \a_3 s_3^{N+1} +\a_4 s_4^{N+1} = 0, \cr
&\;& \hskip -3.1cm  \a_1 s_1^{N+2} + \a_2 s_2^{N+2} + \a_3 s_3^{N+2} +\a_4 s_4^{N+2}  = 0.
\eea

The criteria obtained in the previous section are all very nice but one would see in no time that it is impossible to generalize them directly to the extended SSH models.
Therefore, we will limit ourselves to the special case that there is only one single skin effect factor
\bea\label{skin effect 1A}
r = \left[ t_0^R t_2^R/(t_0^L t_2^L) \right]^{1/4},
\eea
such that $s_1 = r \sb_1, s_2 = r \sb_2, s_3 = r \sb_1^{-1},$ and $s_4=r \sb_2^{-1}.$  This may be achieved when the following condition is satisfied
\bea
&\;& \hskip -3.1cm \left(t_0^R t_1^R + t_1^L t_2^R\right) = r^2\left(t_0^L t_1^L + t_1^R t_2^L\right), \nn
\eea
which may be simplified to
\bea \label{quasi-Hermitian}
t_1^L=t_1^R \sqrt{t_0^R t_{2}^L/\left( t_0^L t_{2}^R \right)}.
\eea
Because of this additional constraint, only five of the six parameters are independent. When this occurs, we expect that the system may be mapped to a Hermitian type 1 extended SSH model, and all the energy eigenvalues would be real. For this reason, we will call this constraint the quasi-Hermitian condition (QHC).  Note that quasi-Hermiticity is only a subset of pseudo-Hermiticity since there obviously exist NH extended systems that do not satisfy the QHC but still have a real energy spectrum.

Now, let's show explicitly how the system may be mapped to a Hermitian type 1 extended SSH model when the QHC is satisfied.  First, we would like to mention that when this is the case, only two of the equations in eqs.~(\ref{Prod and sum 1A}) are independent. They may be reduced to the following forms:
\bea \label{Simplified prod and sum 1A}
&\;& \hskip -3.1cm   \ub_1 +\ub_2 = -\left(t_0^L t_1^L + t_1^R t_2^L\right)/(2r t_0^L t_2^L ), \cr
&\;& \hskip -3.1cm  2 + 4 \ub_1 \ub_2 = \left(t_0^L t_0^R + t_1^L t_1^R + t_2^L t_2^R - E^2\right)/(r^2 t_0^L t_2^L ),
\eea
in terms of $\ub_1=\left(\sb_1 +\sb_1^{-1} \right)/2$ and $\ub_2=\left(\sb_2 +\sb_2^{-1} \right)/2$. Upon using eqs.~(\ref{Simplified SSH EOM 1A}), we see $\a_I = \left\{\left(t_0^L + t_1^R s_I^{-1} + t_2^R s_I^{-2} \right)/E \right\}\b_I$, with $I=1,\ldots, 4$.  After some algebra, the characteristic equation for $\sb$ may be obtained from the condition that the determinant formed by the coefficients of $\b_1, \ldots, \b_4$ in eqs.~(\ref{Simplified BC 2A}) is vanishing:
\bea  \label{Simplified characteristic eq 1A even}
&\;& \hskip -2.6cm  \left\{ \tb_0 U_{N+2}(\ub_1)  + \tb_1 U_{N+1}(\ub_1) + \tb_2  U_{N}(\ub_1) \right\}
\left\{ \tb_0 U_{N}(\ub_2)  + \tb_1  U_{N-1}(\ub_2) + \tb_2  U_{N-2}(\ub_2) \right\} \cr
&\;& \hskip -2.95cm  +\left\{ \tb_0 U_{N+2}(\ub_2)  + \tb_1 U_{N+1}(\ub_2) + \tb_2  U_{N}(\ub_2) \right\}
\left\{ \tb_0 U_{N}(\ub_1)  + \tb_1  U_{N-1}(\ub_1) + \tb_2  U_{N-2}(\ub_1) \right\} \cr
&\;& \hskip -3.1cm  -2  \left\{ \tb_{0} U_{N+1}(\ub_1)  + \tb_1 U_{N}(\ub_1) + \tb_2 U_{N-1}(\ub_1) \right\}
\left\{ \tb_{0} U_{N+1}(\ub_2)  + \tb_1 U_{N}(\ub_2) + \tb_2 U_{N-1}(\ub_2) \right\} \cr
&\;& \hskip -3.1cm  + 2 \left\{ \tb_{0}^2 + \tb_1^2 + \tb_2^2 - 2\tb_0 \tb_2 \left(1 + 2 \ub_1 \ub_2 \right) \right\} = 0.
\eea
Similar to the case of the NH SSH model, we have further introduced $ \tb_2 = \sqrt{t_2^L t_2^R}$. In terms of $ \tb_0,  \tb_1$ and $\tb_2$, eq.~(\ref{Simplified prod and sum 1A}) becomes
\bea \label{Simplified prod and sum 2A}
&\;& \hskip -3.1cm   \ub_1 +\ub_2 = - \tb_1\left(\tb_0 + \tb_2\right)/(2 \tb_0 \tb_2 ), \cr
&\;& \hskip -3.1cm  E^2= \tb_0^2 + \tb_1^2 + \tb_2^2 + 2\tb_0 \tb_2 \left(2 + 4 \ub_1 \ub_2 \right).
\eea
The above characteristic equation is exactly the same as the one obtained in Ref.~\cite{Ext-SSH} if we identify the parameters $(\tb_0, \tb_1, \tb_2)$ with $(t_0, t_1, t_2)$.

By substituting the first part in the above equation into eq.~(\ref{Simplified characteristic eq 1A even}), we obtain a characteristic equation in terms of $\ub_1$ which is a polynomial of degree $2N+2$. By closer examination, we see that there is an additional factor $ \left[ \ub_1 + \tb_1\left(\tb_0 + \tb_2 \right)/(4 \tb_0 \tb_2 )\right]^2$ in the characteristic equation, which does not correspond to any physical states and can be ignored. Since both equations are symmetric with respect to $\ub_1$ and $\ub_2$, we may always require $\ub_1 > \ub_2$ to remove the redundancy in the solutions of $\ub_1$ and $\ub_2$.  Note that here $\ub_1$ and $ \ub_2$ are generally complex, and the relation $\ub_1 > \ub_2$ is defined by ${\rm Re}\{\ub_1\} > {\rm Re}\{\ub_2\}$ or ${\rm Re}\{\ub_1\} = {\rm Re}\{\ub_2\}$ and ${\rm Im}\{\ub_1\} > {\rm Im}\{\ub_2\}$.  Eventually, we find $N$ roots of $u_1$ consistent with what we expected. We would like to mention that by solving the characteristic equation in eq.~(\ref{Simplified characteristic eq 1A even}), we may also obtain the energy spectrum, which agrees completely with that achieved by numerical diagonalization.

Analogous to the Hermitian case, the edge states have almost zero energy, and the characteristic $s$'s associated with these states satisfy the condition in eq.~(\ref{Simplified SSH EOM 1A}) and $\b \approx 0$.  This gives rise to
\bea \label{Edge-state-s}
&\;& \hskip -3.1cm   t_0^R + t_1^L s + t_2^L s^2 \approx 0,  \label{Edge-state-1} \\
&\;& \hskip -3.1cm  t_0^L + t_1^R s^{-1} + t_2^R s^{-2} \approx 0. \label{Edge-state-2}
\eea
In terms of $\sb$, the first one leads to the following quadratic relation
\be
\tb_0 + \tb_1 \sb + \tb_2 \sb^2 \approx 0.
\ee
Thus, the system may be classified according to the modified winding number $\nb$, which dictates in the above quadratic equation the number of roots with absolute values less than 1, just like in the Hermitian case~\cite{Ext-SSH}.  It is known that there are three different phases, and we expect that $\nb$ would also determine the total number of edge states on the system's boundaries.
\bee[label=\arabic*)]
\item $\tb_2+\tb_0>\tb_1$, and  $\tb_2 > \tb_0$:
The system is in the topological phase with $\nb=2$. There are four edge states in total on the boundaries.

\item $\tb_2+\tb_0<\tb_1$:
The system is in the topological phase with $\nb=1$. There are two edge states in total on the boundaries.

\item $\tb_2+\tb_0>\tb_1$, and  $\tb_2 < \tb_0$:
The system is in the trivial phase with $\nb=0$. There are no edge states.
\eee
Again, this may also be obtained using a generalized BZ or modified PBCs~\cite{Skin-effect4, GBZ4, GBZ5, MPBC}.

Because of the skin effect, the edge states may not be evenly distributed over the left and right boundaries. Similar to the NH SSH model, we may use $\n_E$ to determine where they would reside.  In the continuum limit, we have from eq.~(\ref{Secular eq1A})
\bea \label{nu_E 1A}
&\;& \hskip -3.1cm E^2\left(p \right) =  \left(t_0^R  +t_1^L e^{ip} + t_2^L e^{2ip} \right) \left(t_0^L + t_1^R e^{-ip} + t_2^R e^{-2ip} \right).
\eea
Again, we have $\n_E = \n_E^L + \n_E^R$,  with $\n_E^L$ the spectral winding number corresponding to the factor  $\left(t_0^R  +t_1^L  e^{ip} + t_2^L e^{2ip} \right)$, and $\n_E^R$ corresponding to $\left(t_0^L +t_1^R e^{-ip} + t_2^R e^{-2ip} \right)$. Just like in the Hermitian case, the number of solutions to eqs.~(\ref{Edge-state-1}) and (\ref{Edge-state-2}) that have magnitudes smaller or larger than 1 indicates the possible numbers of edge states on the left and right boundaries, respectively. In addition, we must bear in mind that the total number of edge states is $2\nb$. Now, let's summarize the correspondence between $(\n_E^L,  \n_E^R)$ and the locations of the edge states:
\bee[label=\arabic*)]
\item $\nb=2$:
\bee[label=\roman*)]
\item $\n_E = 2$. All four edge states are on the left boundary.

\item $\n_E = 1$. Three edge states are on the left boundary and one on the right.

\item $\n_E = 0$. Two edge states are on the left boundary and two on the right.

\item $\n_E = -1$. One edge state is on the left boundary and three on the right.

\item $\n_E = -2$. All four edge states are on the right boundary.
\eee

\item $\nb=1$:
\bee[label=\roman*)]
\item $(\n_E^L, \n_E^R) = (2, 0), (1, 0), \fbox{(2, -1)}$. Both edge states are on the left boundary.

\item $(\n_E^L, \n_E^R) = \fbox{(2, -1)}, (2, -2),  (1, -1), (0, 0), \fbox{(1, -2)}$. One edge state is on the left boundary and one is on the right.

\item $(\n_E^L, \n_E^R) = \fbox{(1, -2)}, (0, -1), (0, -2)$. Both edge states are on the right boundary.
\eee

\item $\nb=0$: $\n_E = 2, 1, 0, -1, 2$. No edge state.

\eee
Note that we put a box around $(2, -1)$ and $(1, -2)$ to indicate that there is an ambiguity in determining the locations of the edge states when $(\n_E^L, \n_E^R) $ equal to these values.

From eqs.~(\ref{skin effect 1A}) and (\ref{quasi-Hermitian}), we see if $(t_0^L, t_0^R, t_1^L, t_1^R, t_2^L, t_2^R)$ satisfy the QHC and they are scaled in the following way:
\bea \label{scaling t}
&\;& \hskip -3.1cm (t_0^L, t_1^R, t_2^R) \to (\rt^{-1} t_0^L,t_1^R, \rt t_2^R), \nn\\
& \;& \hskip -3.1cm (t_0^R, t_1^L, t_2^L) \to (\rt t_0^R, t_1^L, \rt^{-1} t_2^L),
\eea
then $(\tb_0, \tb_1, \tb_2)$ would remain the same, while the skin effect factor will change by a factor of $\rt$.  Since the characteristic equation and energy spectrum only depend on $(\tb_0, \tb_1, \tb_2)$, they are also invariant under the scaling. Thus, it is a symmetry of the NH extended SSH models. In particular, we may use $\rt$ to scale the magnitude of the $(s_I)'s$ corresponding to the edge states. Hence we may vary the value of $\n_E$ between $-2$ and $2$. In fact, this scaling symmetry remains intact even in general NH extended SSH models. If we can use the scaling to make $\n_E$ zero, the edge states the trajectory of $E$ will enclose all the ``bulk'' states so it gives us another way to identify the edge states in the system, especially when we consider general NH extended SSH models.

\begin{figure}[hbt!]
\centering
\subfloat[]{\includegraphics[width=0.40\textwidth]{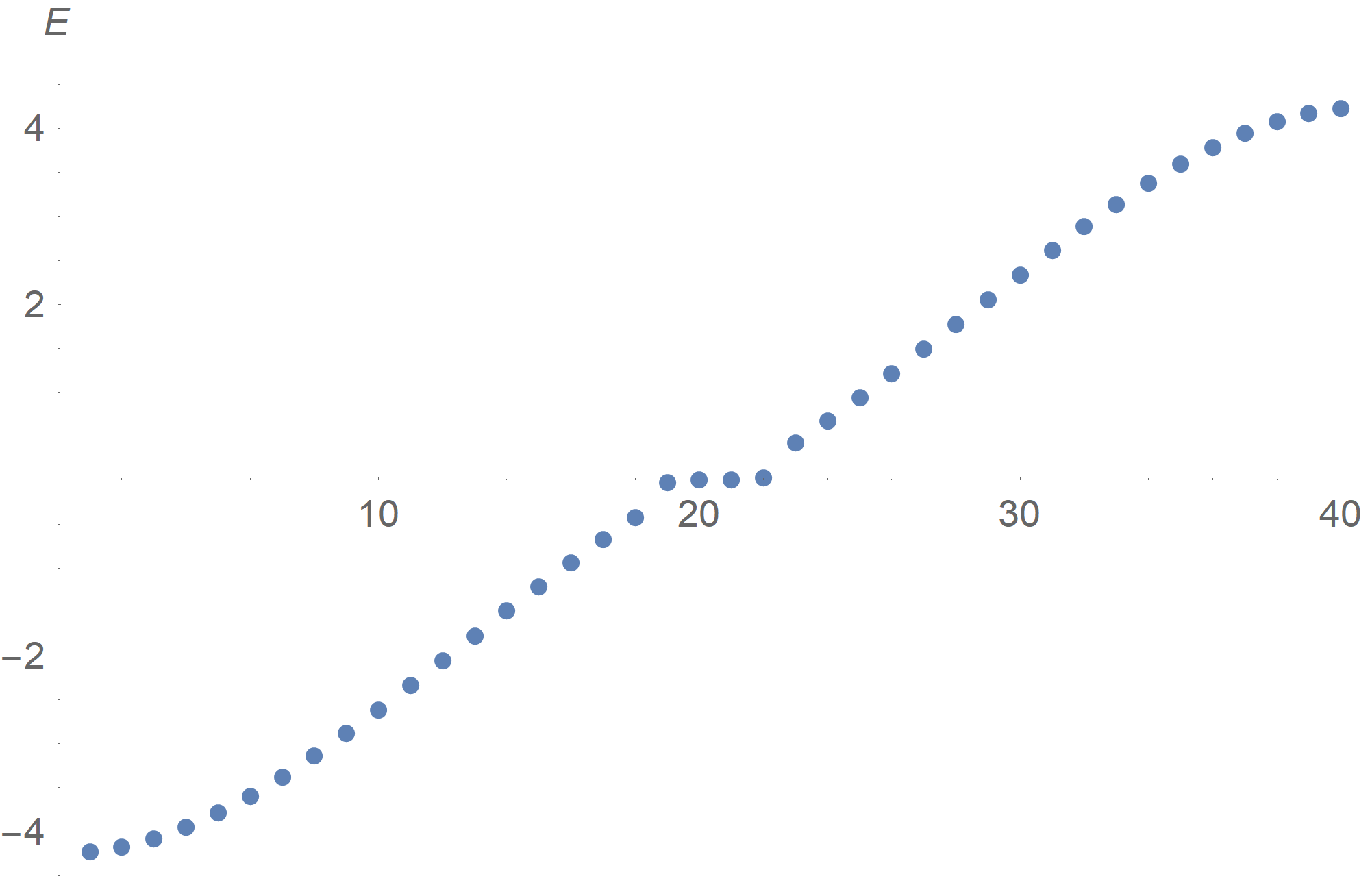}}\\
\subfloat[]{\includegraphics[width=0.30\textwidth]{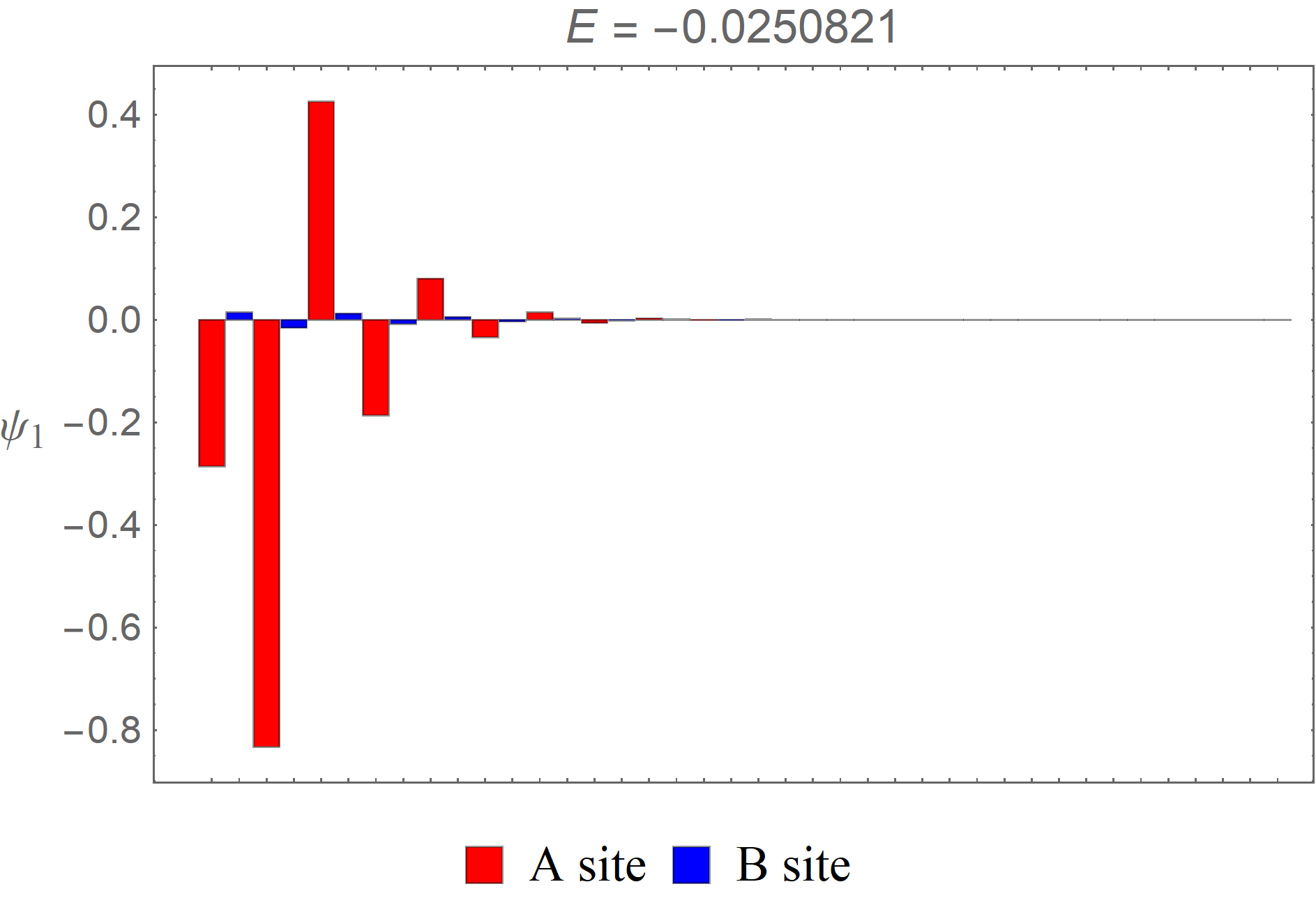}}\hskip 0.5cm
\subfloat[]{\includegraphics[width=0.30\textwidth]{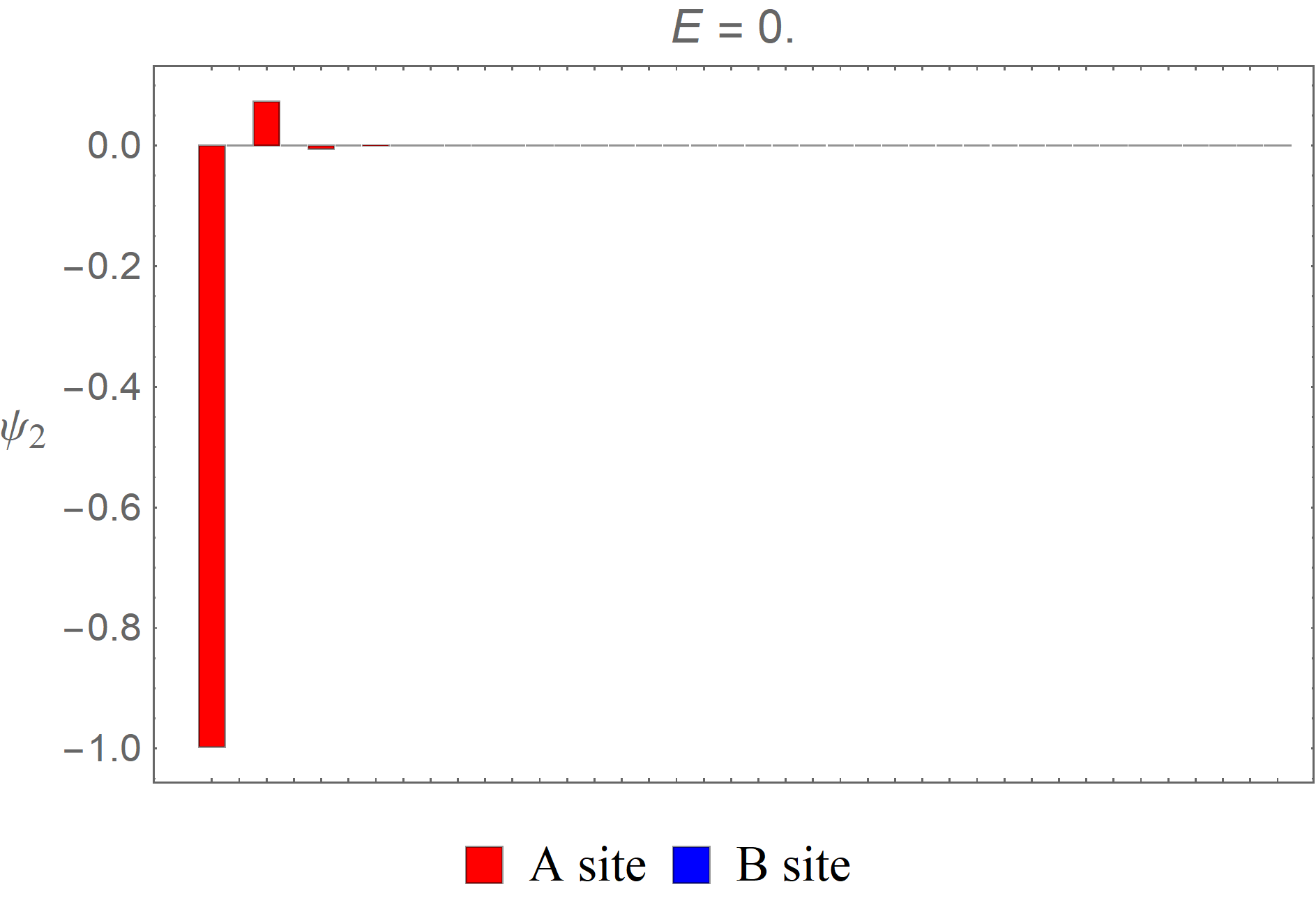}}\\
\subfloat[]{\includegraphics[width=0.30\textwidth]{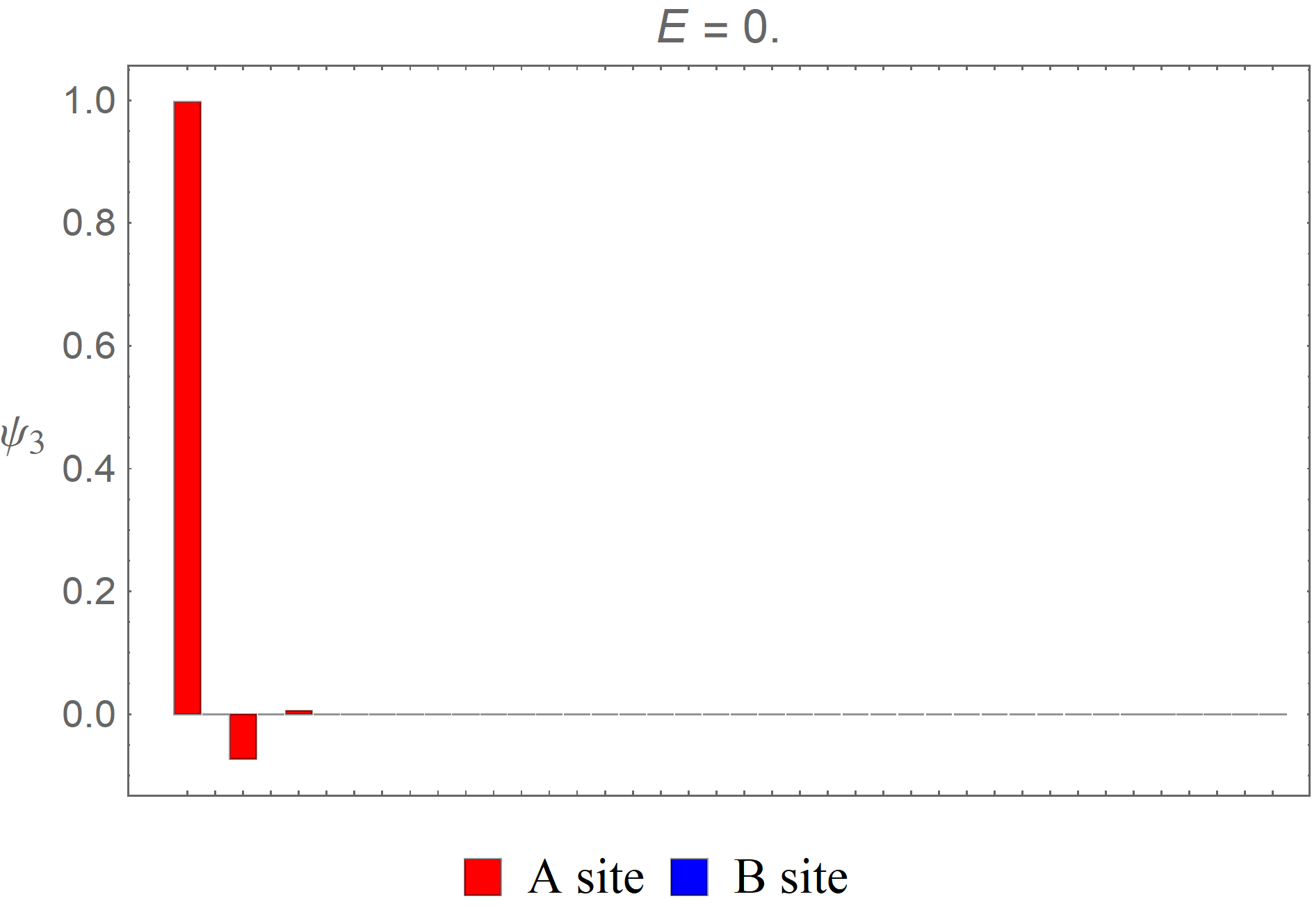}}\hskip 0.5cm
\subfloat[]{\includegraphics[width=0.30\textwidth]{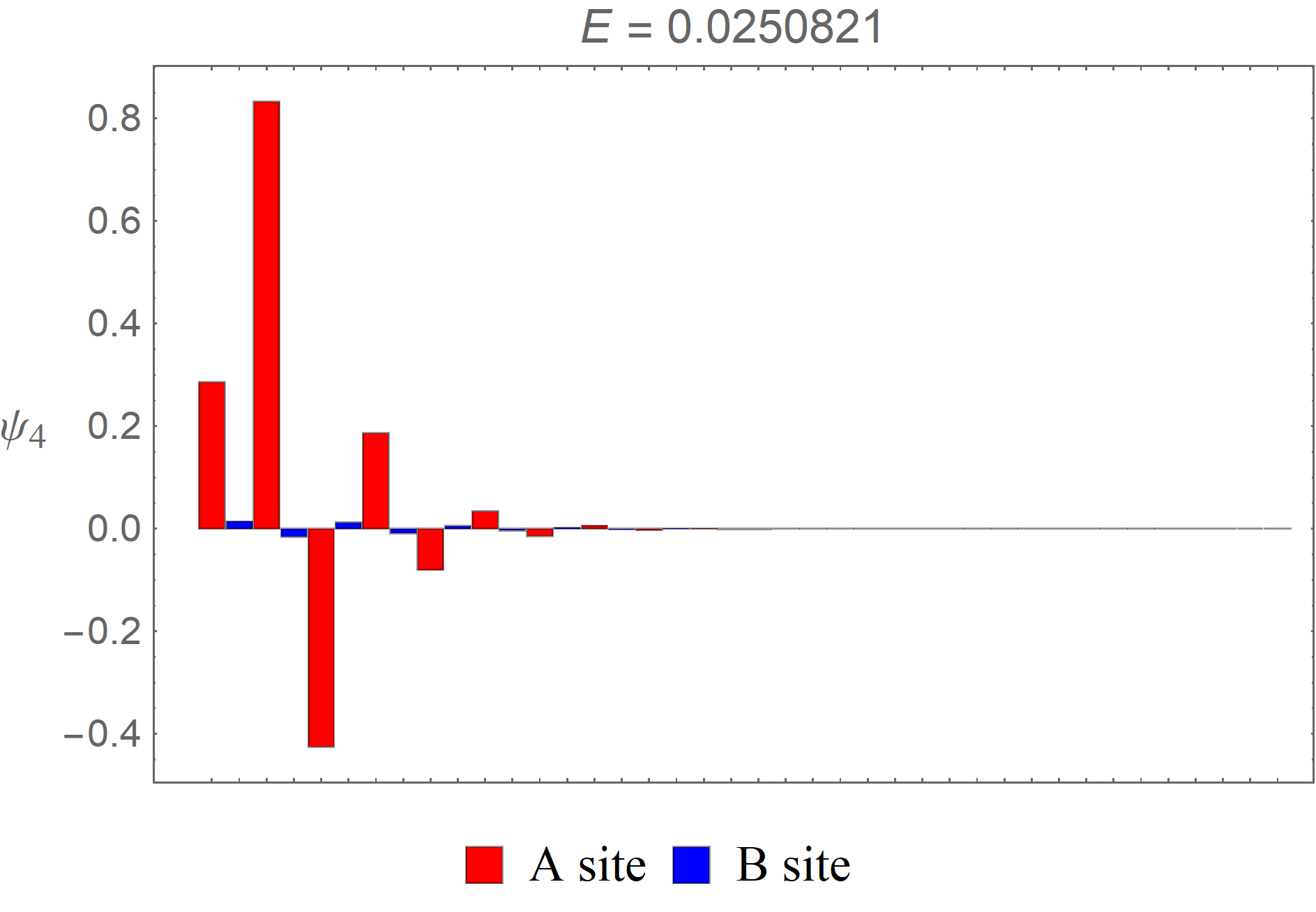}}\\
\subfloat[]{\includegraphics[width=0.30\textwidth]{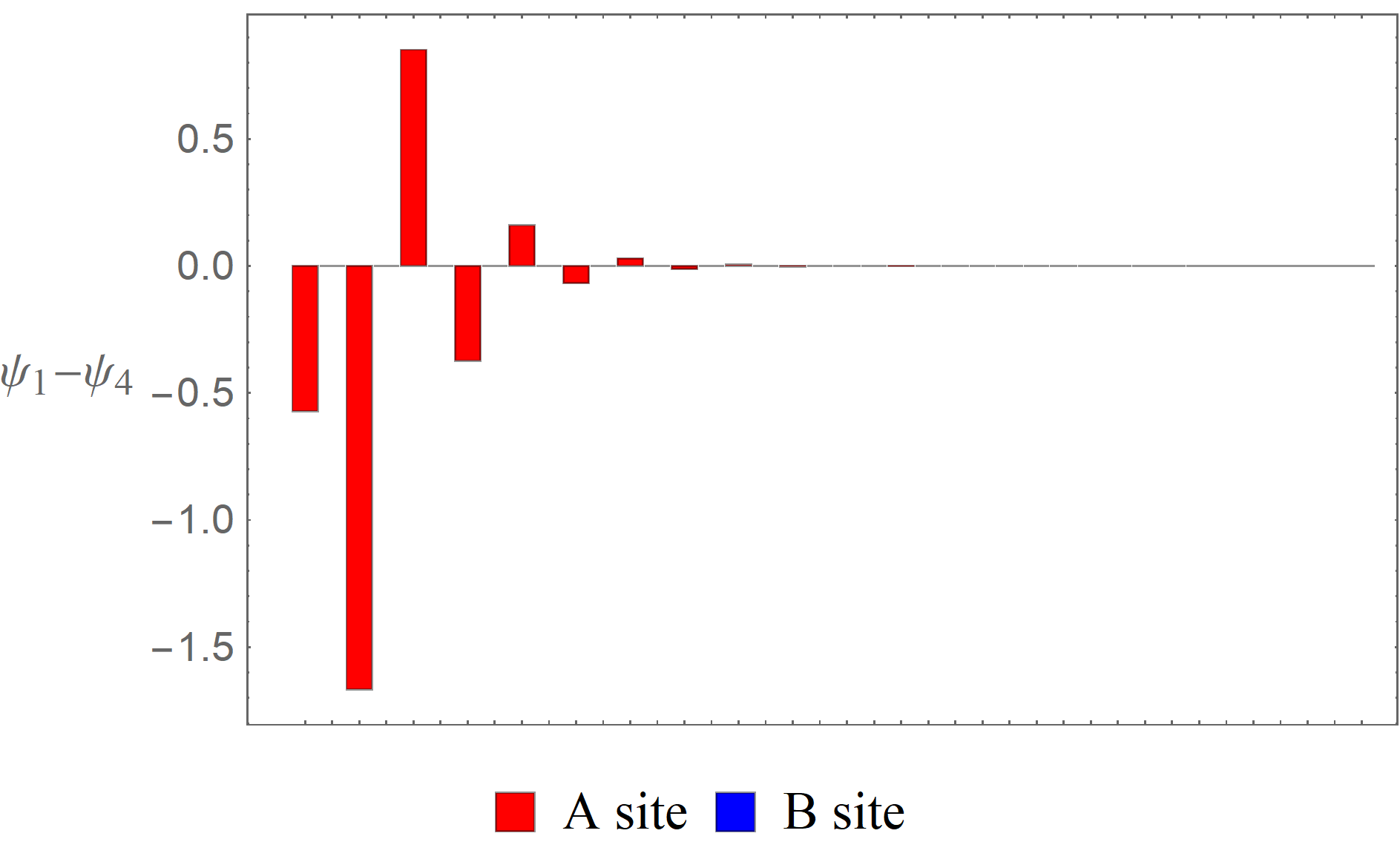}}\hskip 0.5cm
\subfloat[]{\includegraphics[width=0.30\textwidth]{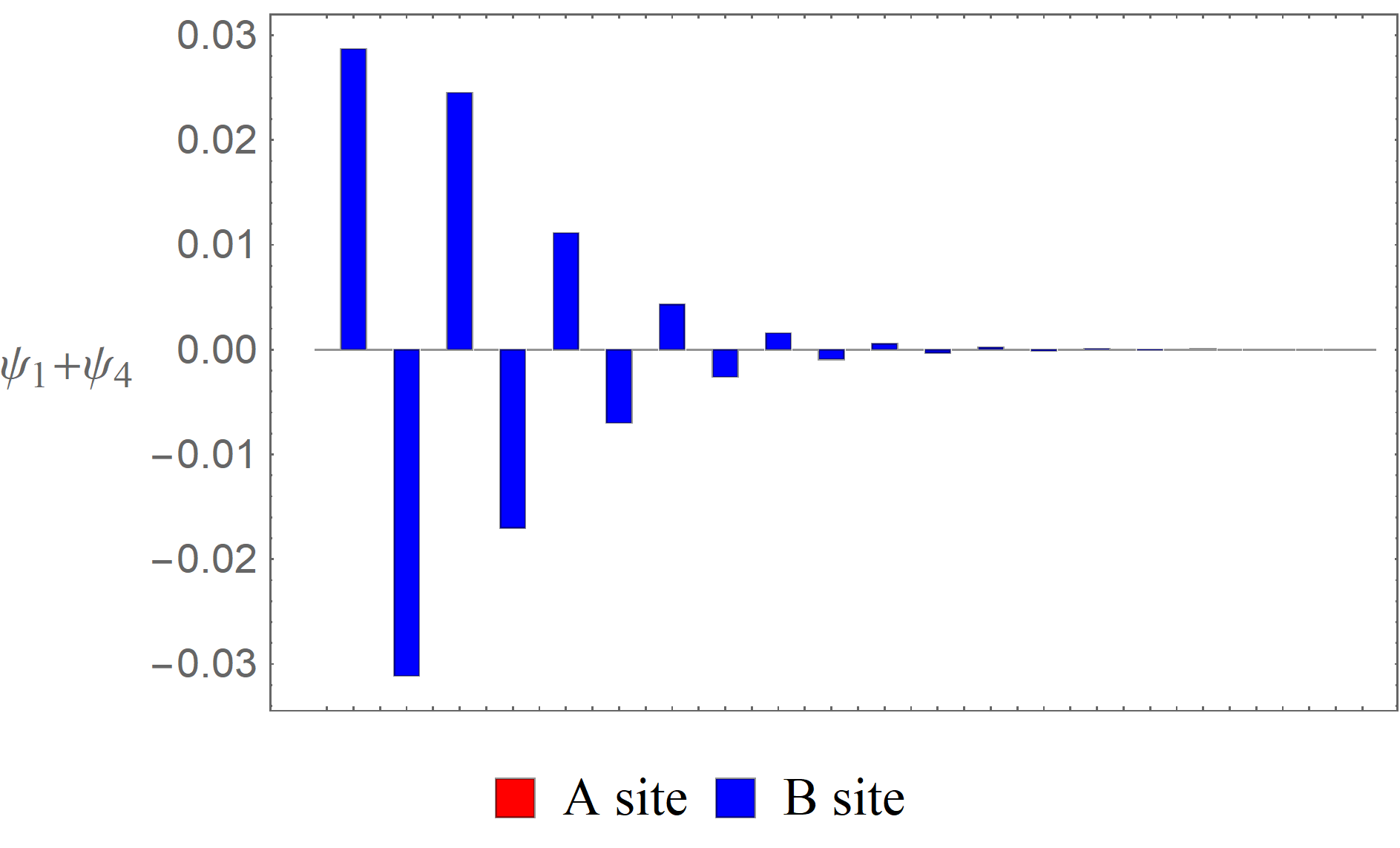}}\\
\subfloat[]{\includegraphics[width=0.30\textwidth]{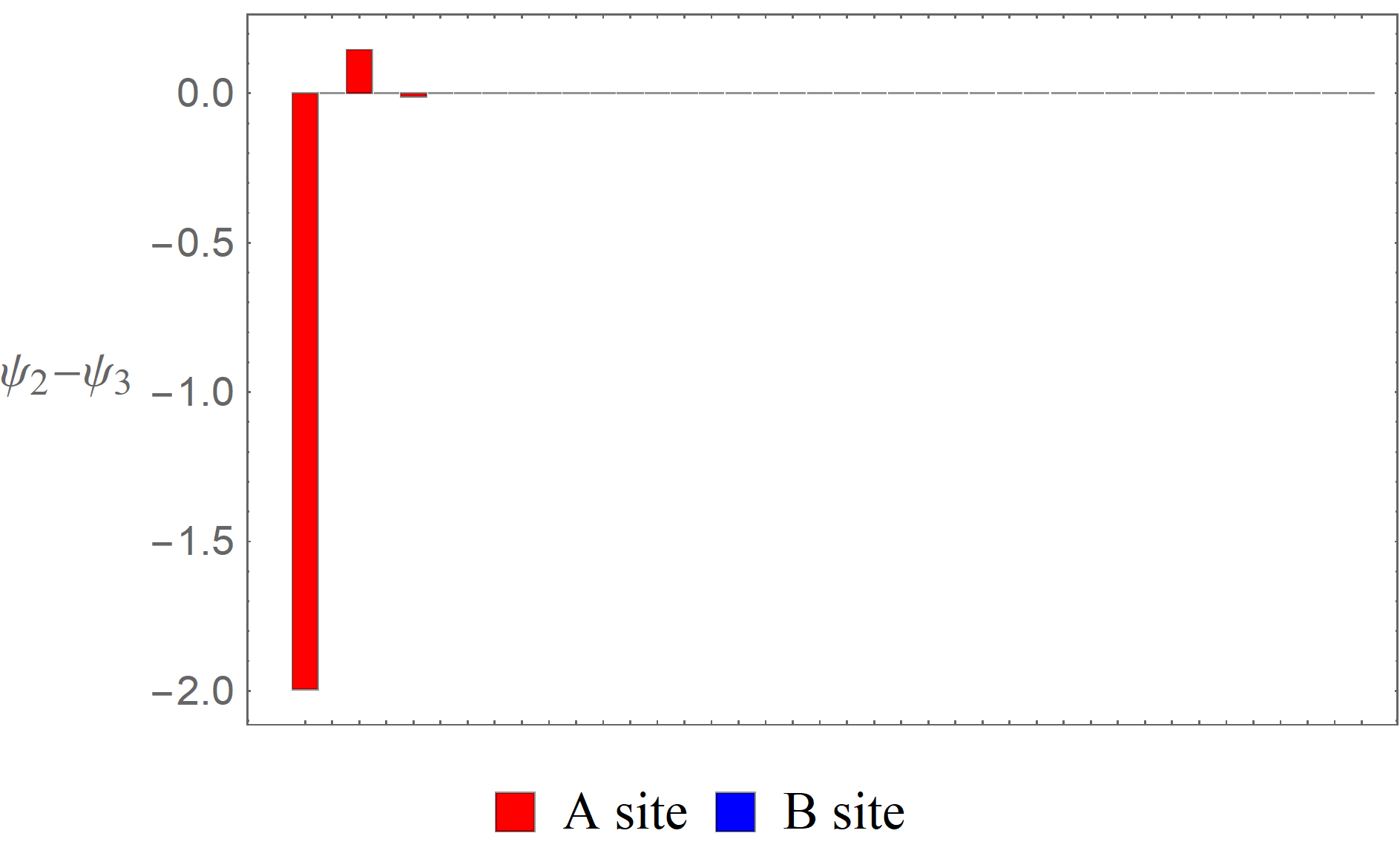}}\hskip 0.5cm
\subfloat[]{\includegraphics[width=0.30\textwidth]{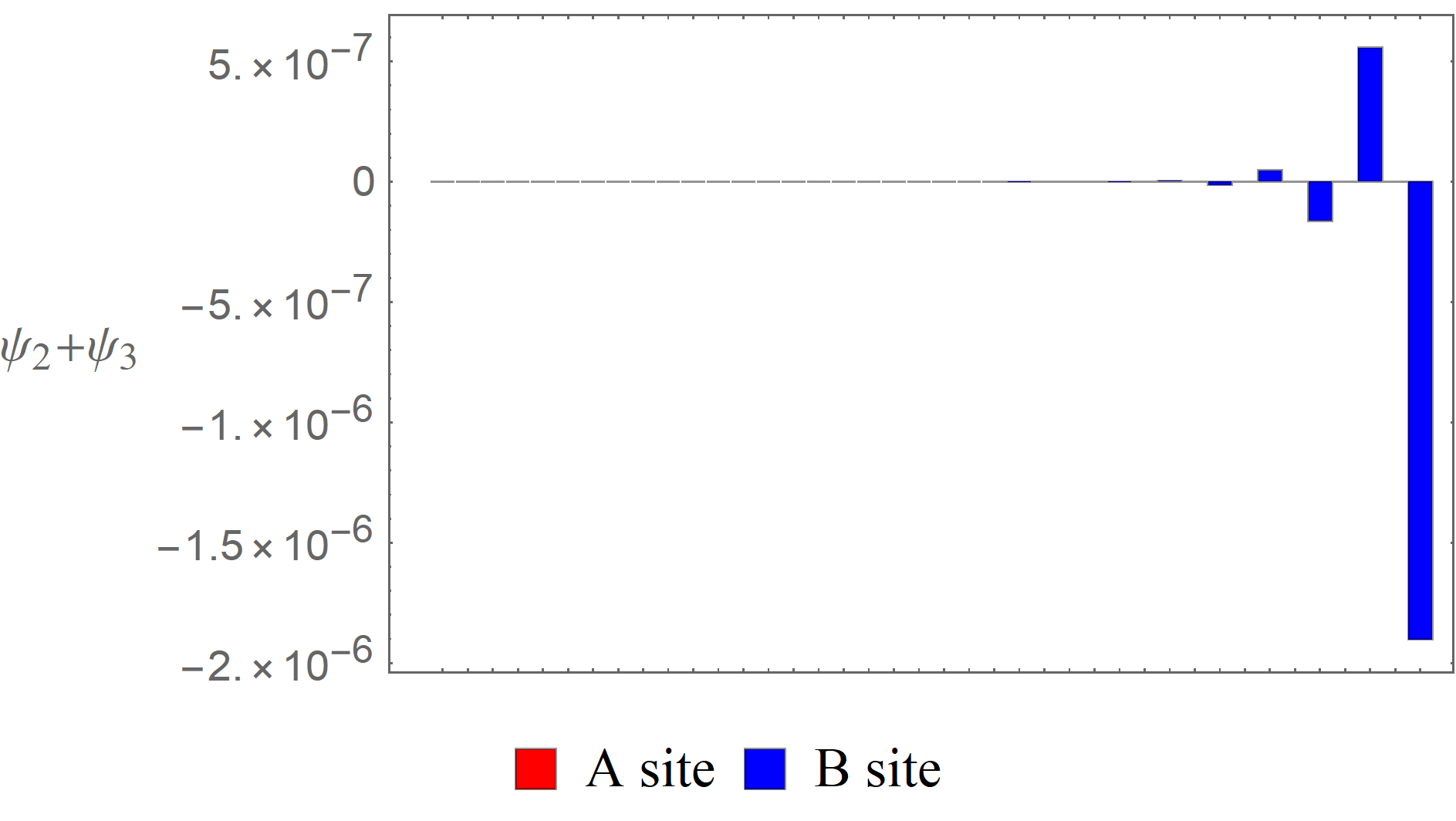}}\\
\caption{(a) The energy spectrum of the NH type 1 extended SSH model with $40$ sites, where $\left( t_0^L, t_0^R, t_1^L, t_1^R,, t_2^L, t_2^R \right) = (1/2, 1/8, 2, 2, 4, 1)$. Thus,  $(\tb_0, \tb_1, \tb_2) =(1/4, 2, 2)$ and it is in the topological phase with $\nb = 2$. (b), (c), (d), (e) The wave functions of the edge states $\psi_1, \psi_2, \psi_3, \psi_4$, which all have approximately zero energies.  (f), (g) The difference and sum and of $(\psi_1, \psi_4)$.  (h), (i) The difference and sum of $(\psi_2, \psi_3)$. Note that there are three and one edge states on the left and right boundaries, respectively. This is consistent with the fact that $\n_E = 1$.}  \label{fig2 nHSSH-ext1}
\end{figure}

With caution, all the previous predictions may be confirmed numerically. First, we consider the case that there are $40$ sites with the parameters chosen to be $\left( t_0^L, t_0^R, t_1^L, t_1^R, t_2^L, t_2^R \right) = (1/2, 1/8, 2, 2, 4, 1)$. Hence, $(\tb_0, \tb_1, \tb_2) =(1/4, 2, 2)$, $\nb = 2$, $r=1/2$ and $(\n_E^L,  \n_E^R) = (2, -1)$. There are four edge states, which we call $\psi_1, \psi_2, \psi_3, \psi_4$ in the order of increasing energy, with $(\psi_1, \psi_4)$ and $(\psi_2, \psi_3)$ forming two chiral conjugate pairs. They all have almost zero energies and there are indeed three on the left boundary and one on the right.  However, due to the limitation of numerical accuracy, it is difficult to see this directly from their wave functions. To see this explicitly, we must consider the following linear combination: $\psi_1 - \psi_4$, $\psi_1 + \psi_4$, $\psi_2 - \psi_3$, $\psi_2 + \psi_3$. From these linear combinations of wave functions, we also see that two of them are non-vanishing on the A sites only and two on the B sites only. Similar to the SSH case, this is evidence that they are derived from the Hermitian extended SSH model. The energy spectrum and wave functions of the edge states are shown in Fig.~\ref{fig2 nHSSH-ext1}.

\begin{figure}[hbt!]
\centering
\subfloat[]{\includegraphics[width=0.50\textwidth]{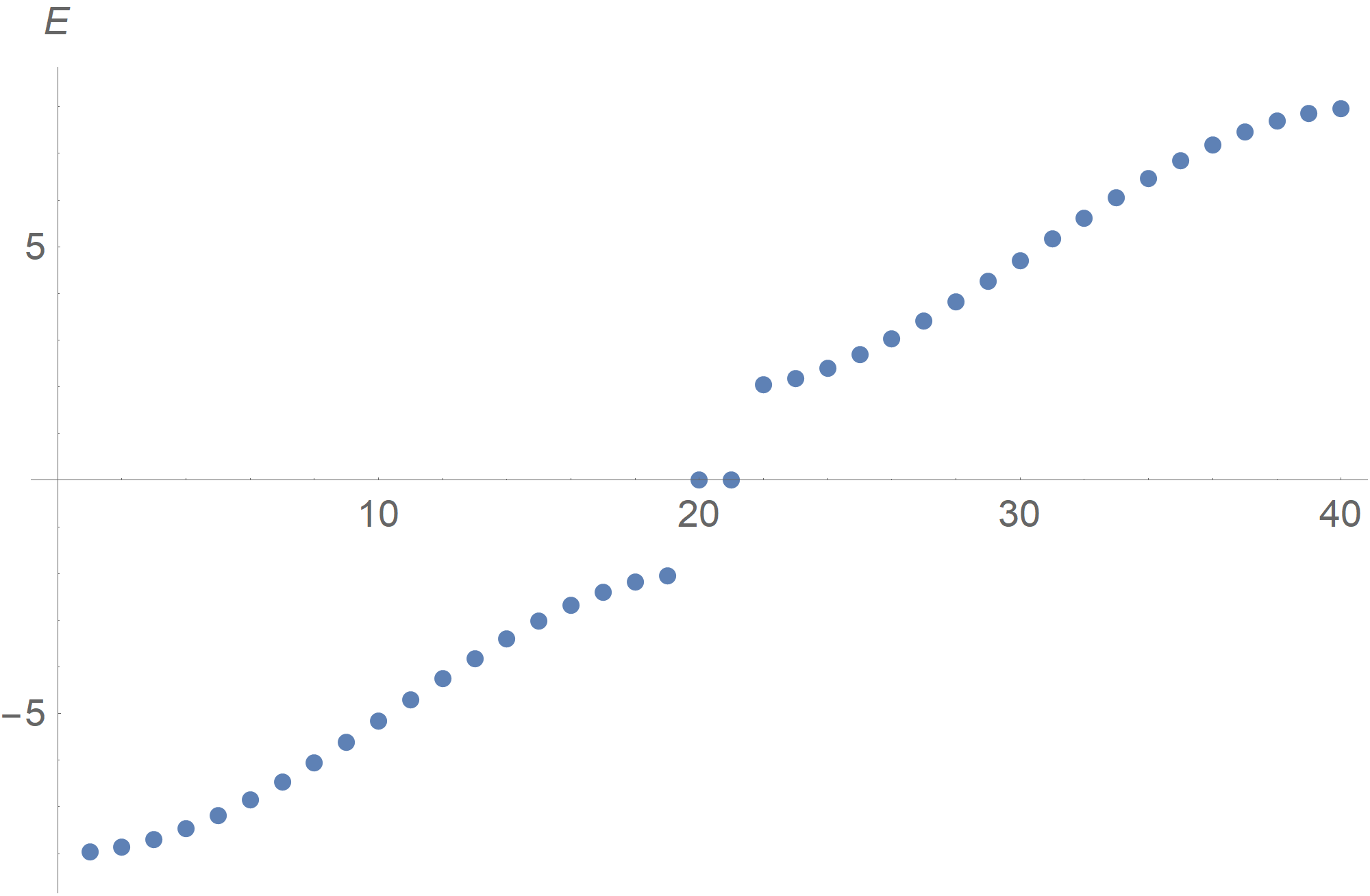}}\\
\subfloat[]{\includegraphics[width=0.30\textwidth]{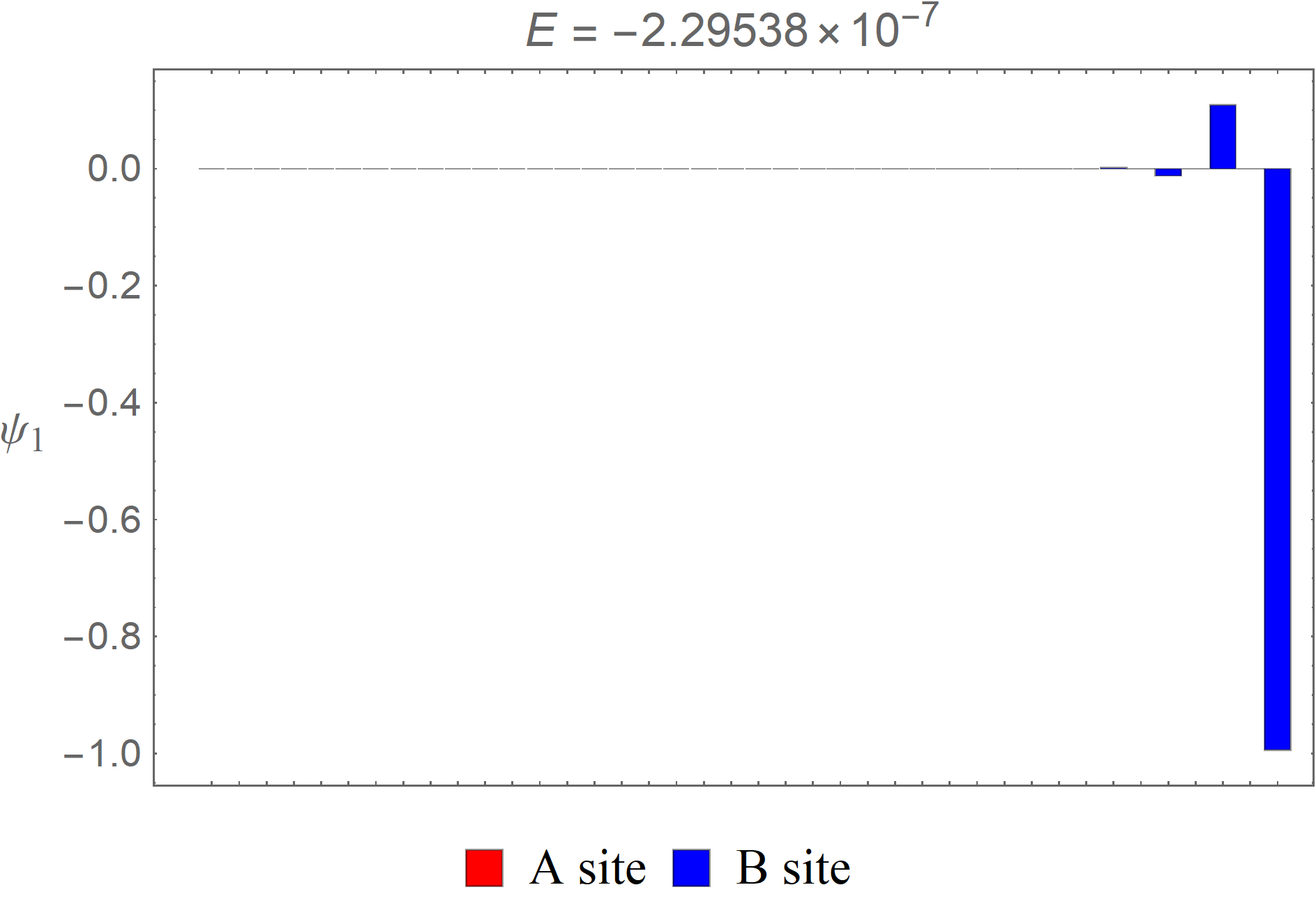}}\hskip 0.5cm
\subfloat[]{\includegraphics[width=0.30\textwidth]{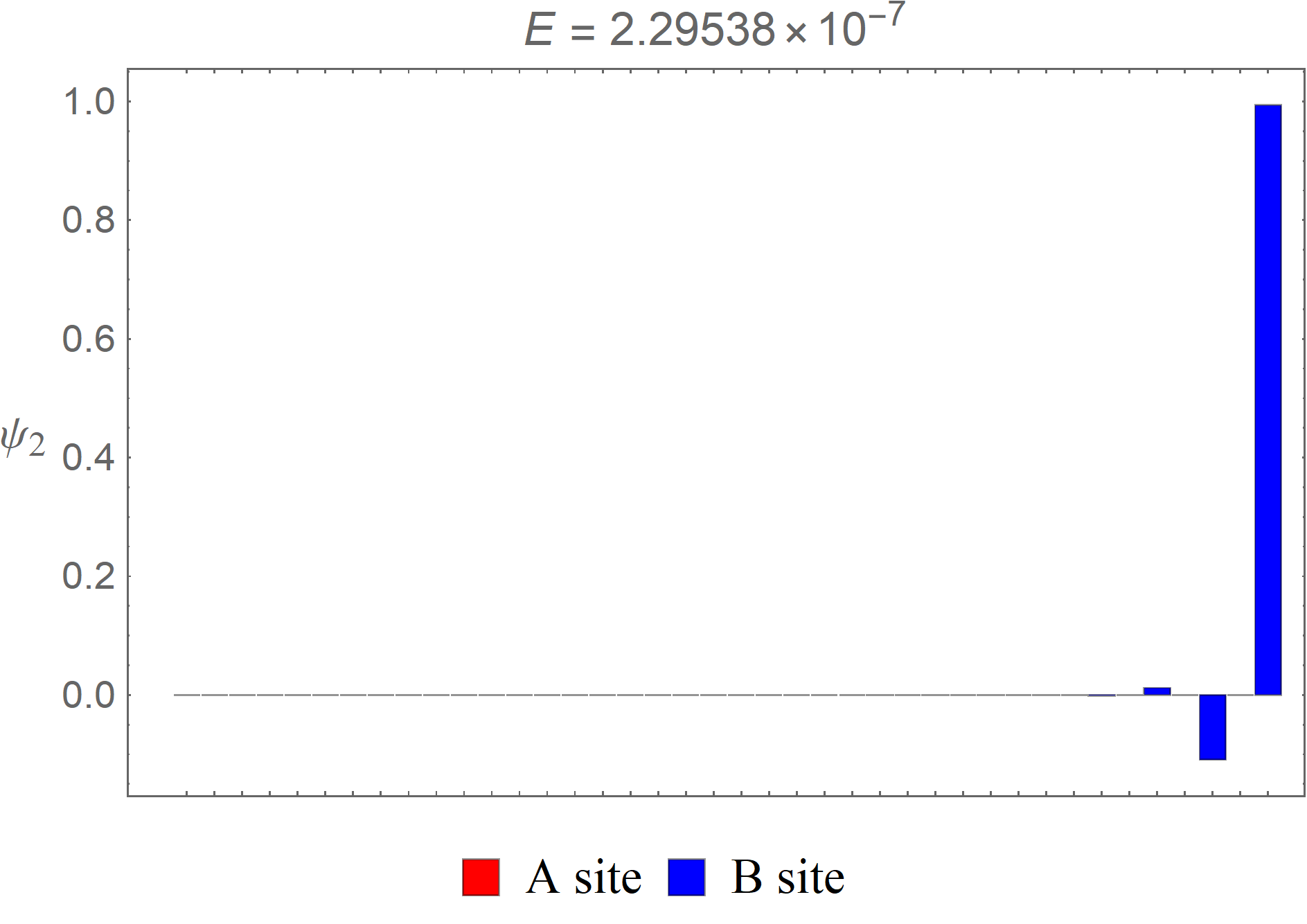}}\\
\subfloat[]{\includegraphics[width=0.30\textwidth]{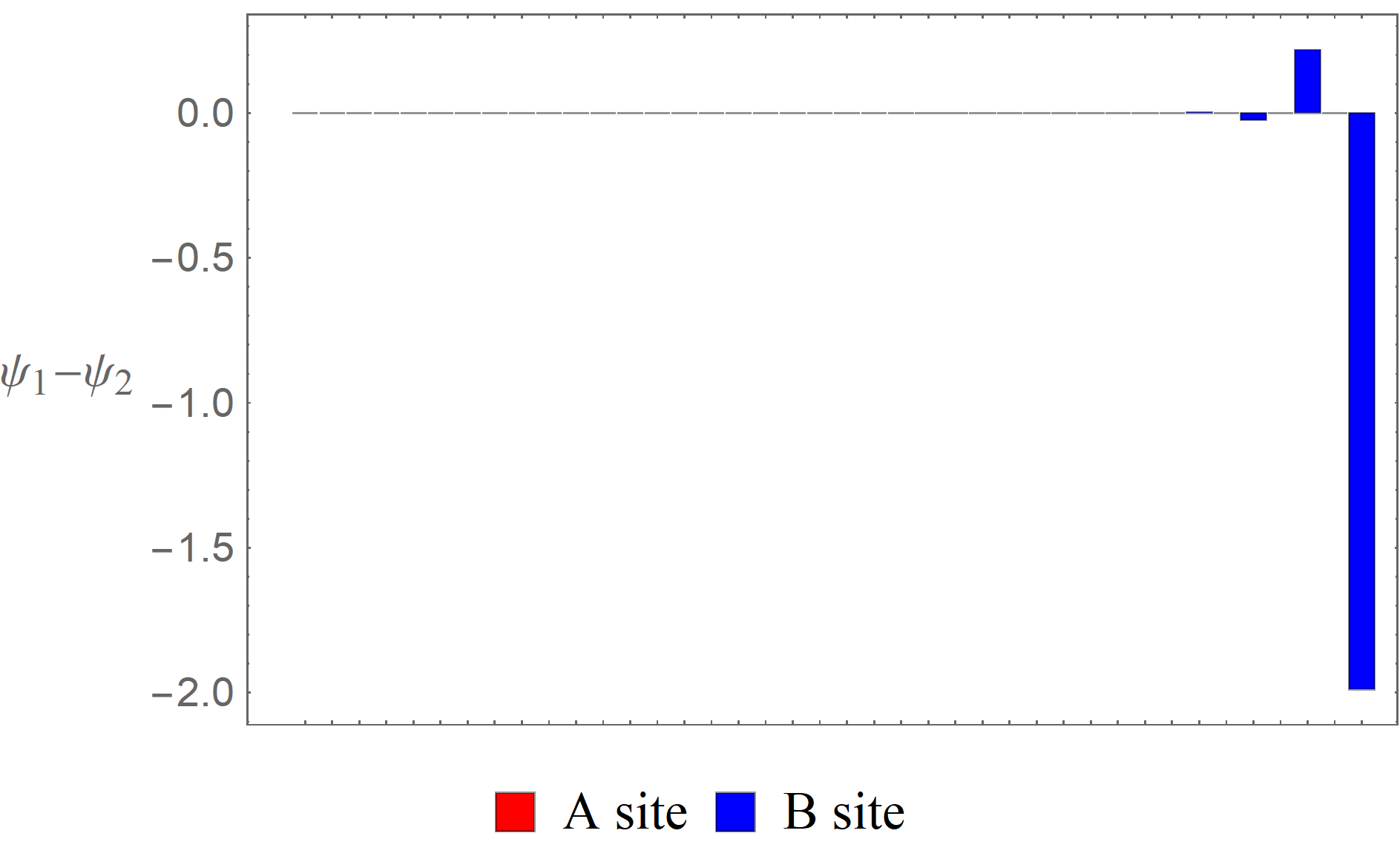}}\hskip 0.5cm
\subfloat[]{\includegraphics[width=0.30\textwidth]{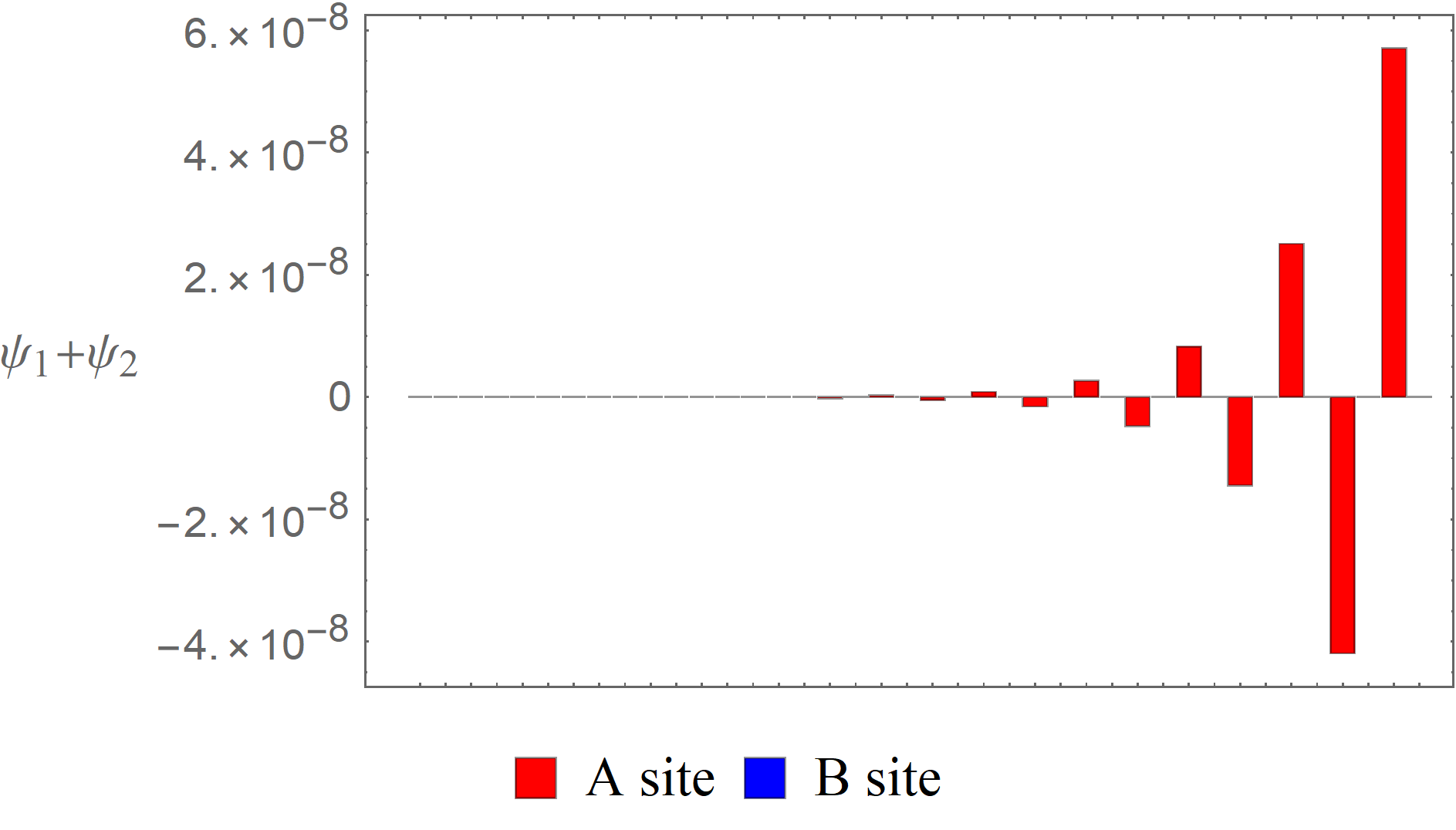}}\\
\caption{(a)The energy spectrum of the NH type 1 extended SSH model with $40$ sites, where $\left( t_0^L, t_0^R, t_1^L, t_1^R, t_2^L, t_2^R \right) = (1/2, 8, 5, 5, 1/4, 4)$. Thus, $(\tb_0, \tb_1, \tb_2) = (2, 5, 1)$ and it is in the topological phase with $\nb = 1$. The energies of the two edge states $\psi_1, \psi_2$ are both approximately zero.  (b), (c) The wave functions of the two edge states in the system. (d), (e) The difference and sum of $\psi_1$ and $\psi_2$.  Note that both edge states are on the right boundary, consistent with the fact that $(\n_E^L, \n_E^R) = (0, -1)$.}  \label{fig3 nHSSH-ext1}
\end{figure}

Next, we choose the parameters to be $\left( t_0^L, t_0^R, t_1^L, t_1^R, t_2^L, t_2^R \right) = (1/2, 8, 5, 5, 1/4, 4)$ so that $(\tb_0, \tb_1, \tb_2) = (2, 5, 1)$, $\nb = 1$, $r=4$, and $(\n_E^L, \n_E^R) = (0, -1)$. There are two edge states, denoted by $\psi_1$ and $\psi_2$, in the order of increasing energy.  Again, they form a chiral conjugate pair.  Due to the skin effect, they are both right edge states and seem to be non-vanishing only on the B sites. By taking the difference and sum of $\psi_1$ and $\psi_2$, we indeed see that both edge states are on the right boundary. Again, one of the resultant wave functions is non-vanishing only on the A sites and the other non-vanishing on the B sites only. The energy spectrum and wave functions of the edge states are shown in Fig.~\ref{fig3 nHSSH-ext1},

A similar analysis may also be carried out for the NH type 2 extended SSH model satisfying the
QHC. This kind of NH extended SSH model has been used extensively to investigate the general BZ in the literature~\cite{Skin-effect4, GBZ4, GBZ5}.  We relegate the details to Appendix A to avoid repetition in the main text.

We may also consider the case that there are an odd number of sites in the NH extended SSH models so that there is one incomplete unit cell by the right boundary.  Since this would change the BC on the right boundary, it would also change the number of edge states.  Similarly, the details of the analysis are deferred to Appendix B.

\subsection{IV. General NH extended SSH models}

Let's now consider a general NH type 1 extended SSH model. By introducing the variables $s_{12}= \sqrt{s_1 s_2}$, \hfill\break
$u_{12}= \left( s_1 + s_2 \right)/(2 s_{12})$, and $s_{34}=\sqrt{s_3 s_4}$, $u_{34}= \left( s_3 + s_4 \right)/(2 s_{34})$, the characteristic equation may be written in terms of these variables:
\bea
&\;& \hskip -3.1cm \biggl\{
- t_0^L s_{12}^{N-3} s_{34}^{N+1} U_{N+1} (u_{34}) \left[ t_0^L s_{12}^2 U_{N+1} (u_{12})
+ t_1^R s_{12} U_{N}(u_{12}) + t_2^R U_{N-1} (u_{12}) \right]
\nonumber \\
&\;& \hskip -2.85cm + t_0^L s_{12}^{N-2} s_{34}^{N} U_{N+2} (u_{34}) \left[ t_0^L s_{12}^2 U_{N} (u_{12})
+ t_1^R s_{12} U_{N-1}(u_{12}) + t_2^R U_{N-2} (u_{12}) \right]
\nonumber \\
&\;& \hskip -2.85cm + s_{12}^{N-3} s_{34}^{N} U_{N} (u_{34}) \left[ (t_0^L)^2 s_{12}^3 U_{N+2} (u_{12})
+ \left(t_0^L t_2^R - (t_1^R)^2 \right) s_{12} U_{N}(u_{12}) - t_1^R t_2^R U_{N-1} (u_{12}) \right]
\nonumber \\
&\;& \hskip -2.85cm - s_{12}^{N-2} s_{34}^{N-1} U_{N+1} (u_{34}) \left[ (t_0^L)^2 s_{12}^3 U_{N+1} (u_{12})
+\left(t_0^L t_2^R - (t_1^R)^2 \right) s_{12} U_{N-1}(u_{12}) - t_1^R t_2^R U_{N-2} (u_{12}) \right]
\nonumber \\
&\;& \hskip -2.85cm + s_{12}^{N-3} s_{34}^{N-1} U_{N-1} (u_{34}) \left[ t_0^L t_1^R  s_{12}^3 U_{N+2} (u_{12})
- \left(t_0^L t_2^R - (t_1^R)^2 \right) s_{12}^2 U_{N+1}(u_{12}) - (t_2^R)^2 U_{N-1} (u_{12}) \right] \nonumber \\
&\;& \hskip -2.85cm - s_{12}^{N-2} s_{34}^{N-2} U_{N} (u_{34}) \left[ t_0^L t_1^R  s_{12}^3 U_{N+1} (u_{12})
- \left(t_0^L t_2^R - (t_1^R)^2 \right) s_{12}^2 U_{N}(u_{12}) - (t_2^R)^2 U_{N-2} (u_{12}) \right]
\nonumber \\
&\;& \hskip -2.85cm + t_2^R s_{12}^{N-2} s_{34}^{N-2} U_{N-2} (u_{34}) \left[ t_0^L s_{12}^2 U_{N+2} (u_{12})
+ t_1^R s_{12} U_{N+1}(u_{12}) + t_2^R U_{N} (u_{12}) \right]
\nonumber \\
&\;& \hskip -2.85cm - t_2^R s_{12}^{N-1} s_{34}^{N-3} U_{N-1} (u_{34}) \left[ t_0^L s_{12}^2 U_{N+1} (u_{12})
+ t_1^R s_{12} U_{N}(u_{12}) + t_2^R U_{N-1} (u_{12}) \right] \biggr\}
\nonumber \\
&\;& \hskip -3.1cm + s_{12}^{-2} s_{34}^{2N-2} \biggl\{ (t_2^R)^2 + 2 t_1^R t_2^R s_{34} u_{34}
+ s_{34}^{2} \bigl[ (t_1^R)^2 + 2 t_0^L t_2^R (2u_{34}^2 -1) \bigr]
+ 2 t_0^L t_1^R s_{34}^{3} u_{34} + (t_0^L)^2 s_{34}^{4} \biggr\}
\nonumber\\
&\;& \hskip -3.1cm + s_{12}^{2N-2} s_{34}^{-2} \biggl\{ (t_2^R)^2 + 2 t_1^R t_2^R s_{12} u_{12}
+ s_{12}^{2} \bigl[ (t_1^R)^2 + 2 t_0^L t_2^R (2u_{12}^2 -1) \bigr]
+ 2 t_0^L t_1^R s_{12}^{3} u_{12} + (t_0^L)^2 s_{12}^{4} \biggr\} = 0.
\eea

From the second and fourth relations in eq.~(\ref{Prod and sum 1A}), we may express $u_{12}, u_{34}$ in terms of $s_{12}, s_{34}$:
\bea \label{u12 and u34 type 1}
&\;& \hskip -3.1cm u_{12} = \frac{s_{12} s_{34} \left\{ t_0^L t_2^L  \left(t_0^R t_1^R  + t_1^L t_2^R  \right) s_{34} - t_0^R t_2^R  \left(t_0^L t_1^L  + t_1^R t_2^L  \right) s_{34} ^{-1} \right\} }{2t_0^L t_0^R t_2^L t_2^R \left(s_{12}^2 - s_{34}^2 \right) }, \nonumber \\
&\;& \hskip -3.1cm u_{34} = \frac{- s_{12} s_{34} \left\{ t_0^L t_2^L  \left(t_0^R t_1^R  + t_1^L t_2^R  \right) s_{12} - t_0^R t_2^R  \left(t_0^L t_1^L  + t_1^R t_2^L  \right) s_{12} ^{-1} \right\} }{2t_0^L t_0^R t_2^L t_2^R \left(s_{12}^2 - s_{34}^2 \right)}.
\eea
By making use of eq.~(\ref{u12 and u34 type 1}), the characteristic equation may be written in terms of $s_{12}$ and $s_{34}$. Furthermore, in terms of the skin effect factor $r$ defined in eq.~(\ref{skin effect 1A}), we introduce  $\sb_{12} = s_{12}/r$ and $\sb_{34} = s_{34}/r=\sb_{12}^{-1}$.  Finally, the characteristic equation may be written in terms of one single variable $\ub =\left(\sb_{12} + \sb_{12}^{-1} \right)/2$.  It may be seen that the resulting expression is a polynomial in $\ub$ of degree $3N+4$. When $N=0$, we obtain a polynomial in $\ub$ of degree $4$, whose solutions do not correspond to any physical states. It is, in fact, a common factor of the characteristic equation for arbitrary $N$, and thus can always be factored out and ignored. The fact that the final result for the characteristic equation is a polynomial in $\ub$ of degree $3N$ is derived from the permutation symmetry of $s_1, \ldots, s_4$. Again, the redundancy may be removed by requiring $s_1 \ge s_2 \ge s_3 \ge s_4$.

\begin{figure}[hbt!]
\centering
\includegraphics[width=0.50\textwidth]{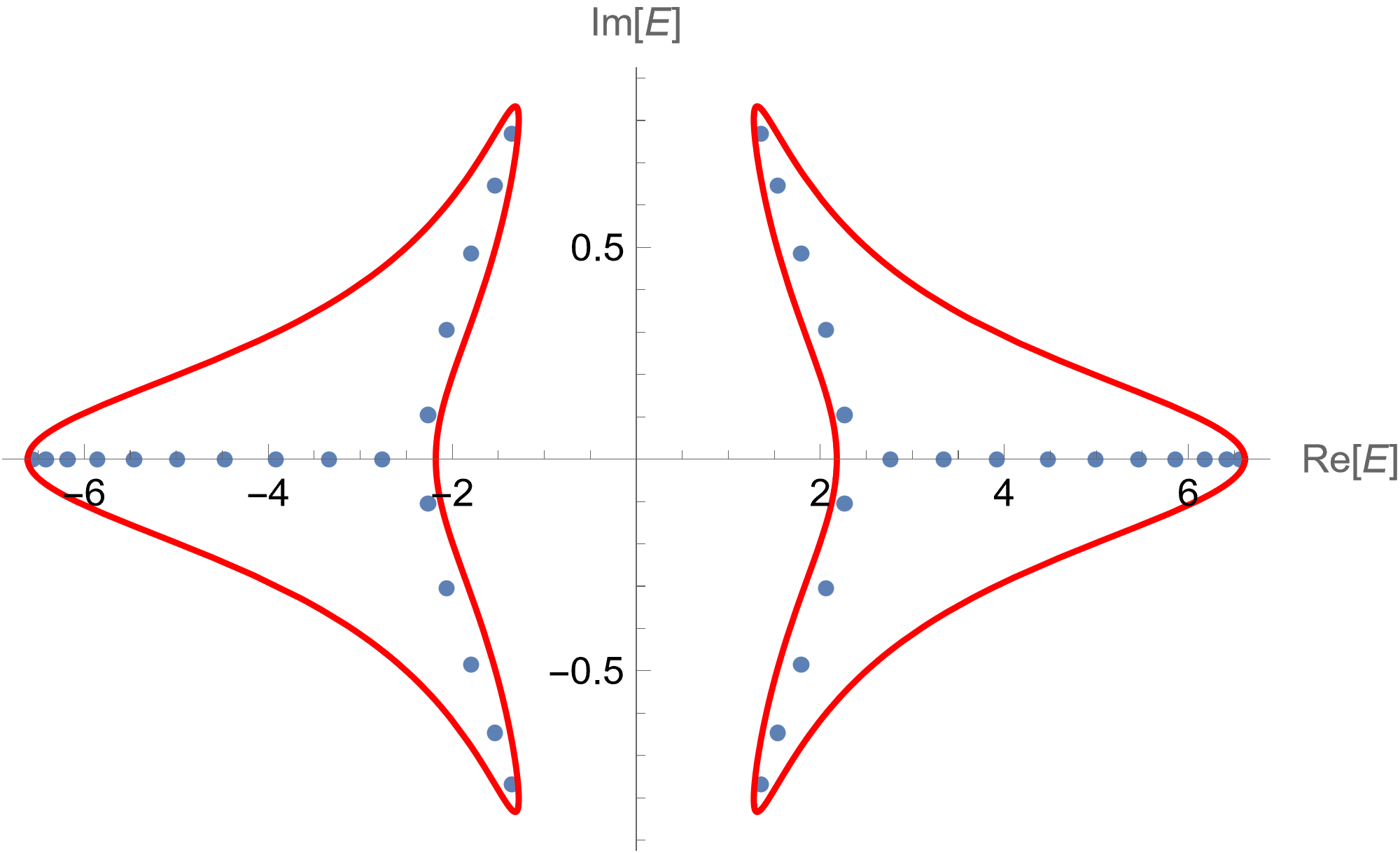}
\caption{The energy spectrum of the NH type 1 extended SSH model with $40$ sites and the trajectory of $E(p)$ on the complex plane. Here, $\left( t_0^L, t_0^R, t_1^L, t_1^R, t_2^L, t_2^R \right)$
$= (9/2, 2, 2, 2,1, 9/4)$. Thus, $(\tb_0, \tb_1, \tb_2) = (3, 2, 3/2)$ and it is in the trivial phase with $\nb = 0$. Note that $\n_E = 0$ and the energies of the eigenstates are encompassed by the trajectories of $E(p)$ on the complex plane.} \label{fig4 nHSSH-ext1}
\end{figure}

Let's first choose the parameters to be $\left( t_0^L, t_0^R, t_1^L, t_1^R, t_2^L, t_2^R \right) = (9/2, 2, 2, 2,1, 9/4)$ so that $(\tb_0, \tb_1, \tb_2) = (3, 2, 3/2)$, $\nb = 0$, $r=1$, and $\n_E = 0$. There are no edge states and we only show the energy spectrum along with the trajectory of $E(p)$ on the complex plane in Fig.~\ref{fig4 nHSSH-ext1}. 

\begin{figure}[hbt!]
\centering
\subfloat[]{\includegraphics[width=0.50\textwidth]{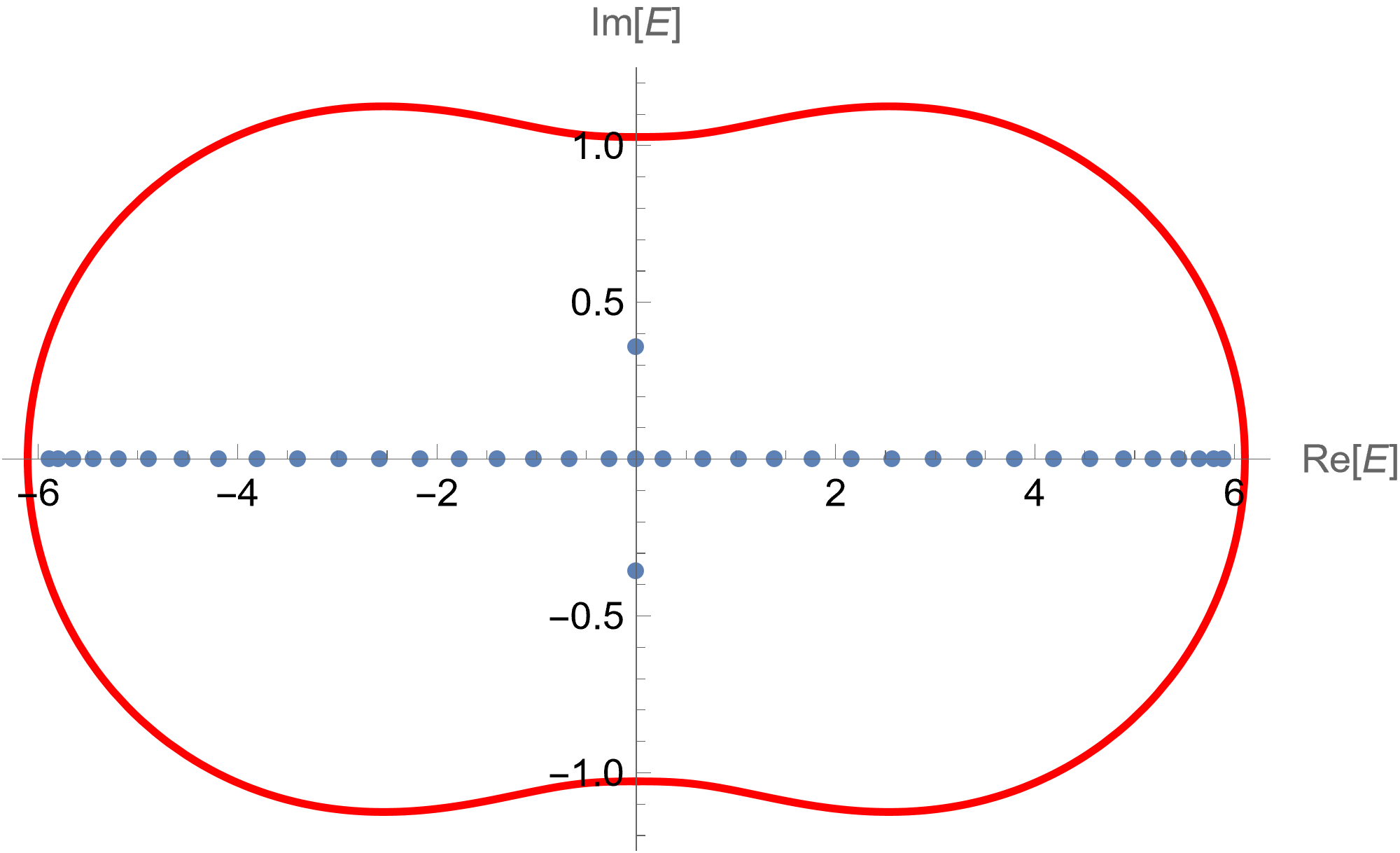}}\\
\subfloat[]{\includegraphics[width=0.30\textwidth]{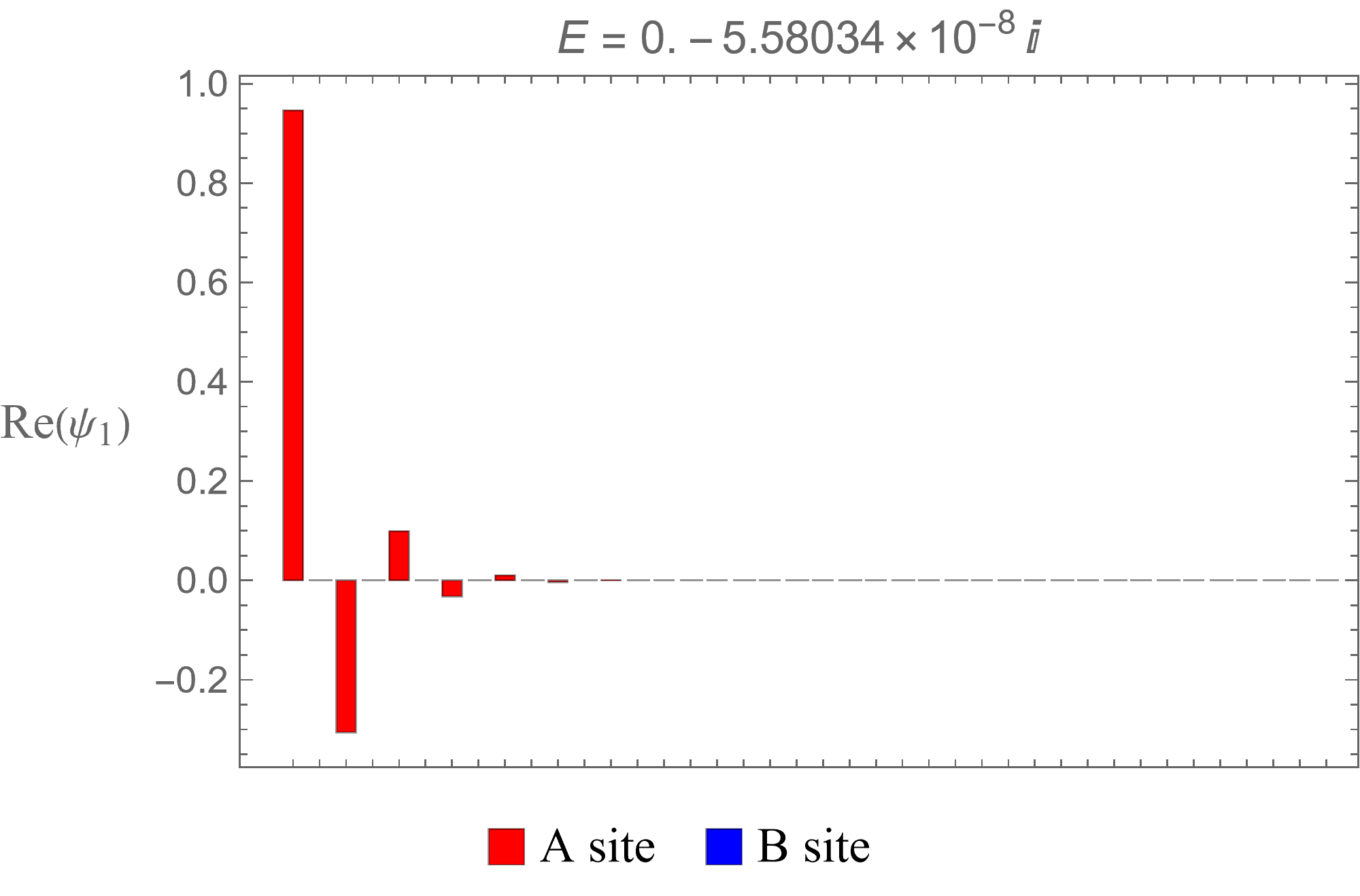}}\hskip 0.5cm
\subfloat[]{\includegraphics[width=0.30\textwidth]{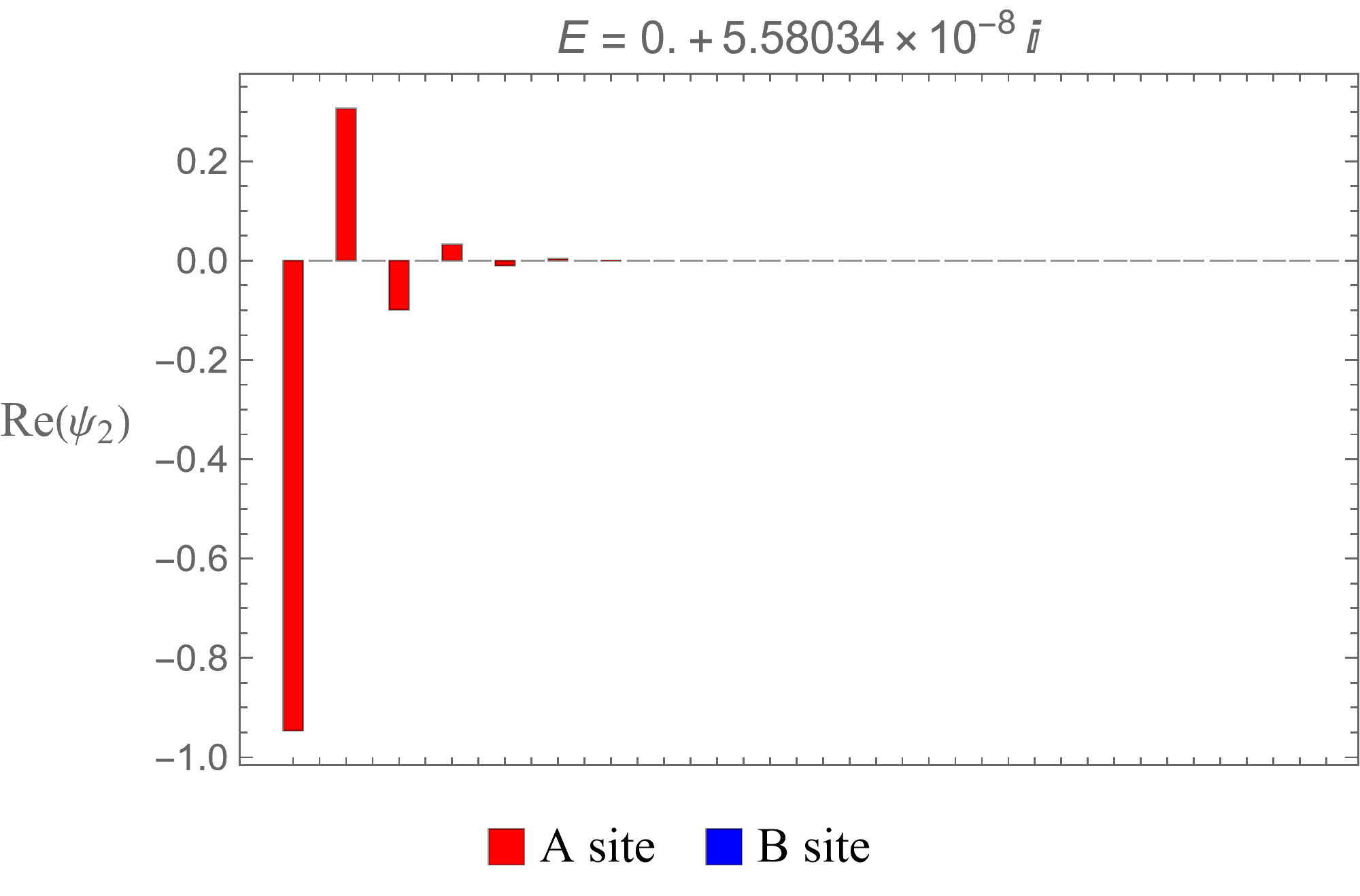}}\\
\subfloat[]{\includegraphics[width=0.30\textwidth]{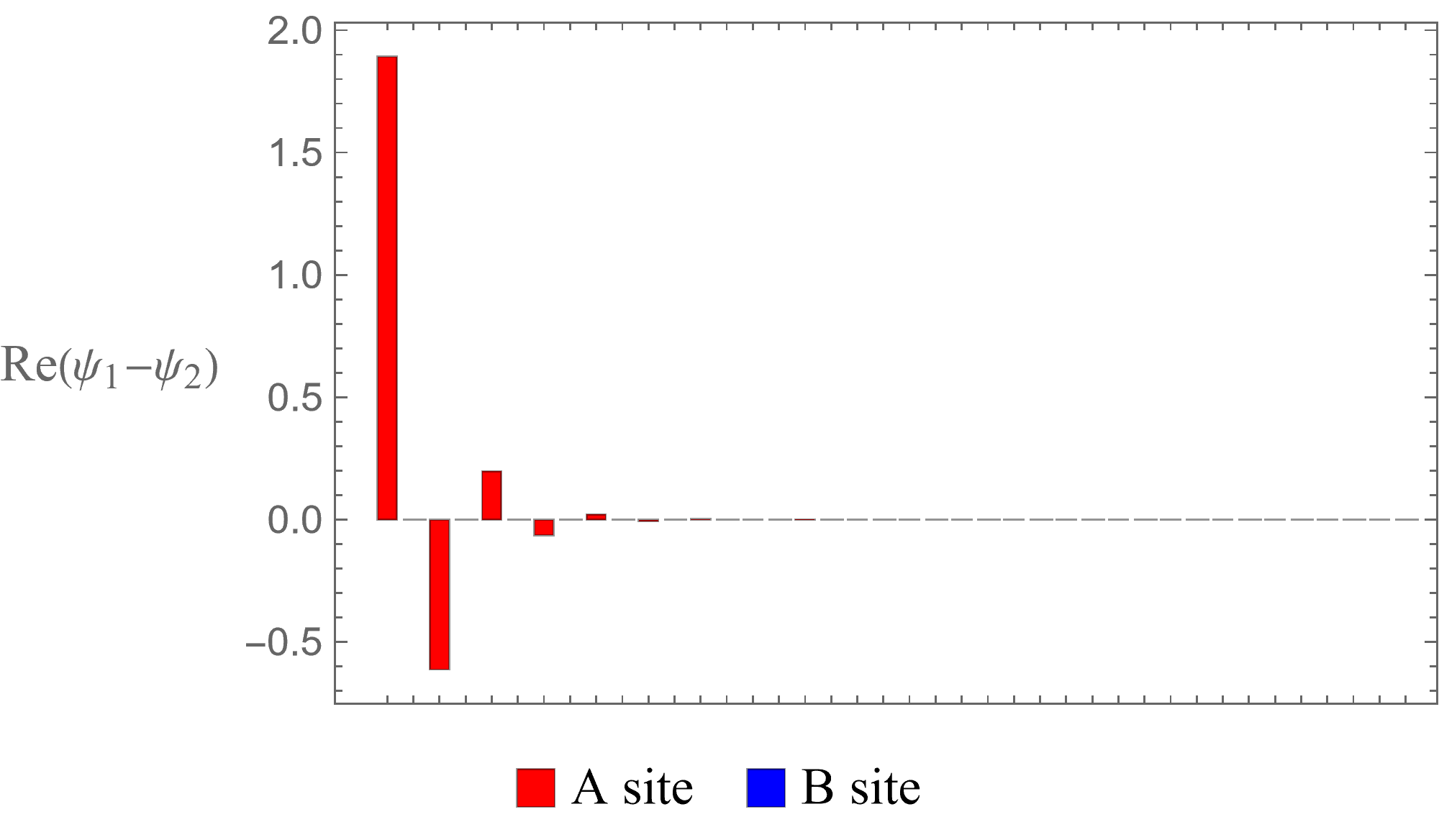}}\hskip 0.5cm
\subfloat[]{\includegraphics[width=0.30\textwidth]{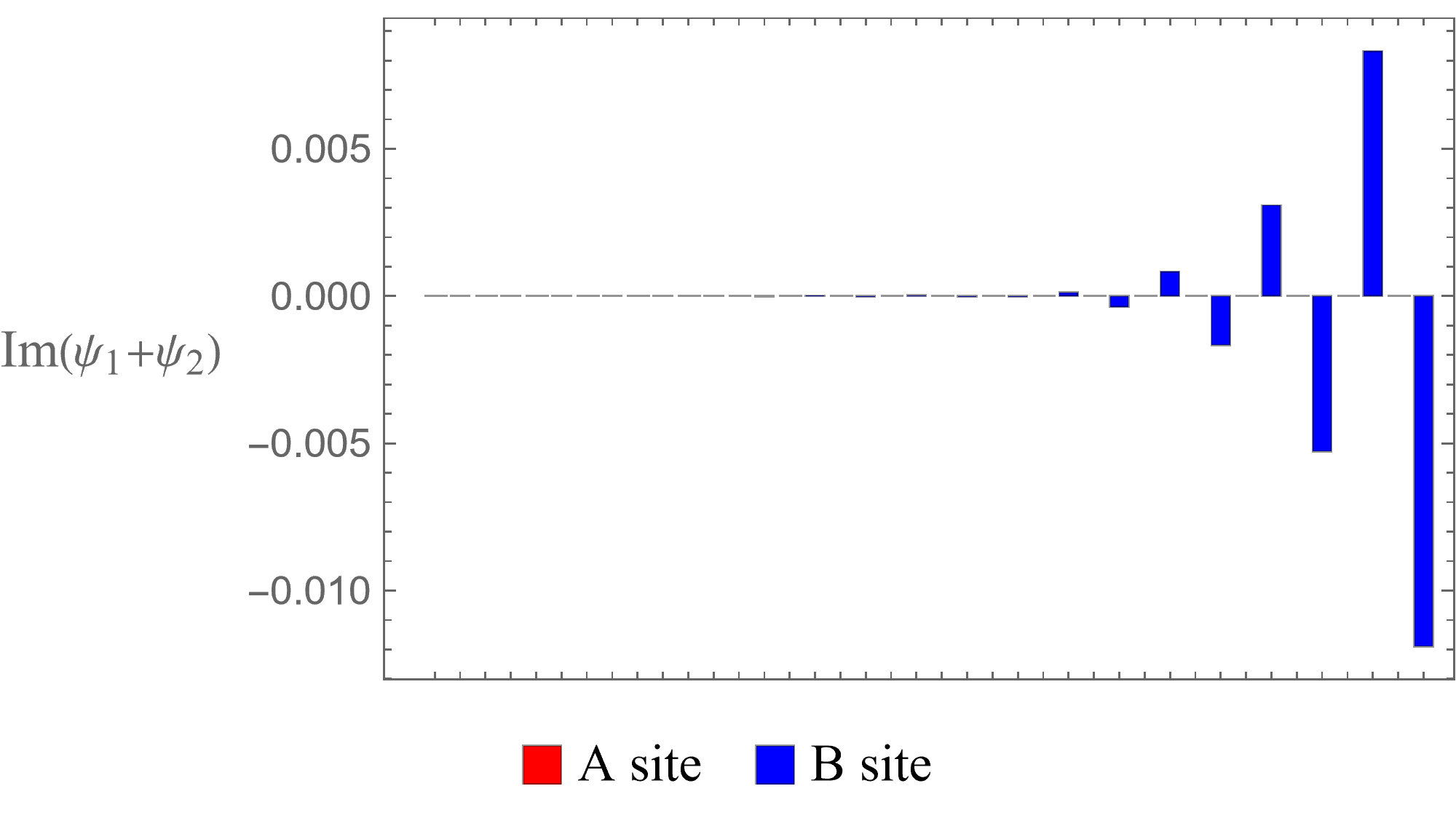}}\\
\caption{(a)The energy spectrum of the NH type 1 extended SSH model with $40$ sites, where $\left( t_0^L, t_0^R, t_1^L, t_1^R, t_2^L, t_2^R \right) = (1, 1, 10/3, 10/3,3/4, 3)$. Thus, $(\tb_0, \tb_1, \tb_2) = (1, 10/3, 3/2)$ and it is in the topological phase with $\nb = 1$. Note that the energies of the two edge states $\psi_1, \psi_2$ are purely imaginary and approximately zero. They are also enclosed by the trajectory to $E(p)$. (b), (c) The wave functions of the two edge states in the system. (d), (e) The difference and sum of $\psi_1$ and $\psi_2$.  Note that $\psi_1 - \psi_2$ is a left edge state and $\psi_1 + \psi_2$ is a right edge state. This is consistent with the fact that $(\n_E^L, \n_E^R) = (1, -2)$.}  \label{fig5 nHSSH-ext1}
\end{figure}

Next, let's consider the case that $\left( t_0^L, t_0^R, t_1^L, t_1^R, t_2^L, t_2^R \right) = (1, 1, 10/3, 10/3,3/4, 3)$ so that $(\tb_0, \tb_1, \tb_2) = (1, 10/3, 3/2)$, $\nb = 1$, $r=\sqrt{2}$, and $(\n_E^L, \n_E^R) = (1, -2)$. Here, we will continue to use the criteria we obtained in the QH case. We expect there would be one left and one right edge state.  By taking the difference and sum of $\psi_1$ and $\psi_2$, we indeed see the former one appears on the left boundary while the latter one is on the right. The energy spectrum, the trajectory of $E(p)$, and the wave functions of the edge states are shown in Fig.~\ref{fig5 nHSSH-ext1}. Since the energies are now generally complex, the wave functions of the eigenstates would also be complex. Consequently, we only show the real part of a wave function, unless its magnitude is too small to be seen.  In such cases, we would show the imaginary part instead. Note that all the energy eigenvalues, including those of the edge states, are enclosed by the trajectory to $E(p)$, which differs from the case that $\n_E = 0$.

\begin{figure}[hbt!]
\centering
\subfloat[]{\includegraphics[width=0.40\textwidth]{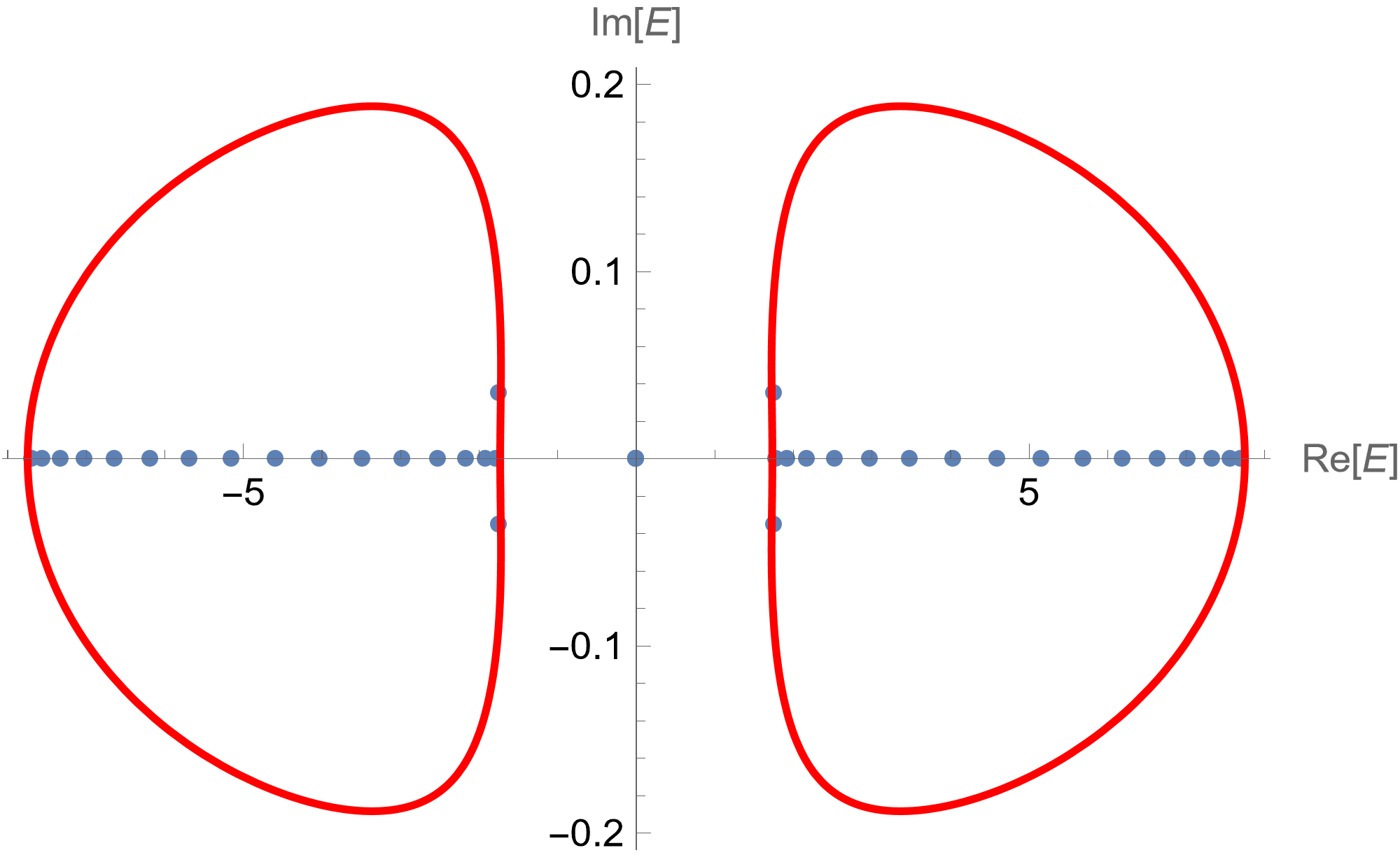}}\\
\subfloat[]{\includegraphics[width=0.30\textwidth]{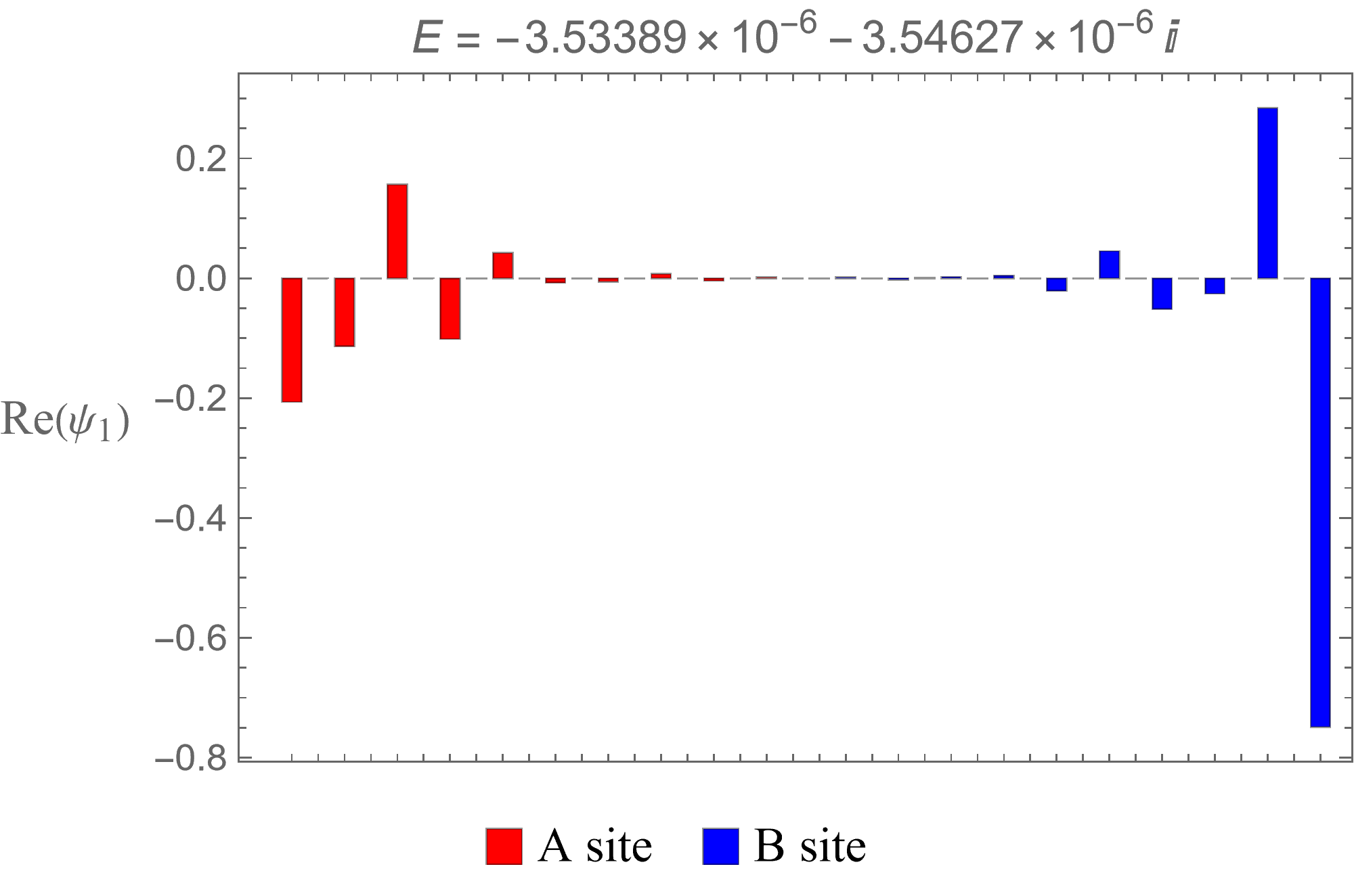}}\hskip 0.5cm
\subfloat[]{\includegraphics[width=0.30\textwidth]{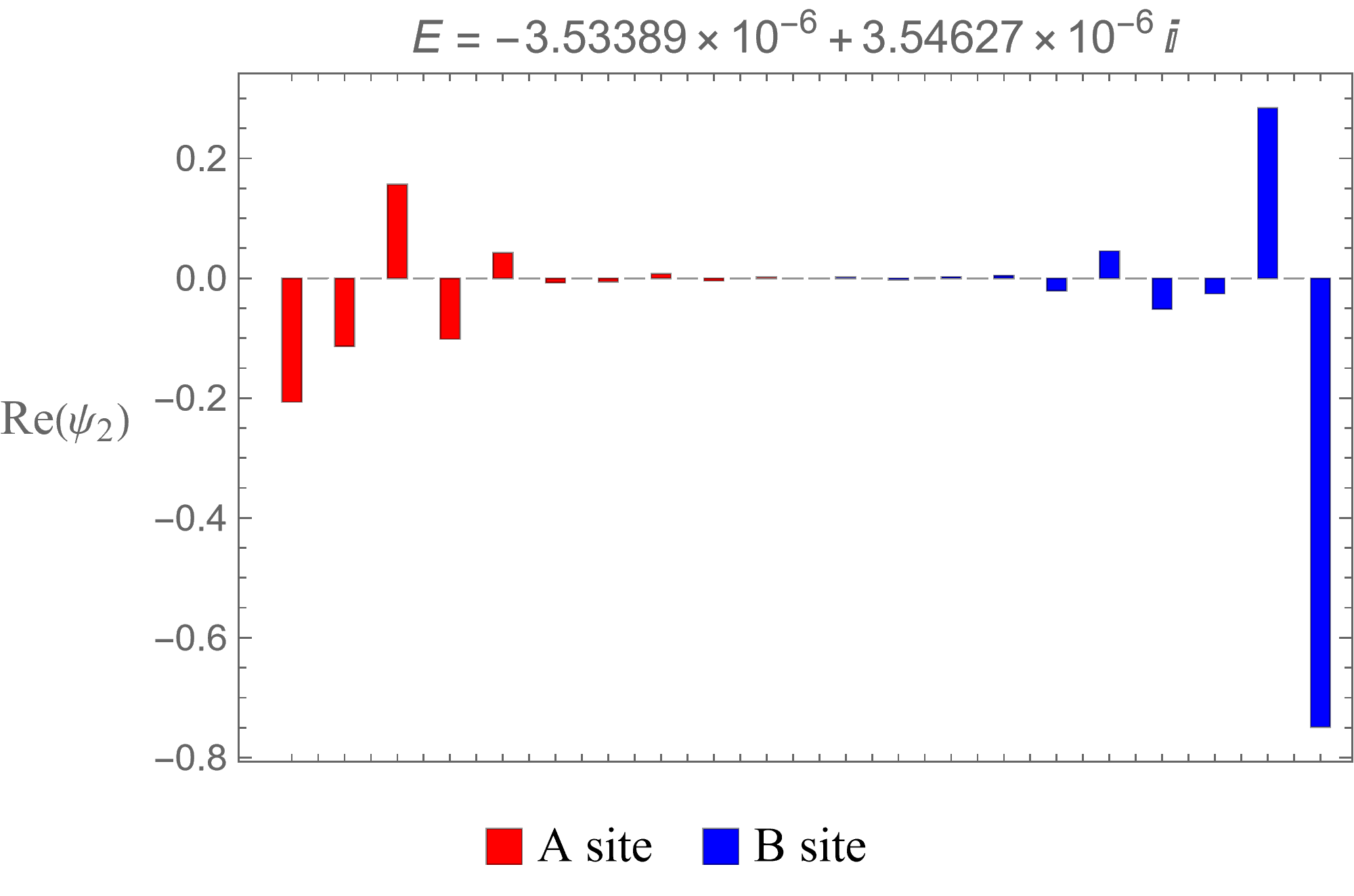}}\\
\subfloat[]{\includegraphics[width=0.30\textwidth]{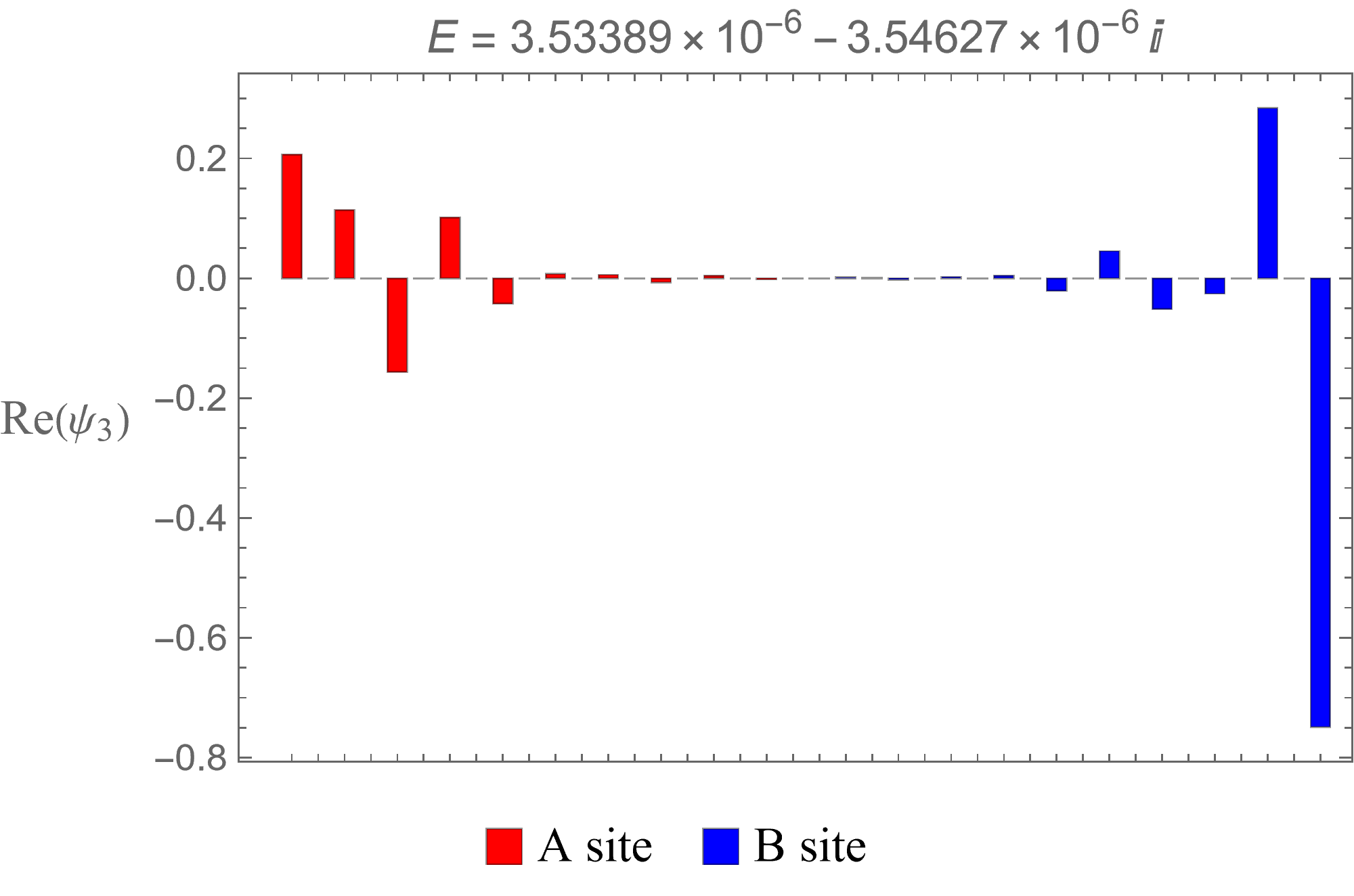}}\hskip 0.5cm
\subfloat[]{\includegraphics[width=0.30\textwidth]{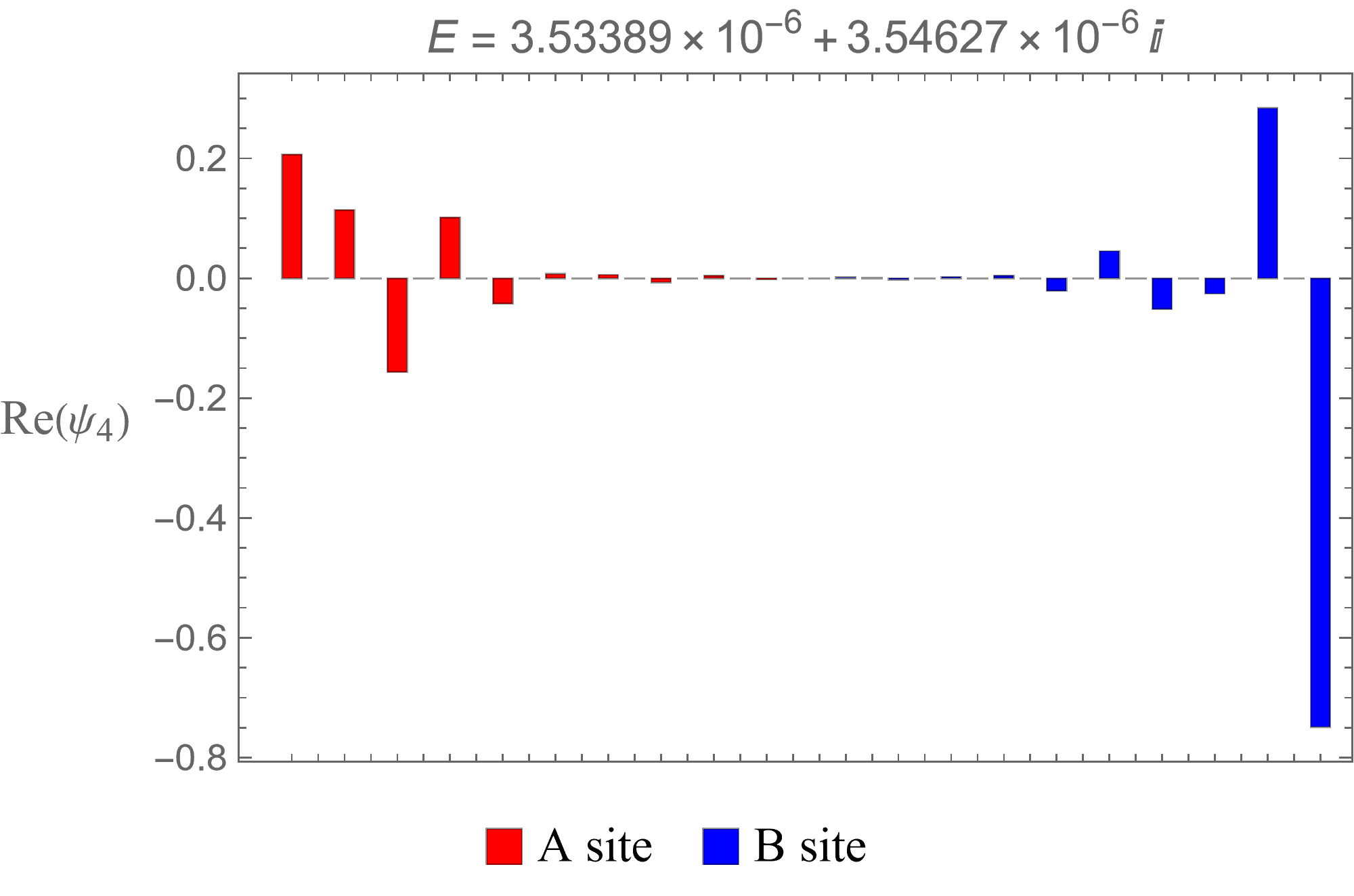}}\\
\subfloat[]{\includegraphics[width=0.30\textwidth]{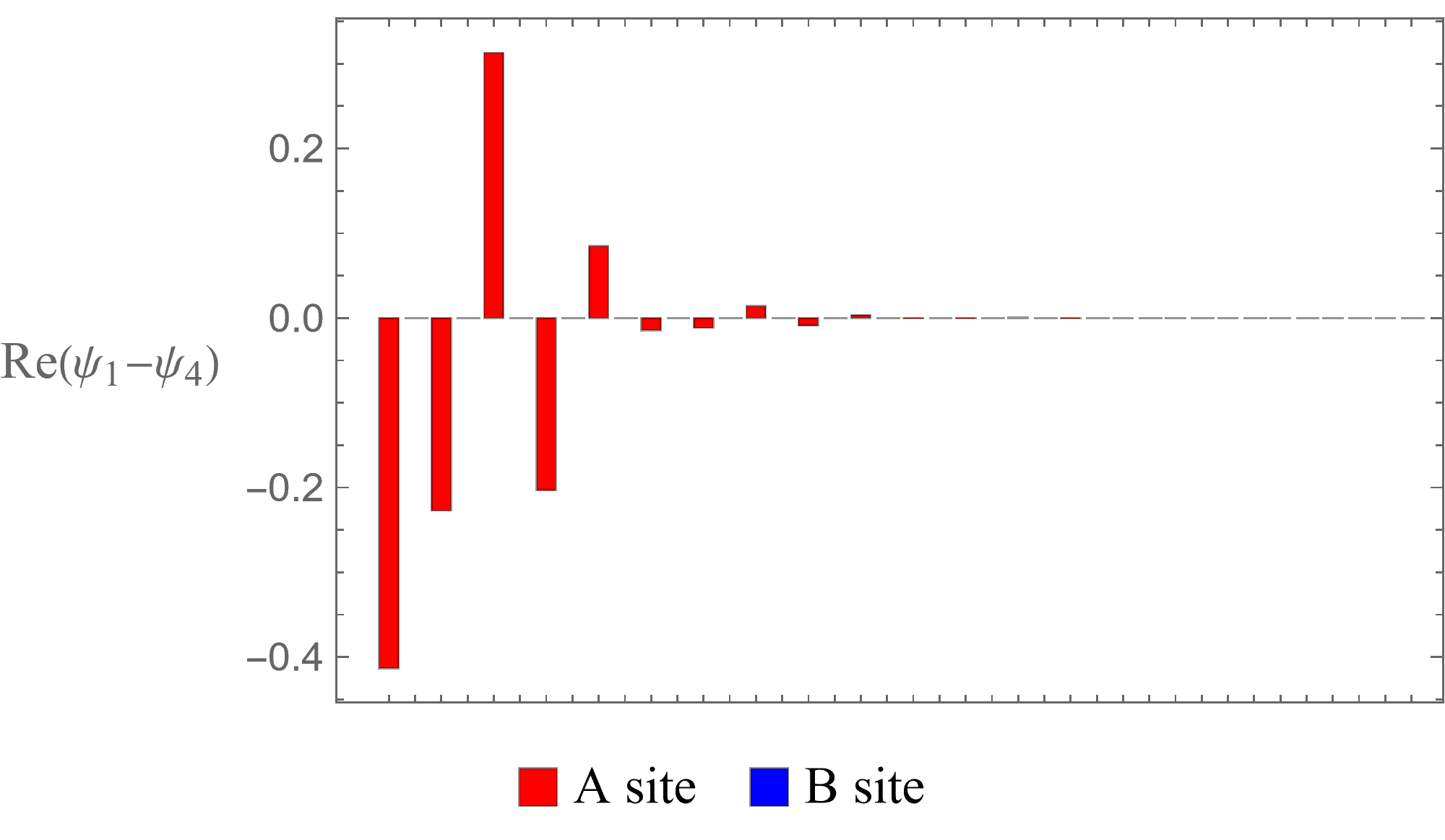}}\hskip 0.5cm
\subfloat[]{\includegraphics[width=0.30\textwidth]{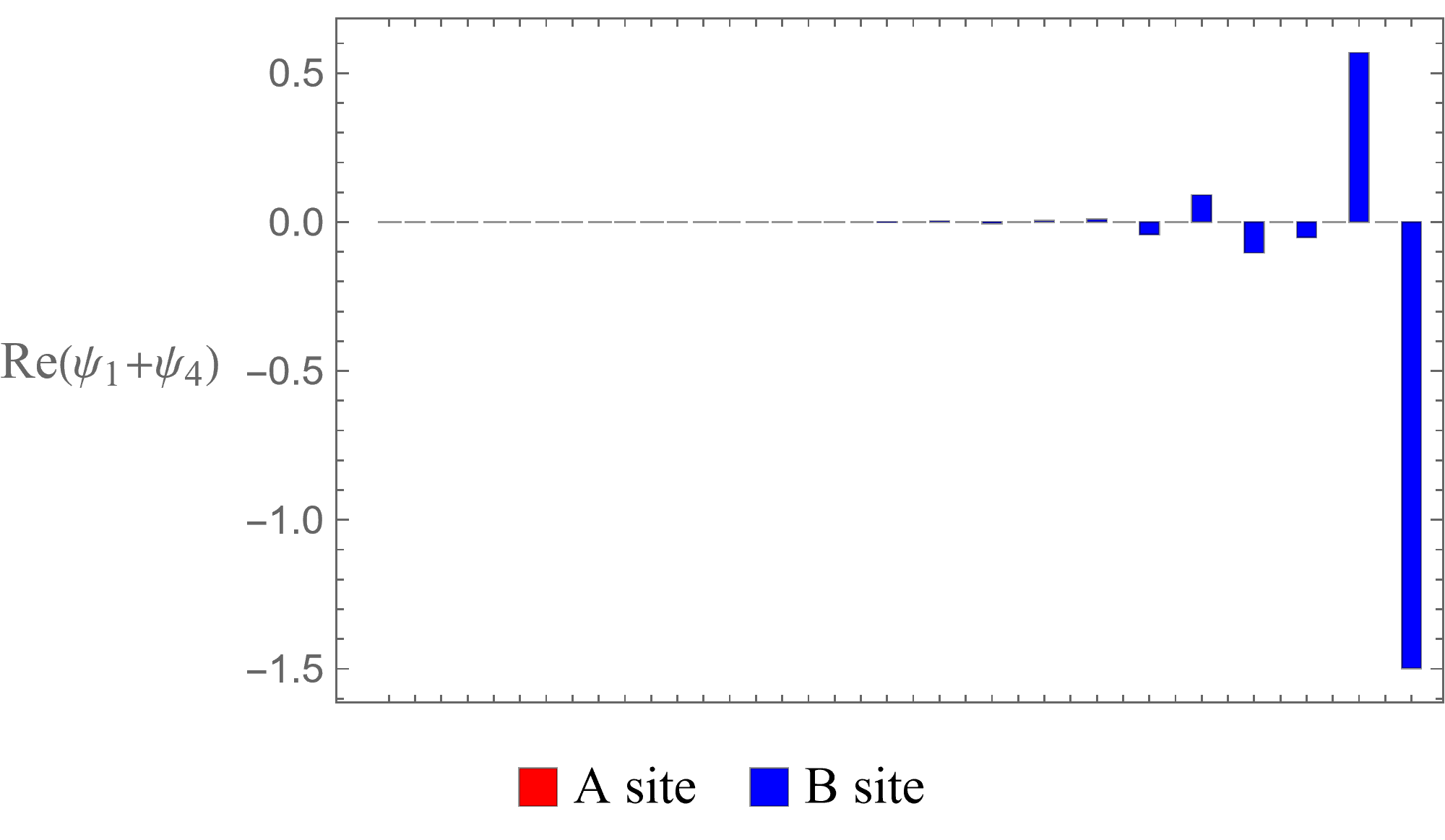}}\\
\subfloat[]{\includegraphics[width=0.30\textwidth]{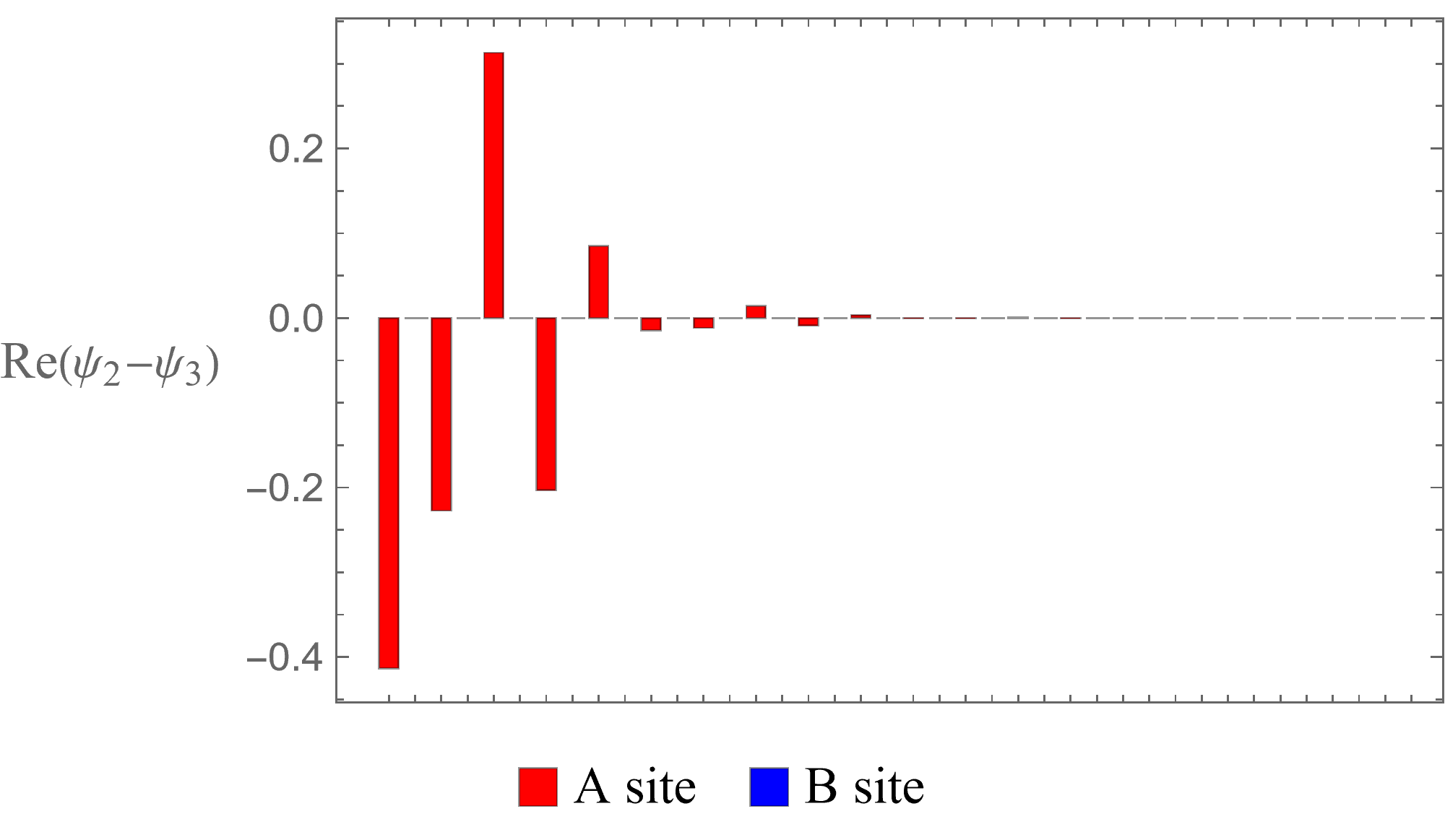}}\hskip 0.5cm
\subfloat[]{\includegraphics[width=0.30\textwidth]{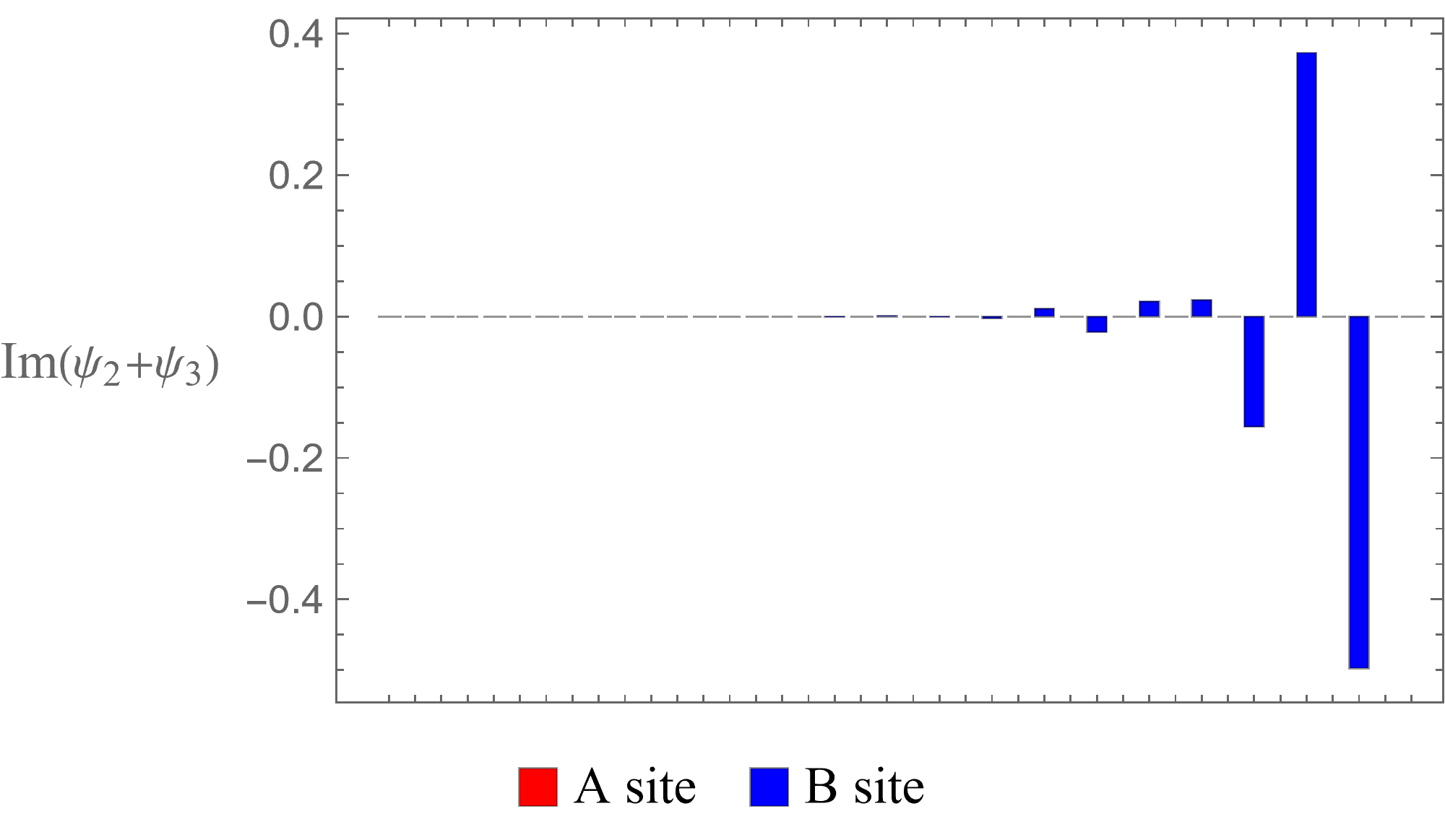}}\\
\caption{(a) The energy spectrum of the NH type 1 extended SSH model with $40$ sites, where $\left( t_0^L, t_0^R, t_1^L, t_1^R,, t_2^L, t_2^R \right) = (1, 1, 3, 3, 7/2, 4)$. Thus,  $(\tb_0, \tb_1, \tb_2) =(1, 3, \sqrt{14})$ and it is in the topological phase with $\nb = 2$. There are four edge states in total. (b), (c), (d), (e) The wave functions of the edge states $\psi_1, \psi_2, \psi_3, \psi_4$ in the system. They all have approximately zero energies.  (f), (g) The difference and sum and of $(\psi_1, \psi_4)$.  (h), (i) The difference and sum of $(\psi_2, \psi_3)$. Note that there are both two edge states on the left and right boundaries, consistent with the fact that $\n_E = 0$.}  \label{fig6 nHSSH-ext1}
\end{figure}

Let's show one more example in which $\left( t_0^L, t_0^R, t_1^L, t_1^R, t_2^L, t_2^R \right) = (1, 1, 3, 3, 7/2, 4)$ so that $(\tb_0, \tb_1, \tb_2) = (1, 3, \sqrt{14})$, $\nb = 2$, $r=(8/7)^{1/4}$, and $(\n_E^L, \n_E^R) = (2, -2)$. The energies of the edge states are all complex with the same absolute value and form two complex conjugate pairs. As we mentioned, their locations all lie outside the trajectories of $E(p)$. By taking the difference and sum of $(\psi_1, \psi_4)$ and $(\psi_2,\psi_3)$, we see there are two left and two right edge states. This is again consistent with the criteria obtained under the QHC. The energy spectrum, the trajectory of $E(p)$, and the wave functions of the edge states are shown in Fig.~\ref{fig6 nHSSH-ext1}.

Similarly, an analogous analysis may be conducted for the NH type 2 extended SSH model.  Again, we leave the details in Appendix C to avoid repetition in the main text.

\subsection{V. Conclusion and discussion}

In this paper, we study the NH extended SSH models.  Just like in the Hermitian case, there are two types of NH extended SSH models.  We first focus on the quasi-Hermitian case, in which the hopping amplitudes obey a specific relation so that the system may be mapped to a corresponding Hermitian one and its energy spectrum is completely real. It is well-known that the conventional BBC may not be naively applied in an NH system. However, we may use the modified winding number $\nb$ to classify the system and determine the total number of edge states on the boundaries analogous to the Hermitian case.  Moreover, we show that the spectral winding numbers, $(\n_E^L, \n_E^R)$, may be used to classify the system further. It dictates how the edge states would be distributed over the left and right boundaries.

By analyzing the EOM and BCs of a finite chain of such a system, we are able to find the characteristic equation in terms of $\ub_1$ and  $\ub_2$, which in turn satisfy a linear relation implied by the secular equation. By solving numerically the characteristic and secular equations, we may find the energy spectrum of the system.  The results are then compared to those obtained by numerically solving the matrix eigenvalue problem.  They agree well with each other.

We then relax the QHC and consider general NH extended SSH models. Although we are still able to express the characteristic equation in terms of one single variable $\ub =\left(\sb_{12} + \sb_{12}^{-1} \right)/2$, the energy eigenvalues of such a system are complex in general. We continue to employ the classification in terms of $\nb$ and $(\n_E^L, \n_E^R) $ that we obtained under the QHC to check whether it is still applicable for general NH extended SSH models. For all the cases we considered, no inconsistency has been found. In principle, one may use the generalized BZ to calculate the modified winding number in general NH extended SSH models but we are not able to obtain an analytic expression at the moment.  It is worth mentioning that similar to the Hermitian case, the two types of extended SSH models are related by a duality and may be mapped to each other by re-grouping the ``atoms'' in a unit cell.

It is known that non-Hermiticity also opens up the possibility that the Hamiltonian matrix becomes defective, and Jordan decomposition must be exploited to diagonalize such systems. As a result, generalized eigenvectors would appear. In the literature, this is referred to as an exceptional point in the parameter space of the hopping amplitudes.  Finding the condition for exceptional points and their implications in NH systems remains an intriguing problem.  Another interesting direction is to extend our analysis of 1D NH systems to 2D or higher dimensions. It is known that a relation between the 1D winding number and the 2D Chern number may be obtained if there is a certain symmetry in the system~\cite{Ext-SSH}. It would be interesting to carry out a similar study of NH systems to see whether such a relation still exists.

\section*{Acknowledgments}
The authors would like to thank Prof. Ming-Che Chang for stimulating discussions. J. S. Y. and H. C. K  are partly supported by the Grants MOST-110-2112-M-003-008-MY3 and Grants MOST-110-2112-M-A49 -012 -MY2 of the Ministry of Science and Technology, Taiwan, respectively.  

\appendix
\section{Appendix A:  QH type 2 extended SSH models}\label{QH-Ext-SSH-2}

The Hamiltonian of the NH type 2 extended SSH model with OBCs is given by:
\bea
&\;& \hskip -1.1cm H_{\rm ext2} =\sum_{j=2}^{N-1}\left\{ \left(t_0^L  A_j^\dag  + t_1^R A_{j+1}^\dag + t_{-1}^L A_{j-1}^\dag \right) B_j  + \left(t_0^R  A_j  +t_1^L A_{j+1} + t_{-1}^R A_{j-1} \right) B_j^\dag  \right\}  \\
&\;& \hskip 0.9cm + \left(t_0^L  A_1^\dag  + t_1^R A_2^\dag \right) B_1+ \left(t_0^R  A_1  + t_1^L A_2\right) B_1^\dag + \left(t_0^L  A_N^\dag  + t_{-1}^L A_{N-1}^\dag\right) B_N + \left(t_0^R  A_N  + t_{-1}^R A_{N-1}\right) B_N^\dag. \nn
\eea
Here, $t_{-1}^{L,R}$ are another type of third-nearest-neighbor left and right hopping amplitudes. It is again straightforward to achieve the EOM
\bea \label{SSH EOM 1B}
&\;& \hskip -3.1cm E A_j - \left(t_0^L B_j + t_1^R B_{j-1} + t_{-1}^L B_{j+1} \right) =0; \cr
&\;& \hskip -3.1cm E B_j - \left(t_0^R  A_j +t_1^L A_{j+1} + t_{-1}^R A_{j-1} \right) =0, \quad \mbox{\rm for }  j = 2, \ldots, N-1.
\eea
and BCs
\bea \label{Simplified BC 1B}
&\;& \hskip -3.1cm B_0 = A_0 =0; \cr
&\;& \hskip -3.1cm  B_{N+1} = A_{N+1} =0.
\eea

As pointed out in Ref. \cite{Ext-SSH}, type 1 and 2 extended SSH models are closely related.  By renaming $B_j$ and $A_{j+1}$ as $\tA_j$ and $\tB_{j}$,  a bulk NH type 1 extended SSH model may be mapped to a bulk NH type 2 extended SSH model with the following correspondence:
\bea \label{Type 1 and type 2 mapping}
&\;& \hskip -3.1cm(\tt_0^R , \tt_1^L, \tt_{-1}^R) = (t_1^R , t_0^L, t_2^R); \cr
&\;& \hskip -3.1cm  (\tt_0^L , \tt_1^R, \tt_{-1}^L) = (t_1^L , t_0^R, t_2^L).
\eea
Substituting the above results into eq.~(\ref{nu_E 1A}), we have
\bea \label{nu_E 1B}
&\;& \hskip -3.1cm E^2\left(p \right) =  \left(\tt_0^R  +\tt_1^L e^{ip} + \tt_{-1}^R e^{-ip} \right) \left(\tt_0^L + \tt_1^R e^{-ip} + \tt_{-1}^L e^{ip} \right).
\eea
Similarly, we have $\nt_E = \nt_E^L + \nt_E^R$,  with $\nt_E^L$ and $\nt_E^R$ the spectral winding number corresponding to the former and latter factors, respectively.  It is obvious that $\nt_E^R= \n_E^L-1$  and $\nt_E^L= \n_E^R+1$. Consequently, the winding numbers have the following relation:
\bea
\nt_E = \n_E,  {\rm and }\; \tilde{\nb} = 1 - \nb.
\eea

From now on, we will discard the tilde symbol to simplify expressions, as there is little chance of causing confusion. Following procedures similar to those in type1, we may have one single skin effect factor
\bea\label{skin effect 1B}
r = \left[ t_1^R t_{-1}^R/(t_1^L t_{-1}^L) \right]^{1/4},
\eea
if the following QHC is satisfied
\be
t_0^L=t_0^R \sqrt{t_1^R t_{-1}^L/\left( t_1^L t_{-1}^R \right)}.
\ee
This would translate into $\left( t_1 - \g_1/2 \right)= \left( t_1 + \g_1/2 \right) \sqrt{\left( t_2 + \g_2/2 \right)/\left( t_2 - \g_2/2 \right)}$ in the literature~\cite{Skin-effect4, GBZ4, GBZ5}.

Similarly, the system may be mapped to a Hermitian type 2 extended SSH model with a real energy spectrum. In addition to $\tb_0, \tb_1$, we introduce $\tb_{-1} = \sqrt{t_{-1}^L t_{-1}^R}$, and the resultant characteristic equation may be expressed in terms of these parameters in the following form:
\bea  \label{Simplified characteristic eq 1B even}
&\;& \hskip -2.6cm   \left\{ U_{N+2}(\ub_1) U_{N}(\ub_2) - 2 U_{N+1}(\ub_1) U_{N+1}(\ub_2) + U_{N}(\ub_1) U_{N+2}(\ub_2) +2 \right\} \cr
&\;& \hskip -3.1cm  - \left[ \frac{\tb_1}{\tt_{-1}} + \frac{\tb_{-1}}{\tb_1} \right]  \left\{ U_{N+1}(\ub_1) U_{N-1}(\ub_2) - 2 U_{N}(\ub_1) U_{N}(\ub_2) + U_{N-1}(\ub_1) U_{N+1}(\ub_2) +2 \right\}  \cr
&\;& \hskip -3.1cm +   \left\{ U_{N}(\ub_1) U_{N-2}(\ub_2) - 2 U_{N-1}(\ub_1) U_{N-1}(\ub_2) + U_{N-2}(\ub_1) U_{N}(\ub_2) +2 \right\}  = 0.
\eea
Similar to the type 1 case, the above characteristic equation is exactly the same as the one obtained in Ref.~\cite{Ext-SSH}, if we identify the parameters $(\tb_0, \tb_1, \tb_{-1})$ with $(t_0, t_1, t_{-1})$.

Analogously, here we have
\bea \label{Simplified prod and sum 2B}
&\;& \hskip -3.1cm   \ub_1 +\ub_2 = - \tb_0\left(\tb_1 + \tb_{-1}\right)/(2 \tb_1 \tb_{-1} ), \cr
&\;& \hskip -3.1cm  E^2= \tb_0^2 + \tb_1^2 + \tb_{-1}^2 + 2\tb_1 \tb_{-1} \left(1 + 2 \ub_1 \ub_2 \right).
\eea
Again, they are symmetric with respect to $\ub_1$ and $\ub_2$ and we may remove the redundancy in the solutions of $\ub_1$ and $\ub_2$ by requiring $\ub_1 > \ub_2$.

Similar to the type 1 case, we may use the modified winding number $\nb$ to classify such a system into three different phases:
\bee[label=\arabic*)]
\item $\tb_1+\tb_{-1}>\tb_0$, and  $\tb_1 > \tb_{-1}$:
The system is in the topological phase with $\nb=1$ and two edge states on the boundaries.

\item $\tb_1+\tb_{-1}<\tb_0$:
The system is in the trivial phase with $\nb=0$ and no edge states.

\item $\tb_1+\tb_{-1}>\tb_0$, and  $\tb_1 < \tb_{-1}$:
The system is in the topological phase with $\nb=-1$ and two edge states on the boundaries.
\eee

Just like in the type 1 case, the edge states are generally not evenly distributed over the left and right boundaries due to the skin effect. Similarly, we may use $\n_E^L$ and $\n_E^R$ to determine where they would appear:
\bee[label=\arabic*)]
\item $\nb=1$ or $\nb=-1$:
\bee[label=\roman*)]
\item $(\n_E^L, \n_E^R) = (1, 1), (0, 1), \fbox{(1, 0)}$. Both edge states are on the left boundary.

\item $(\n_E^L, \n_E^R) = \fbox{(1, 0)}, (1, -1), (0, 0), (-1, 1), \fbox{(0, -1)}$. One edge state is on the left boundary and one on the right.

\item $(\n_E^L, \n_E^R) = \fbox{(0, -1)}, (-1, 0), (-1, -1)$. Both edge states are on the right boundary.
\eee

\item $\nb=0$: $\n_E = 2, 1, 0, -1, 2$. No edge state.

\eee
Note that we put a box around $(1, 0)$ and $(0, -1)$ to indicate that there is an ambiguity in determining the locations of the edge states when $(\n_E^L, \n_E^R) $ equal to these values.

\begin{figure}[htb!]
\centering
\subfloat[]{\includegraphics[width=0.50\textwidth]{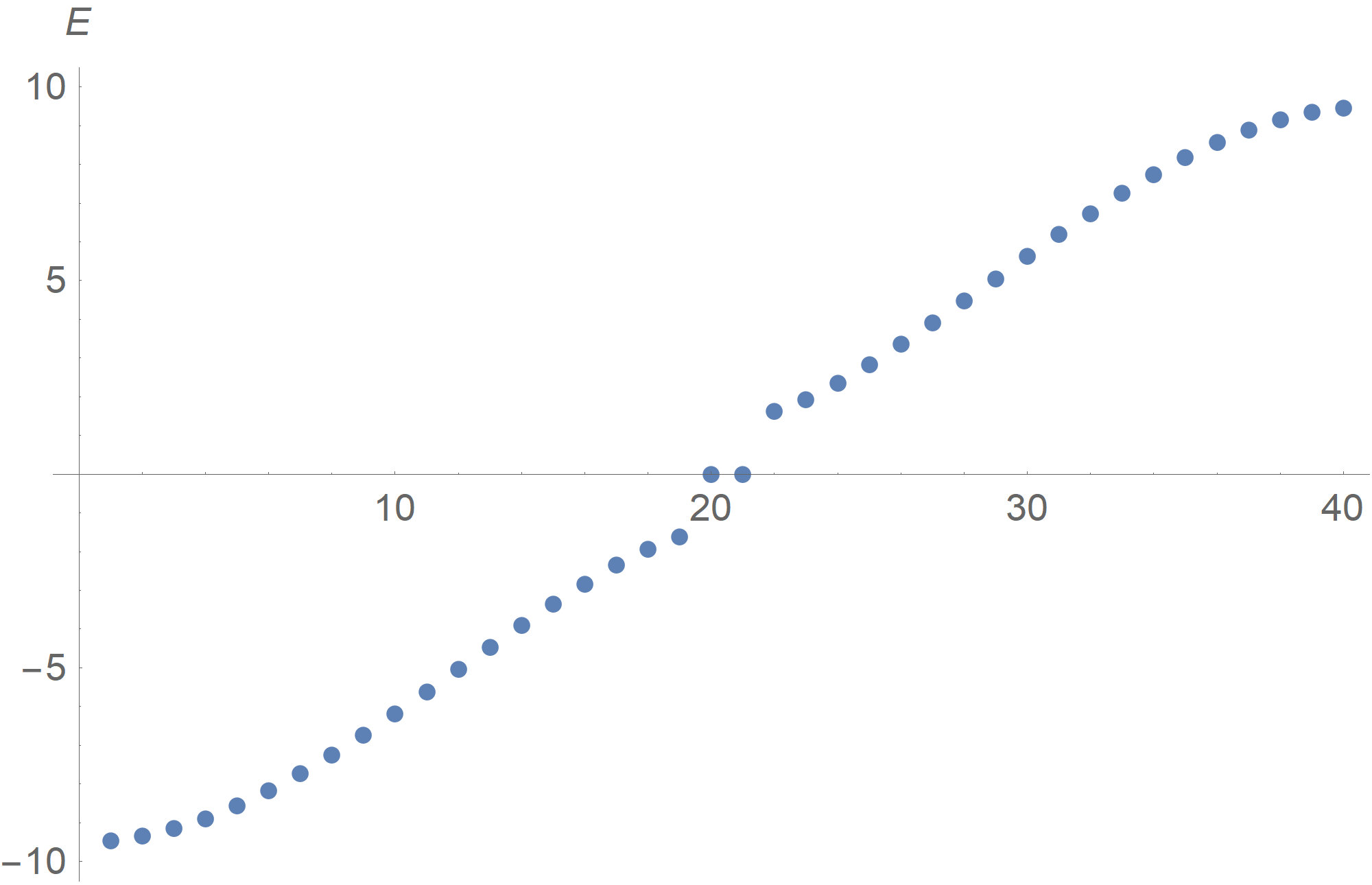}}\\
\subfloat[]{\includegraphics[width=0.30\textwidth]{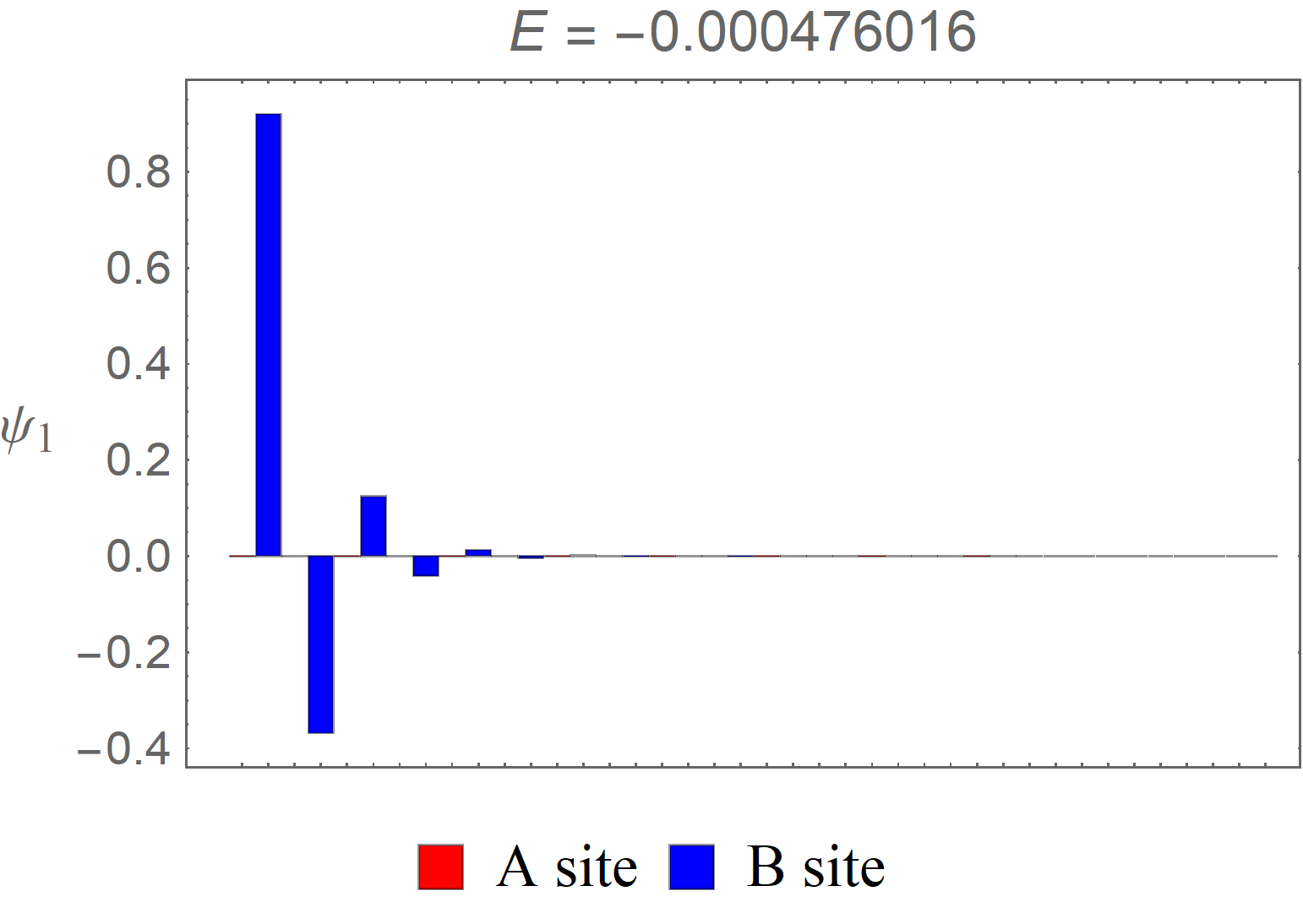}}\hskip 0.5cm
\subfloat[]{\includegraphics[width=0.30\textwidth]{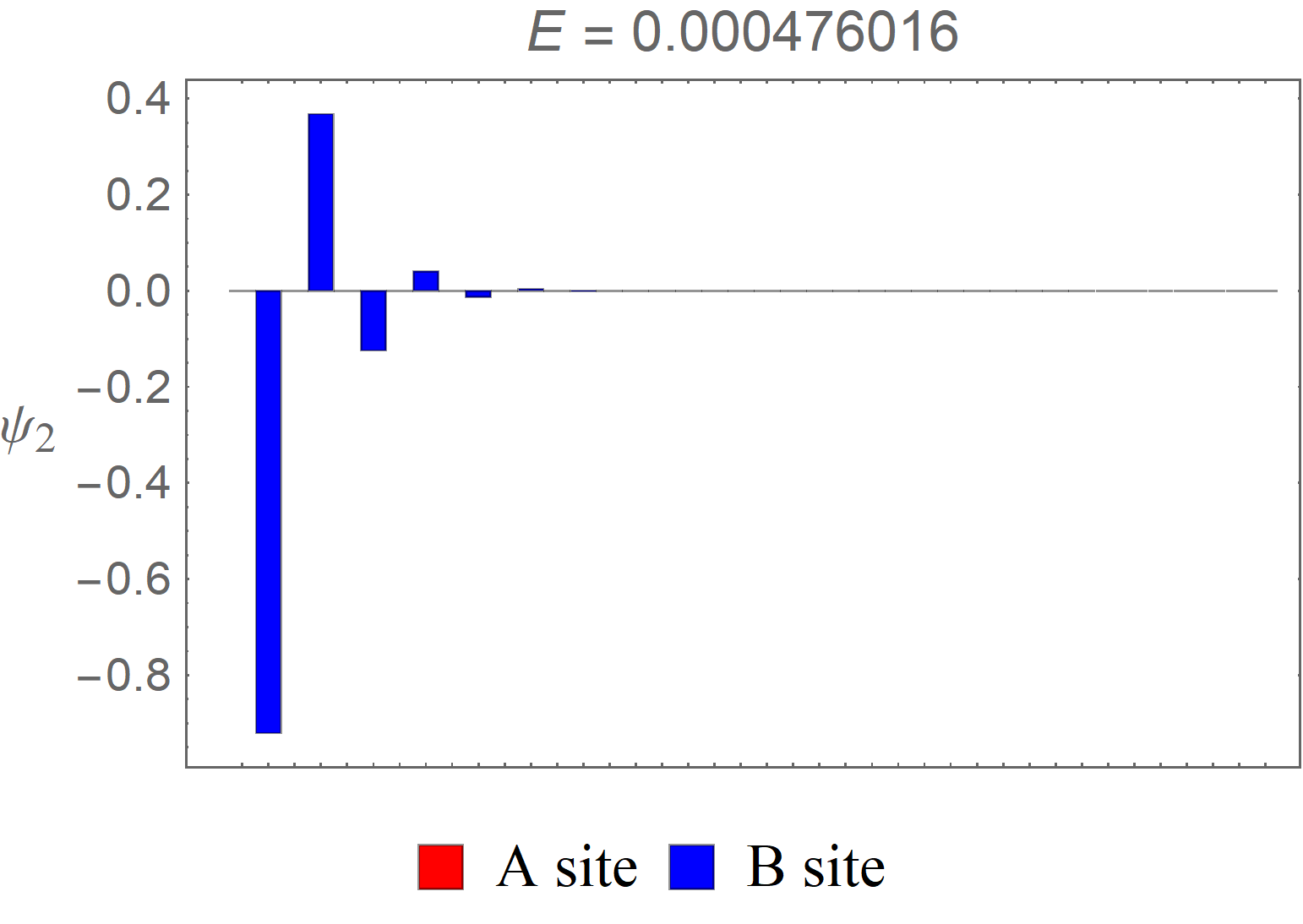}}\\
\subfloat[]{\includegraphics[width=0.35\textwidth]{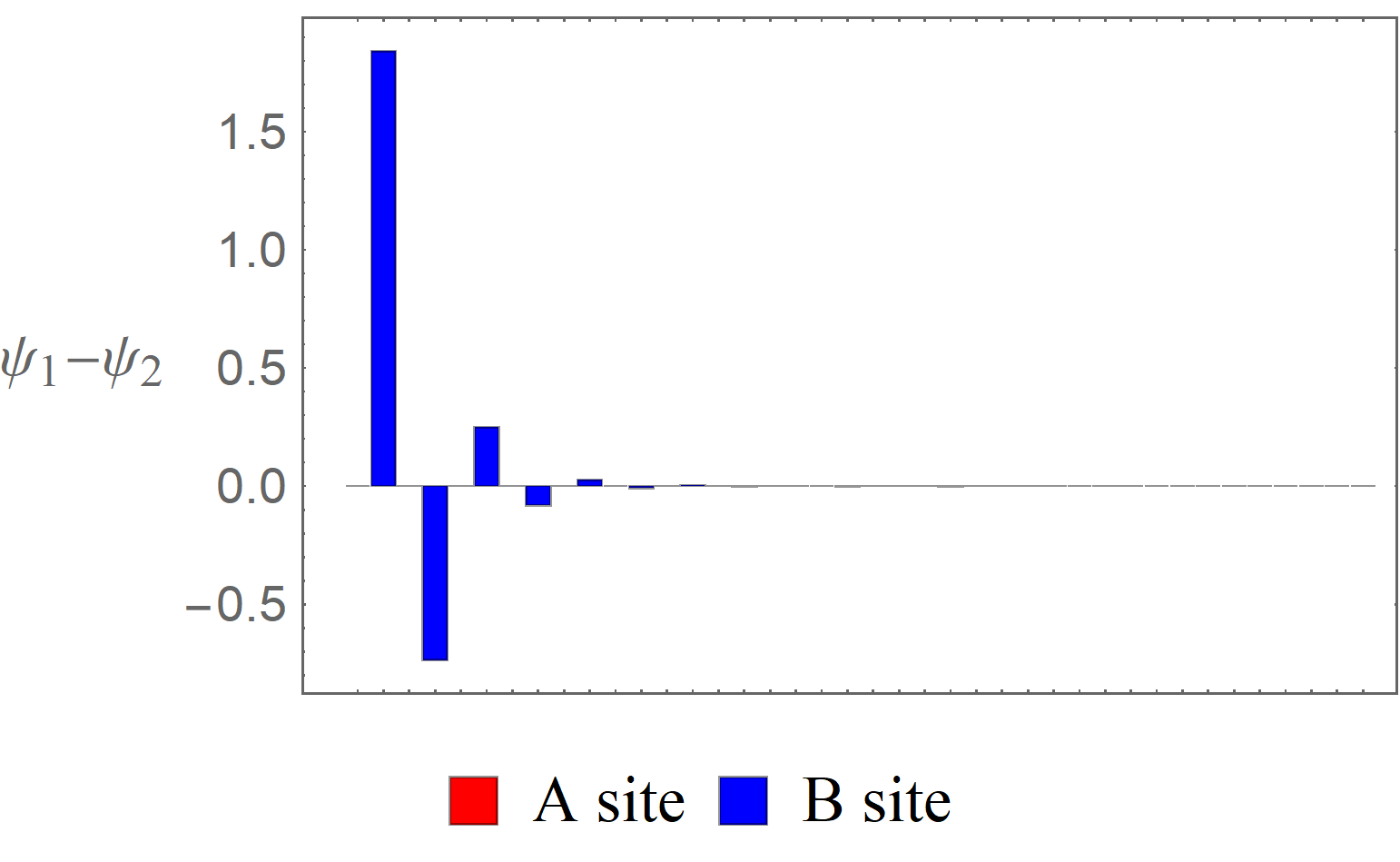}}\hskip 0.5cm
\subfloat[]{\includegraphics[width=0.38\textwidth]{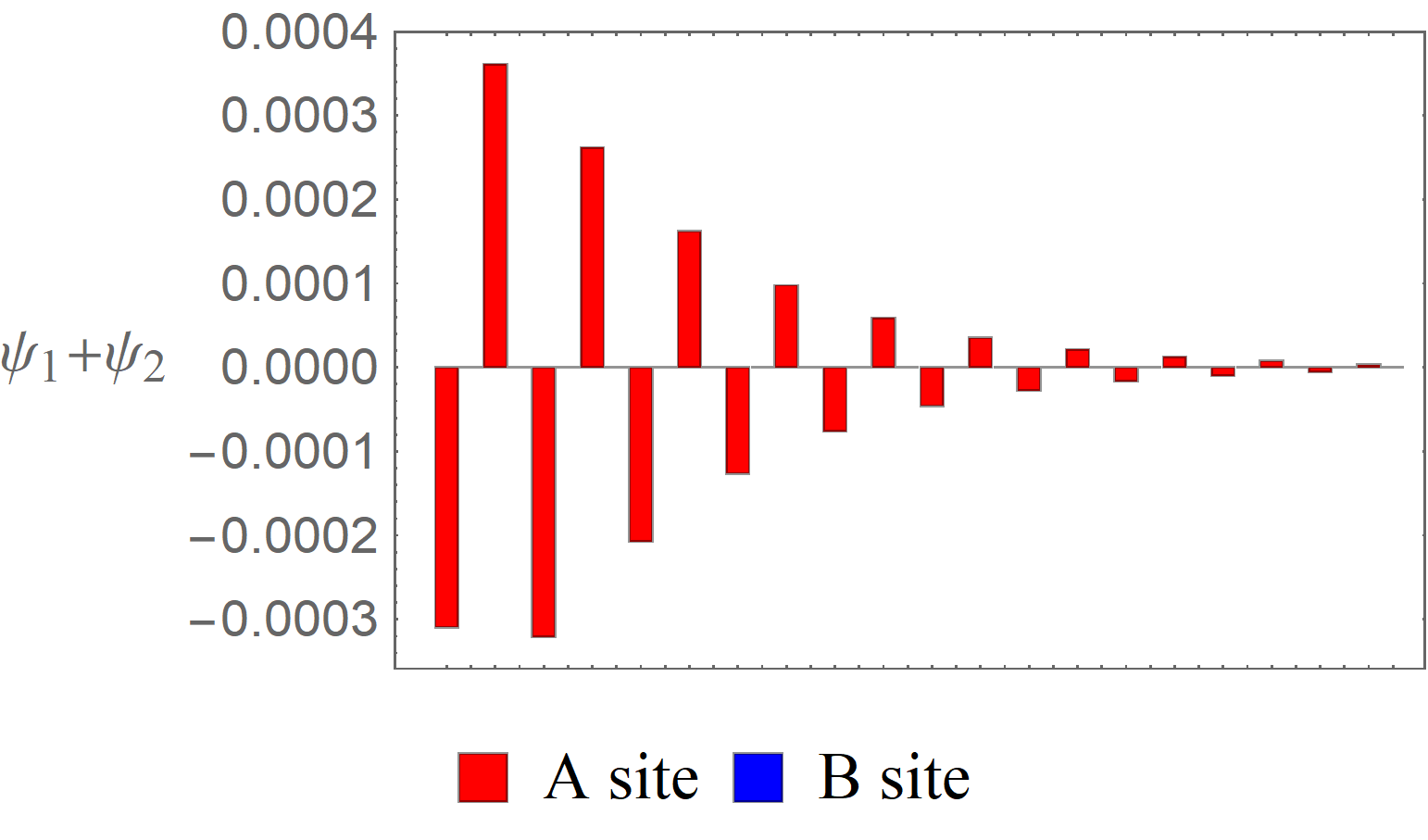}}
\caption{(a)The energy spectrum of the NH type 2 extended SSH model, where $\left( t_0^L, t_0^R, t_1^L, t_1^R, t_{-1}^L, t_{-1}^R \right) = (4, 4, 1, 1/4, 10, 5/2)$.  As a result, $(\tb_0, \tb_1,  \tb_{-1}) = (4, 1/2, 5)$ and the system is in the topological phase $\nb = -1$. The energies of the two edge states $\psi_1, \psi_2$ are both approximately zero.  (b), (c) The wave functions of the two edge states in the system.  (d), (e)  The difference and sum of $\psi_1$ and $\psi_2$.  Note that both left edge states appear on the left boundary as $(\n_E^L, \n_E^R) = (0, 1)$.}  \label{fig7 nHSSH-ext2}
\end{figure}

For demonstration, we consider the case $\left( t_0^L, t_0^R, t_1^L, t_1^R, t_{-1}^L, t_{-1}^R \right) = (4, 4, 1, 1/4, 10, 5/2)$ so that $(\tb_0, \tb_1,  \tb_{-1}) = (4, 1/2, 5)$, $\nb = -1$ and $r=1/2$. There are two edge states, which we call $\psi_1$ and $\psi_2$. Since $(\n_E^L, \n_E^R) = (0, 1)$, both edge states show up on the left boundary due to the skin effect. Again, this may be seen explicitly by taking the difference and sum of $\psi_1$ and $\psi_2$. In particular, $\psi_1+\psi_2$ is non-vanishing on the B sites only.  We show the results in Fig.~\ref{fig7 nHSSH-ext2} .

We would like to point out that although the $\nb=1$ and  $\nb=-1$ cases both have two edge states in total, we may still tell the difference between them by looking at the wave functions of the edge states.  In particular, if we compare the results of the case that $\left( t_0^L, t_0^R, t_1^L, t_1^R, t_{-1}^L, t_{-1}^R \right) = (4, 4, 10, 5/2, 1, 1/4)$ so that $(\tb_0, \tb_1,  \tb_{-1}) = (4, 5, 1/2)$, $\nb = 1$, $(\n_E^L, \n_E^R) = (1, 0)$ and $r=1/2$ to those of the previous case, we will see that now both edge states are mainly on the A sites instead of B sites.

\section{Appendix B:  QH extended SSH models with an odd number of sites}\label{QH-Ext-SSH-odd}

In the type 1 case, if there are $2N+1$ sites so that the chain ends with $A_{N+1}$ rather than $B_N$ on the right boundary, then the corresponding BCs on the right becomes
\bea
&\;& \hskip -3.1cm  B_{N+1} = A_{N+2} =0.
\eea
Following similar steps carried out previously, we may again find the corresponding characteristic equation for $\sb$:
\bea  \label{Characteristic eq 1A odd}
&\;& \hskip -2.6cm  \left\{ U_{N+2}(\ub_1) U_{N}(\ub_2) - 2 U_{N+1}(\ub_1) U_{N+1}(\ub_2) + U_{N}(\ub_1) U_{N+2}(\ub_2) +2 \right\} \cr
&\;& \hskip -3.1cm  - \left(\tb_2/\tb_0 \right) \left\{ U_{N+1}(\ub_1) U_{N-1}(\ub_2) - 2 U_{N}(\ub_1) U_{N}(\ub_2) + U_{N-1}(\ub_1) U_{N+1}(\ub_2) +2 \right\} = 0.
\eea
Since the unit cell defined by the two sites by the right boundary differs from the one by the left boundary, the corresponding winding numbers would also be different.  From eq.~(\ref{Type 1 and type 2 mapping}), we see the right boundary of the type 1 extended SSH model would be mapped to that of the type 2 extended SSH model. Hence, from the unit cell by the right boundary we have $\nt_E^R= \n_E^L-1$  and $\nt_E^L= \n_E^R+1$ and $\tilde{\nb} =1 - \nb$.

\begin{figure}[hbt!]
\centering
\subfloat[]{\includegraphics[width=0.50\textwidth]{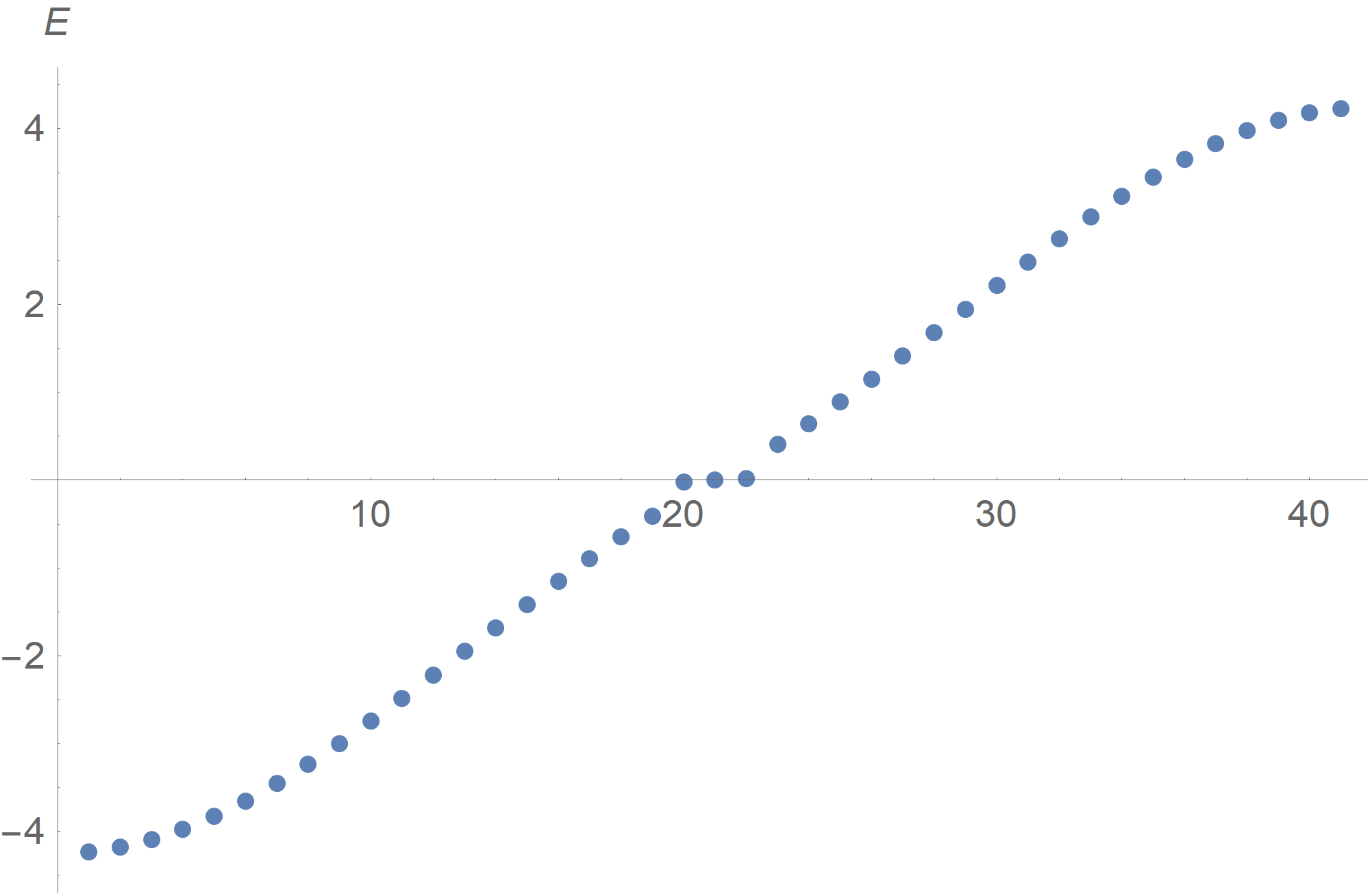}}\\
\subfloat[]{\includegraphics[width=0.30\textwidth]{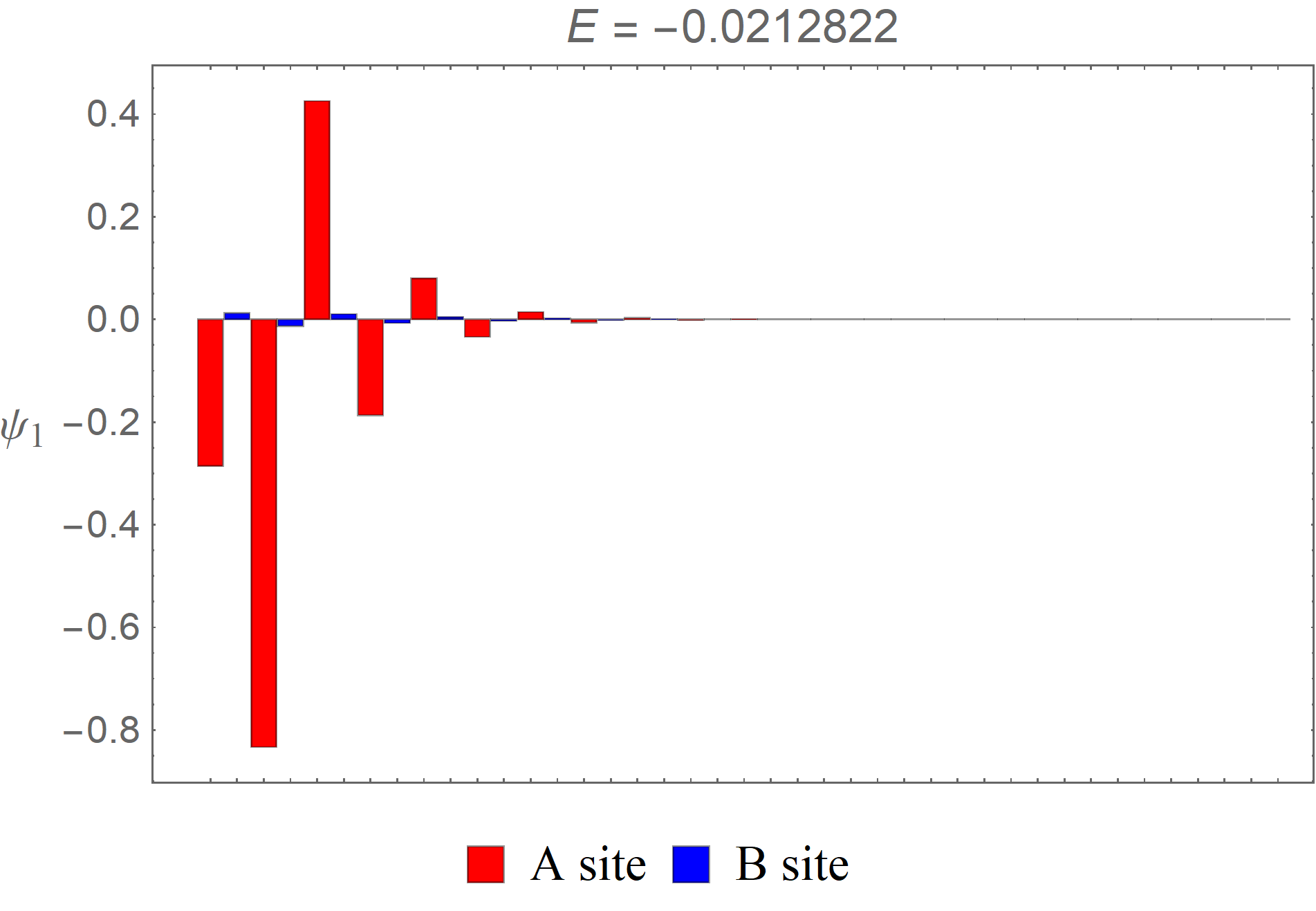}}\hskip 0.5cm
\subfloat[]{\includegraphics[width=0.30\textwidth]{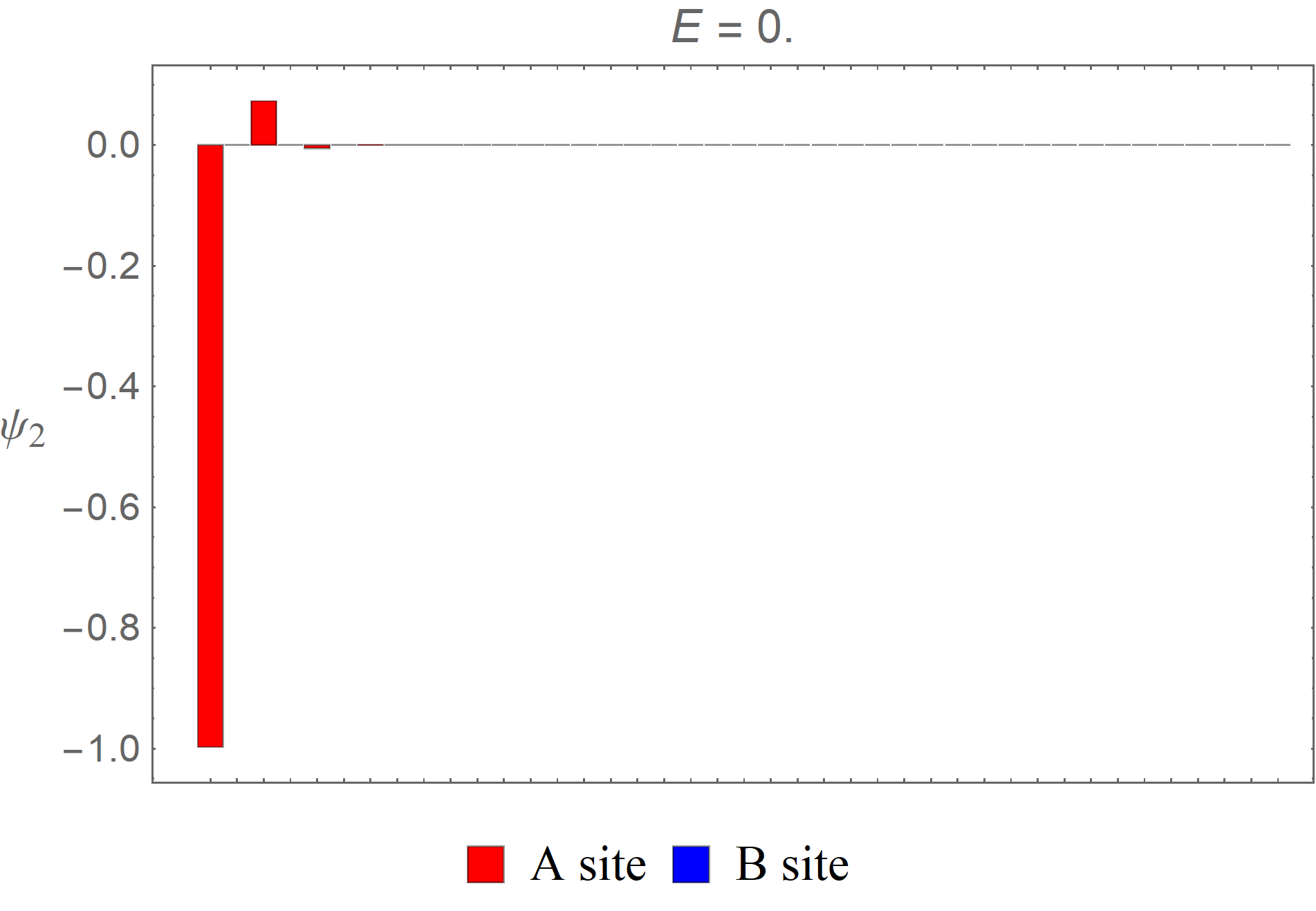}}\\
\subfloat[]{\includegraphics[width=0.30\textwidth]{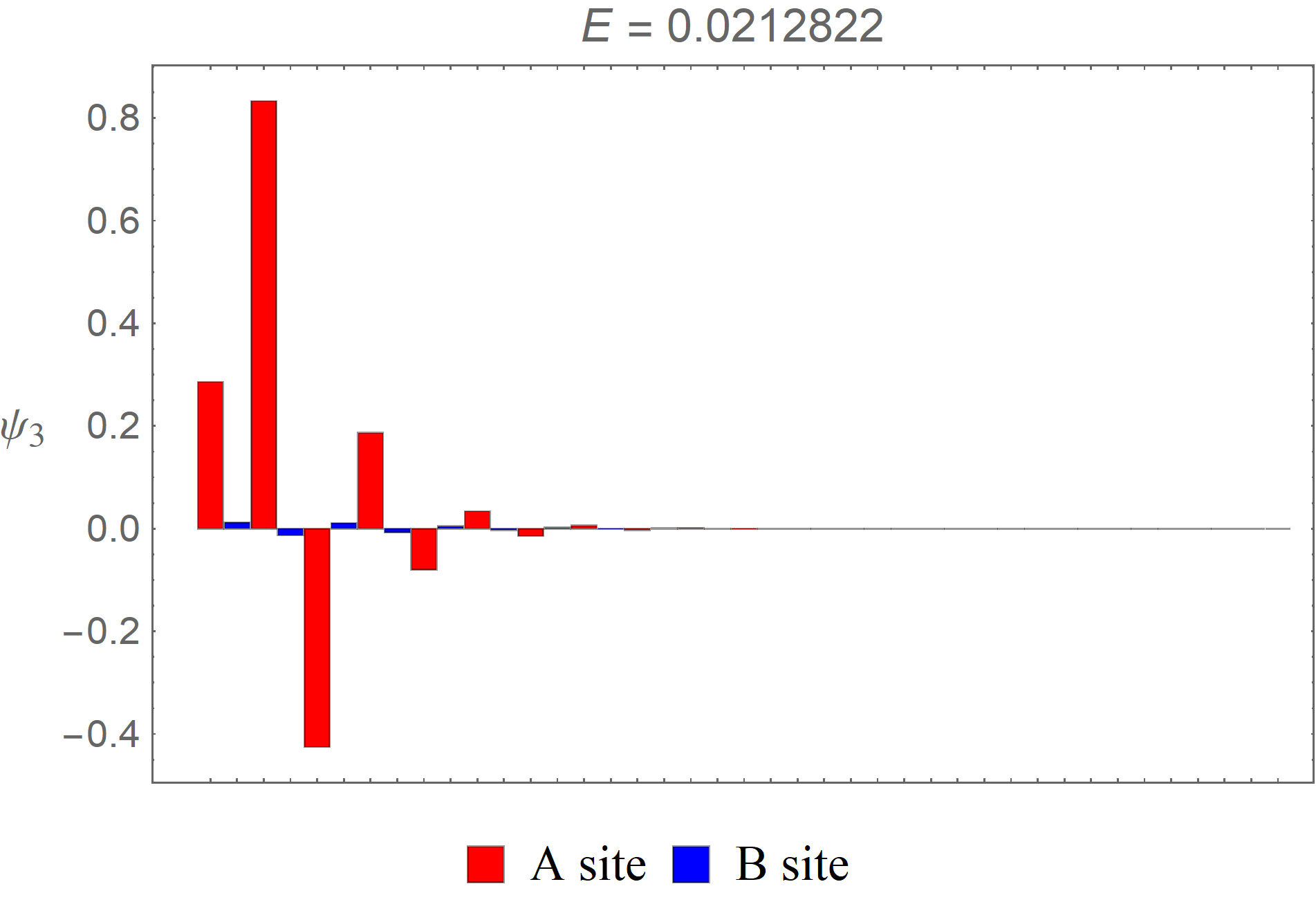}}\hskip 0.5cm
\subfloat[]{\includegraphics[width=0.33\textwidth]{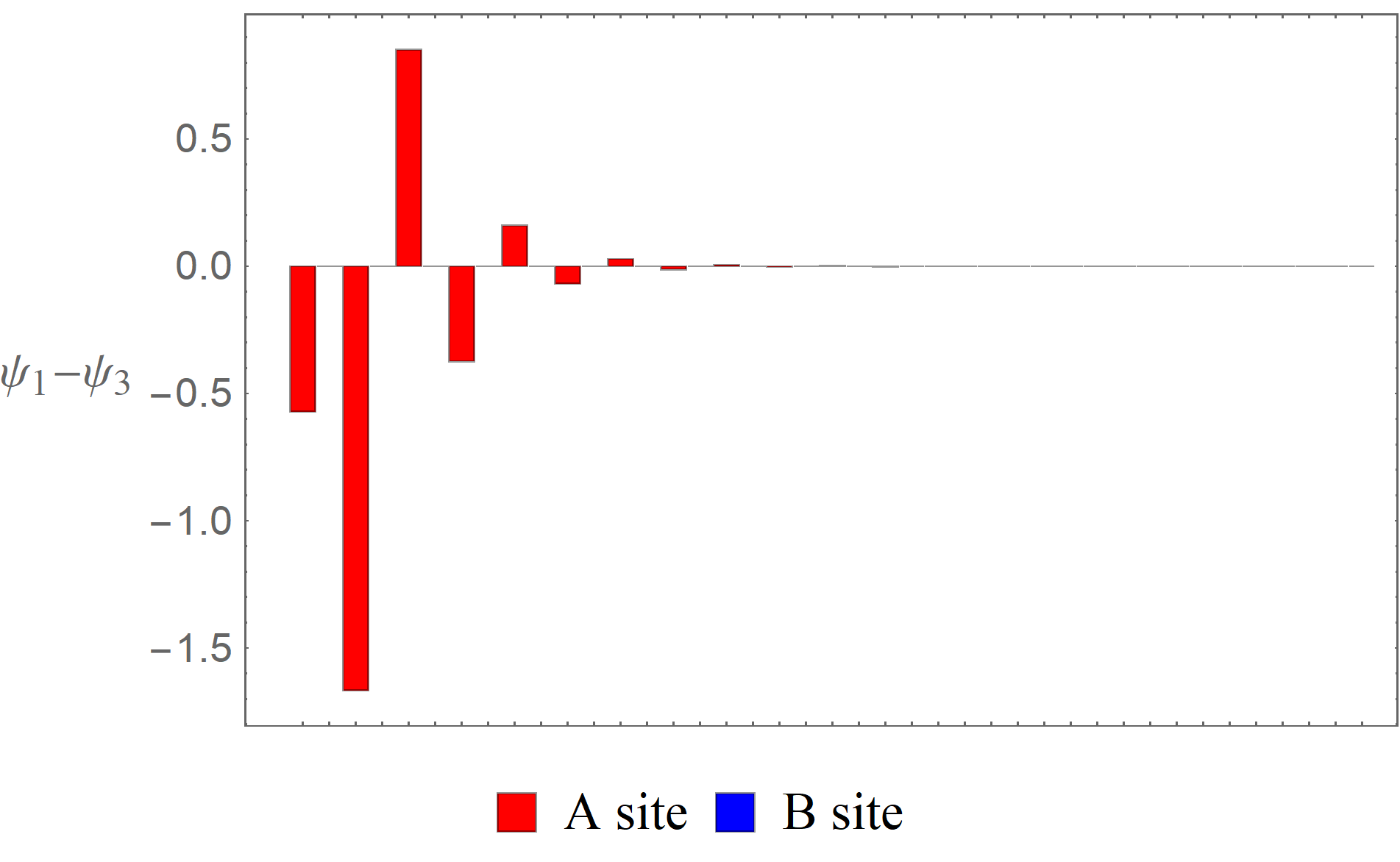}}\\
\subfloat[]{\includegraphics[width=0.38\textwidth]{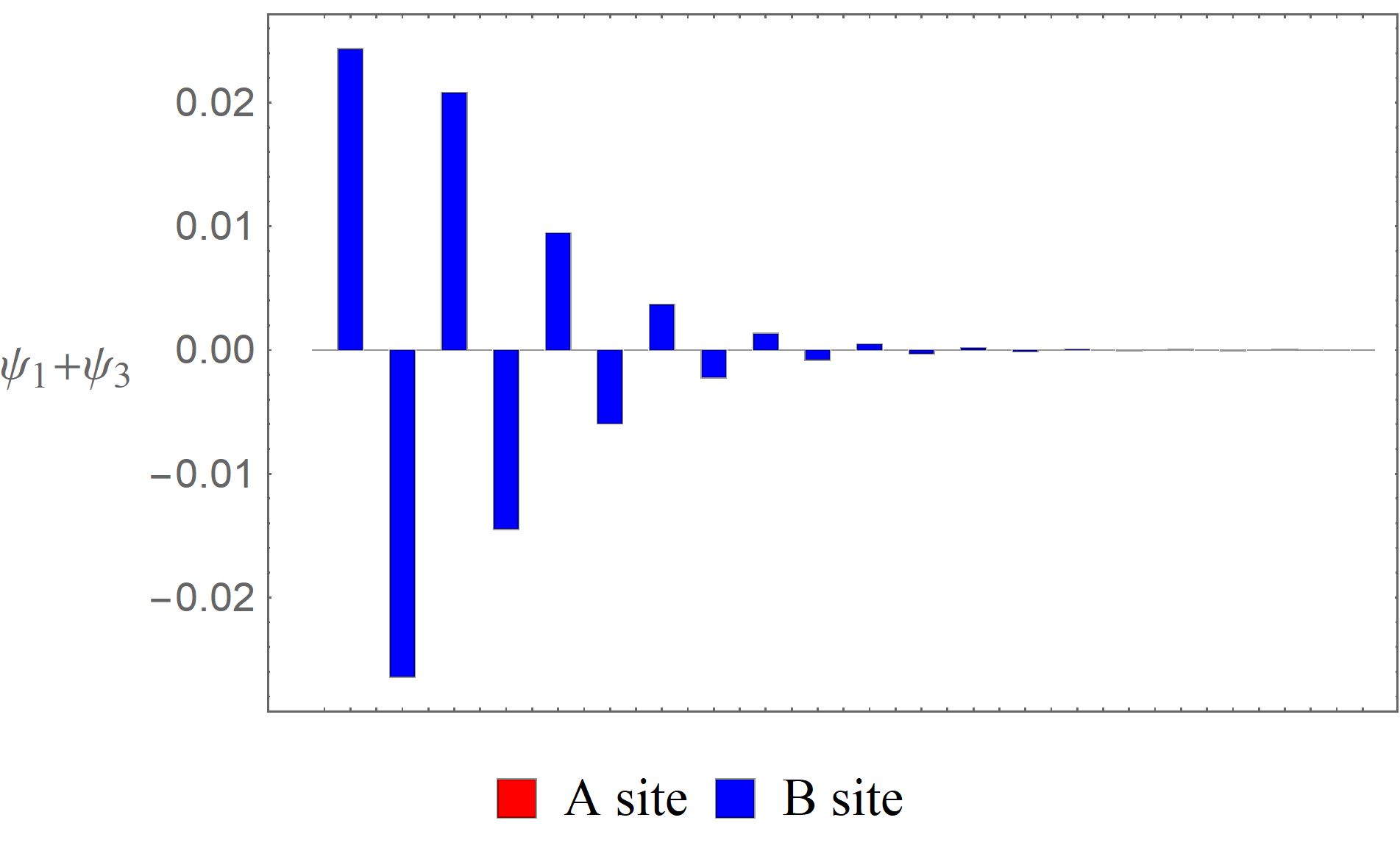}}\\
\caption{(a) The energy spectrum of the NH  type 1 extended SSH model with $41$ sites. The values of the parameters remain exactly the same as those in Fig.~\ref{fig2 nHSSH-ext1} and thus $\nb = 2$.  However, since the number of sites in the system is odd, the modified winding number corresponding to the unit cell by the right boundary is now $\nb_{\rm R}= 1-\nb_{\rm L} = -1$. (b), (c), (d) Wave functions of the edge states $\psi_1, \psi_2, \psi_3$. The energies of the edge states $\psi_1$ and $\psi_3$ are approximately zero, while $\psi_2$ is a state with exact zero energy.  Note that they are all left edge states. (e), (f) The difference and sum of $\psi_1$ and $\psi_3$.  Note that $\psi_1 + \psi_3$ is non-vanishing on the B sites only.}  \label{fig8 nHSSH-ext1}
\end{figure}

In Fig.~\ref{fig8 nHSSH-ext1}, we show the energy spectrum and wave functions of the edge states for the case that all the parameters remain the same as those considered in Fig.~\ref{fig2 nHSSH-ext1}, but there are now $41$ sites in the system.  Using the unit cell by the left boundary, we have $\nb = 2$. In contrast, using the unit cell by the right boundary, we have $\tilde{\nb} = -1$.  Hence, there are three edge states, which we call $\psi_1, \psi_2$, and $\psi_3$.  Moreover, since $(\n_E^L, \n_E^R) = (2, -1)$ the left boundary contributes two left edge states and $(\nt_E^L, \nt_E^R) = (0, 1)$ so the right boundary contributes one left edge states. This is a reflection of the skin effect. Among them, $\psi_1$ and $\psi_3$ have almost zero energies and form a chiral conjugate pair. In contrast, $\psi_2$ is an exact zero energy state protected by the chiral symmetry and it is non-vanishing only on the A sites~\cite{CZM}. Taking the sum of $\psi_1$ and $\psi_3$, we however achieve a wave function that is non-vanishing only on the B sites.

For the case of the NH type 2 extended SSH model with $2N+1$ sites, we may obtain the corresponding characteristic equation by making the following substitution $\left( \tb_0, \tb_1, \tb_2 \right) \to \left( \tb_1, \tb_0, \tb_{-1} \right)$ in eq.~(\ref{Characteristic eq 1A odd}). The reason is that when there are an odd number of sites in the system, the right boundary of a type 2 system is equivalent to that of a type 1 system as discussed in Appendix A.  Therefore, it would be nothing but a space inversion of the system that we just considered.

\section{Appendix C:  General NH type 2 extended SSH model}\label{GNH-Ext-SSH-2}

Similarly for a general NH type 2 extended SSH model, the characteristic equation may be expressed in terms of $s_{12}, s_{34}, u_{12}$, and $u_{34}$:
\bea
&\;& \hskip -3.1cm \biggl\{
- (t_{-1}^L)^2 s_{12}^{N} s_{34}^{N+2} U_{N} (u_{34}) U_{N} (u_{12})
+ t_{-1}^L s_{12}^{N-1} s_{34}^{N+1} U_{N+1} (u_{34})
\left[ t_{-1}^L s_{12}^2  U_{N-1}(u_{12}) + t_1^R U_{N+1} (u_{12}) \right]
\nonumber \\
&\;& \hskip -2.85cm - t_1^R t_{-1}^L s_{12}^{N} s_{34}^{N} U_{N+2} (u_{34}) U_{N} (u_{12})
+ t_{-1}^L s_{12}^{N-1} s_{34}^{N+1} U_{N-1} (u_{34})
\left[ t_{-1}^L s_{12}^2 U_{N+1} (u_{12}) + t_1^R s_{12} U_{N-1}(u_{12})  \right]
\nonumber \\
&\;& \hskip -2.85cm - s_{12}^{N-2} s_{34}^{N} U_{N} (u_{34})
\left[ (t_0^L)^2 s_{12}^4 U_{N} (u_{12})  + t_1^R t_{-1}^L s_{12}^2 U_{N+2}(u_{12})
+ t_1^R t_{-1}^L s_{12}^2 U_{N-2}(u_{12}) + (t_1^R)^2 U_{N} (u_{12}) \right]
\nonumber \\
&\;& \hskip -2.85cm + t_1^R s_{12}^{N-1} s_{34}^{N-1} U_{N+1} (u_{34})
\left[ t_{-1}^L s_{12}^2 U_{N+1} (u_{12}) + t_1^R U_{N-1} (u_{12}) \right]
- (t_1^R)^2 s_{12}^{N} s_{34}^{N-2} U_{N} (u_{34}) U_{N} (u_{12})
\nonumber \\
&\;& \hskip -2.85cm  -t_1^R t_{-1}^L s_{12}^{N} s_{34}^{N} U_{N-2} (u_{34}) U_{N} (u_{12})
+ t_1^R s_{12}^{N-1} s_{34}^{N-1} U_{N-1} (u_{34})
\left[ t_{-1}^L  s_{12}^2 U_{N-1} (u_{12}) + t_1^R U_{N+1} (u_{12}) \right] \biggr\}\nonumber \\
&\;& \hskip -3.1cm + ( t_1^R -t_{-1}^L s_{12}^{2}) ( t_1^R -t_{-1}^L s_{34}^{2})
\bigl[s_{12}^{-4} s_{34}^{2N} + s_{12}^{2N} s_{34}^{-4} \bigr] = 0.
\eea

Again, we may express $u_{12}, u_{34}$ in terms of $s_{12}, s_{34}$:
\bea \label{u12 and u34 type 2}
&\;& \hskip -3.1cm u_{12} = \frac{s_{12} s_{34} \left\{ t_1^L t_{-1}^L  \left(t_0^R t_1^R  + t_0^L t_{-1}^R  \right) s_{34} - t_1^R t_{-1}^R  \left(t_0^L t_1^L  + t_0^R t_{-1}^L  \right) s_{34} ^{-1} \right\} }{2t_1^L t_1^R t_{-1}^L t_{-1}^R \left(s_{12}^2 - s_{34}^2 \right) }, \nonumber \\
&\;& \hskip -3.1cm u_{34} = \frac{- s_{12} s_{34} \left\{ t_1^L t_{-1}^L  \left(t_0^R t_1^R  + t_0^L t_{-1}^R  \right) s_{12} - t_1^R t_{-1}^R  \left(t_0^L t_1^L  + t_0^R t_{-1}^L  \right) s_{12} ^{-1} \right\} }{2t_1^L t_1^R t_{-1}^L t_{-1}^R  \left(s_{12}^2 - s_{34}^2 \right)}.
\eea
By making use of eq.~(\ref{u12 and u34 type 2}), the characteristic equation may be written in terms of $s_{12}$ and $s_{34}$.  Similarly, the characteristic equation may be expressed in terms of one single variable $\ub =\left(\sb_{12} + \sb_{12}^{-1} \right)/2$.  Again, the resultant characteristic equation is a polynomial in $\ub$ of degree $3N$ after simplification.

\begin{figure}[hbt!]
\centering
\subfloat[]{\includegraphics[width=0.50\textwidth]{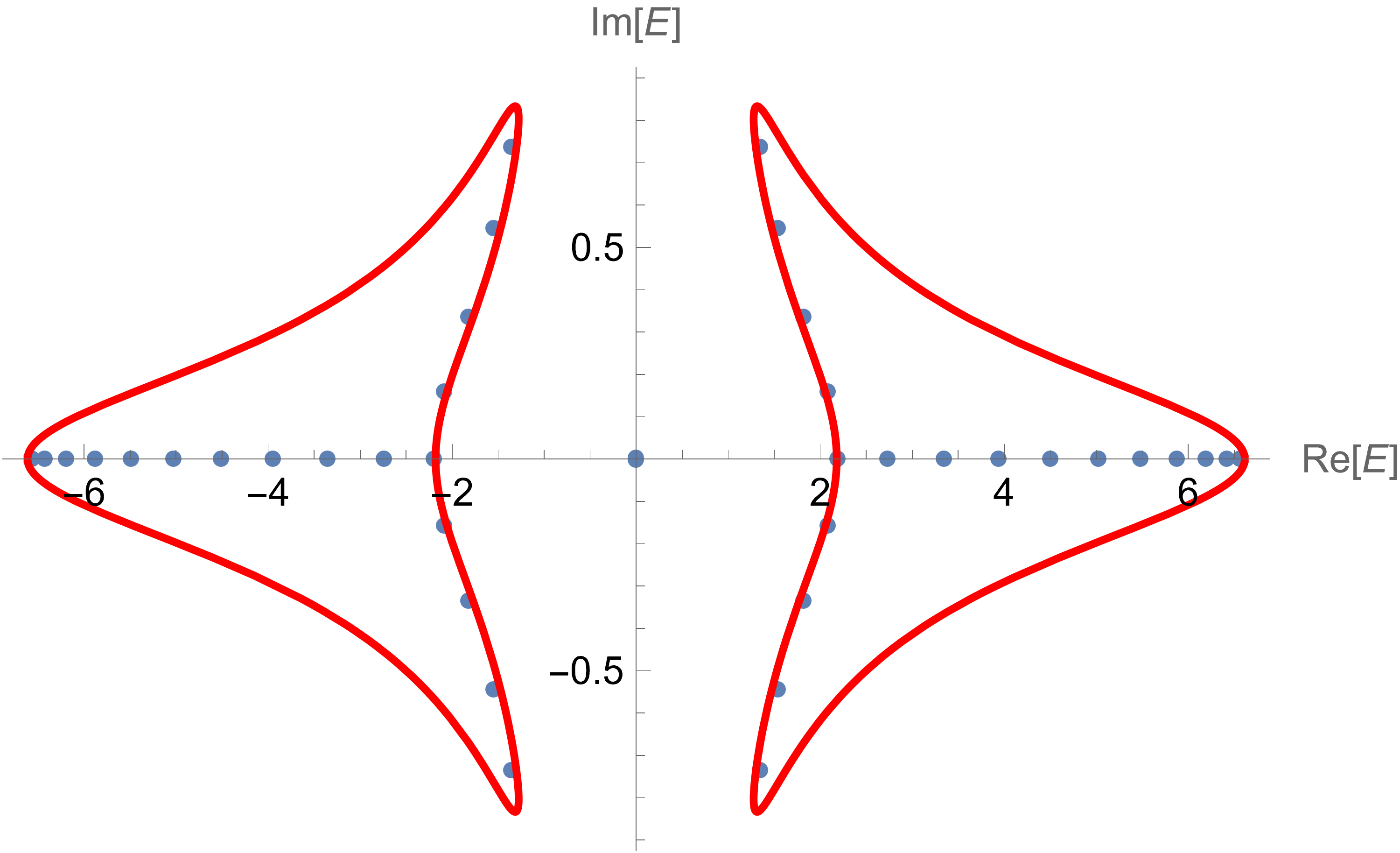}}\\
\subfloat[]{\includegraphics[width=0.30\textwidth]{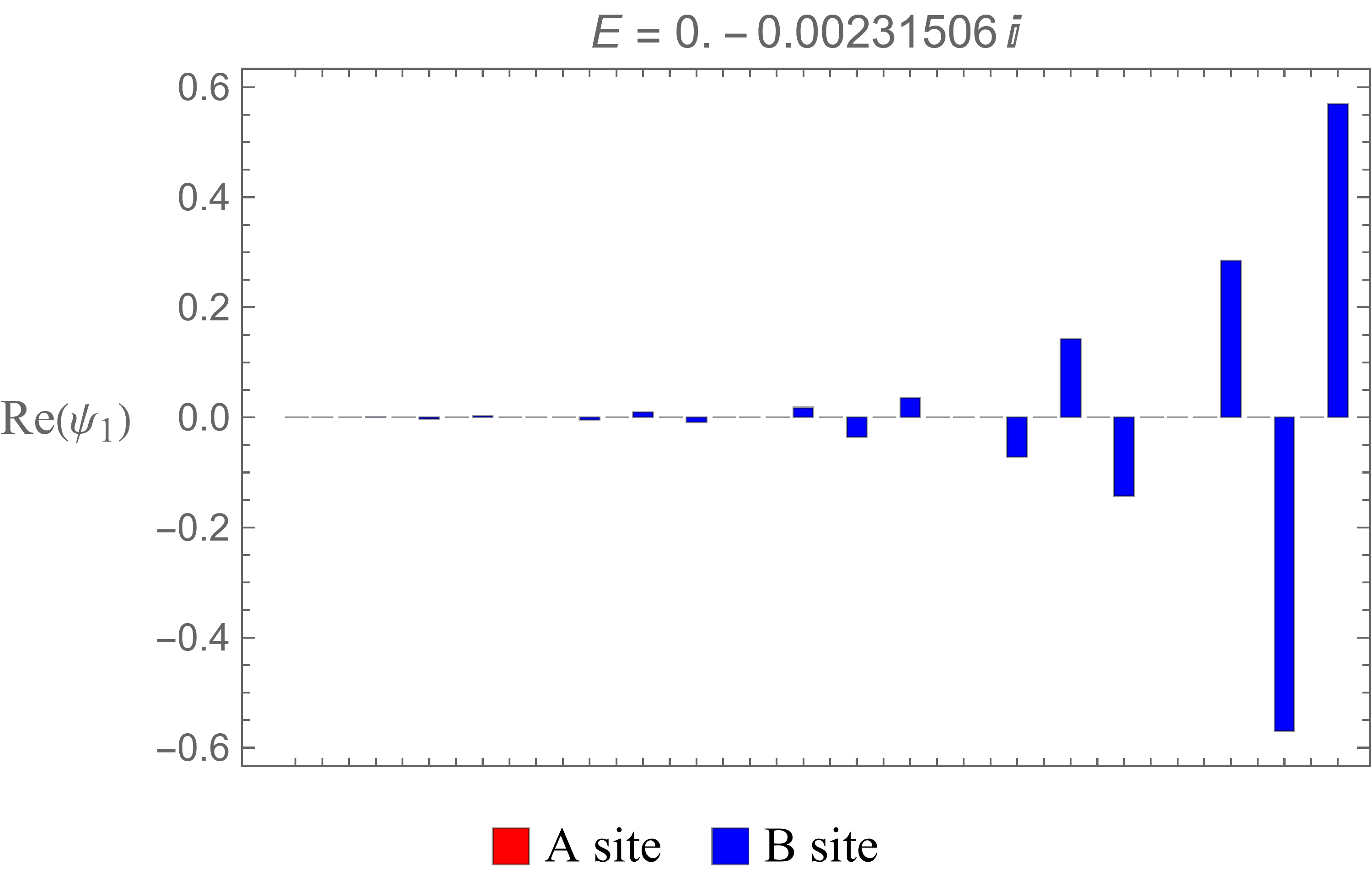}}\hskip 0.5cm
\subfloat[]{\includegraphics[width=0.30\textwidth]{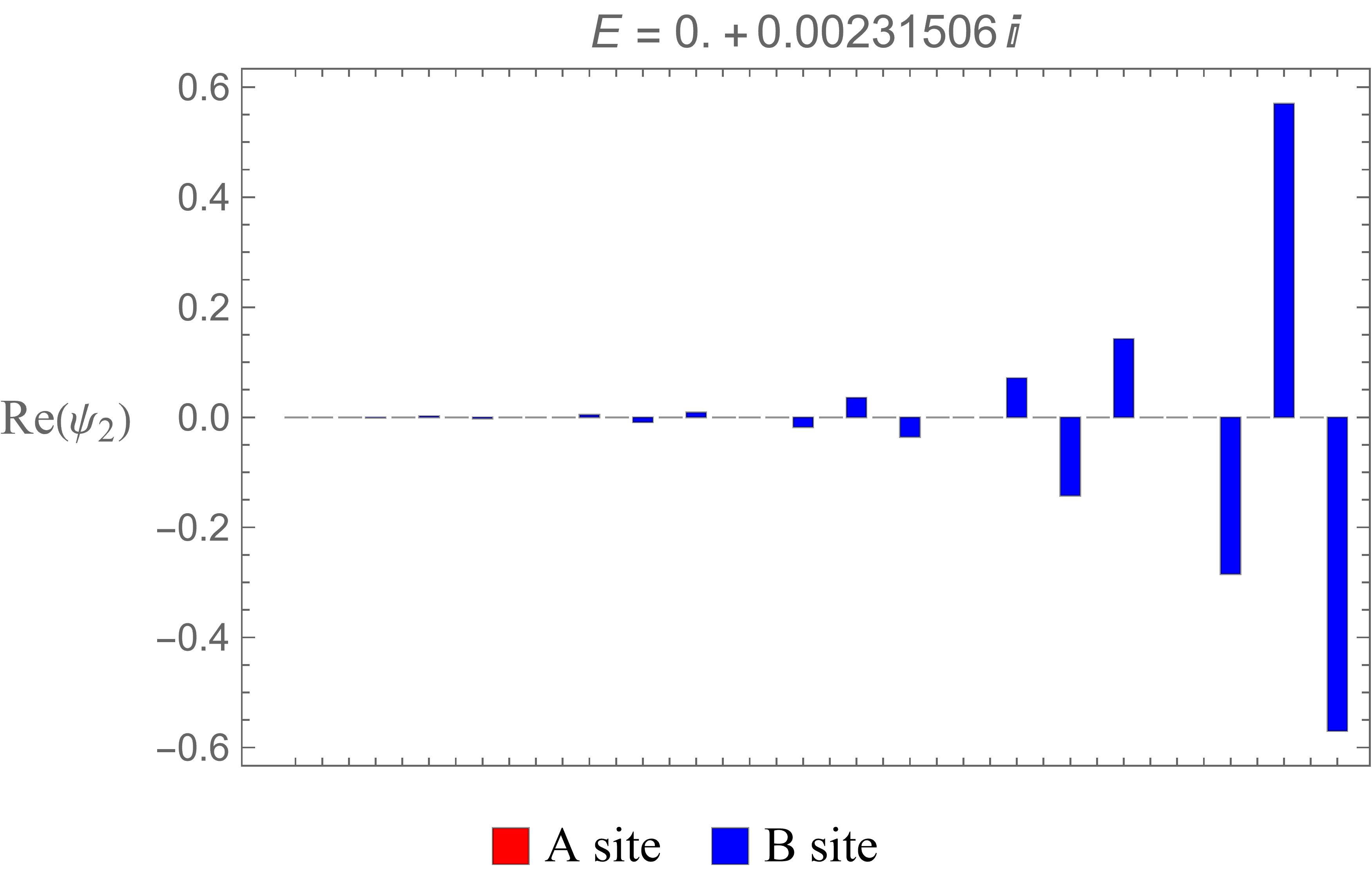}}\\
\subfloat[]{\includegraphics[width=0.30\textwidth]{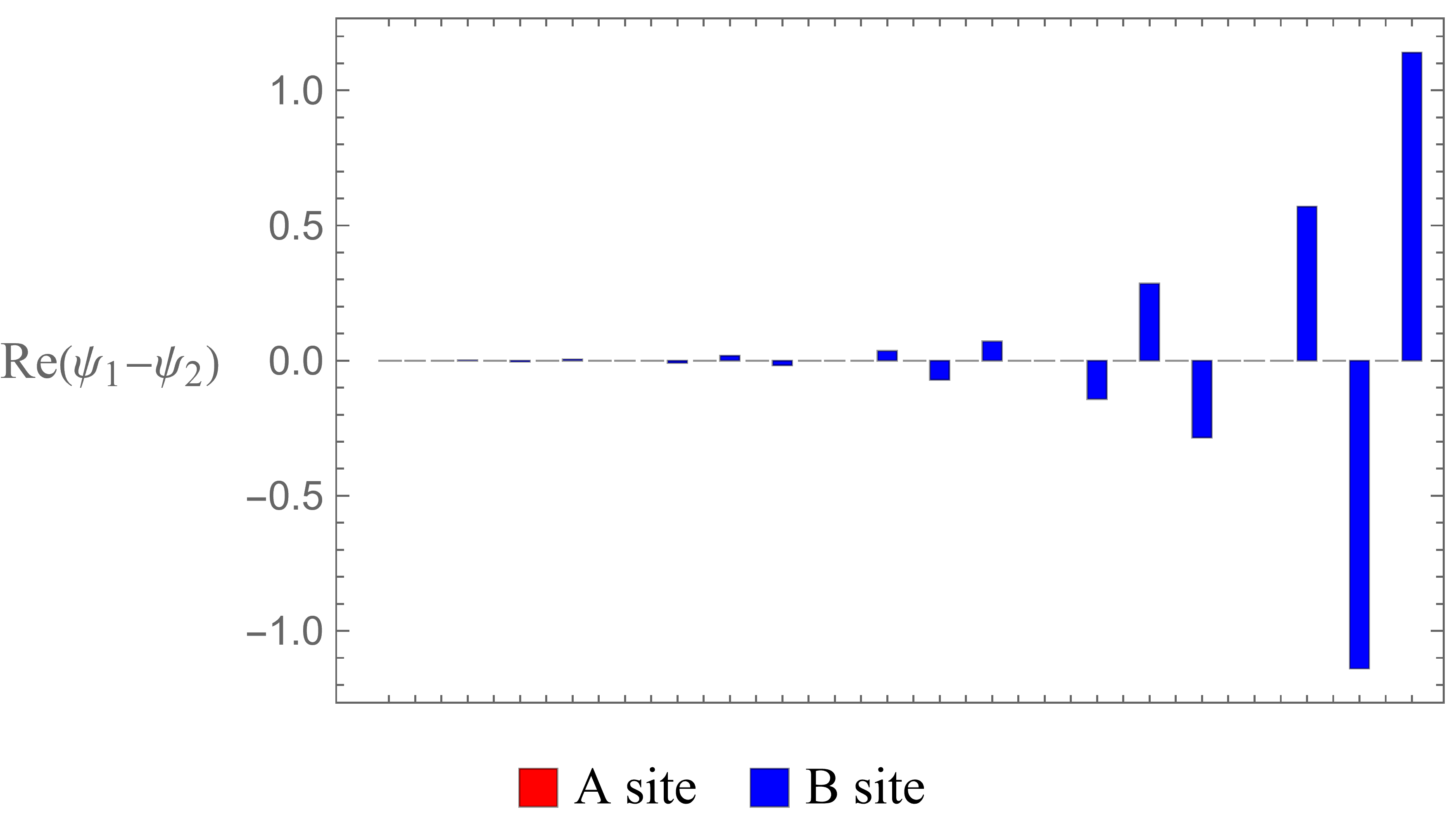}}\hskip 0.5cm
\subfloat[]{\includegraphics[width=0.33\textwidth]{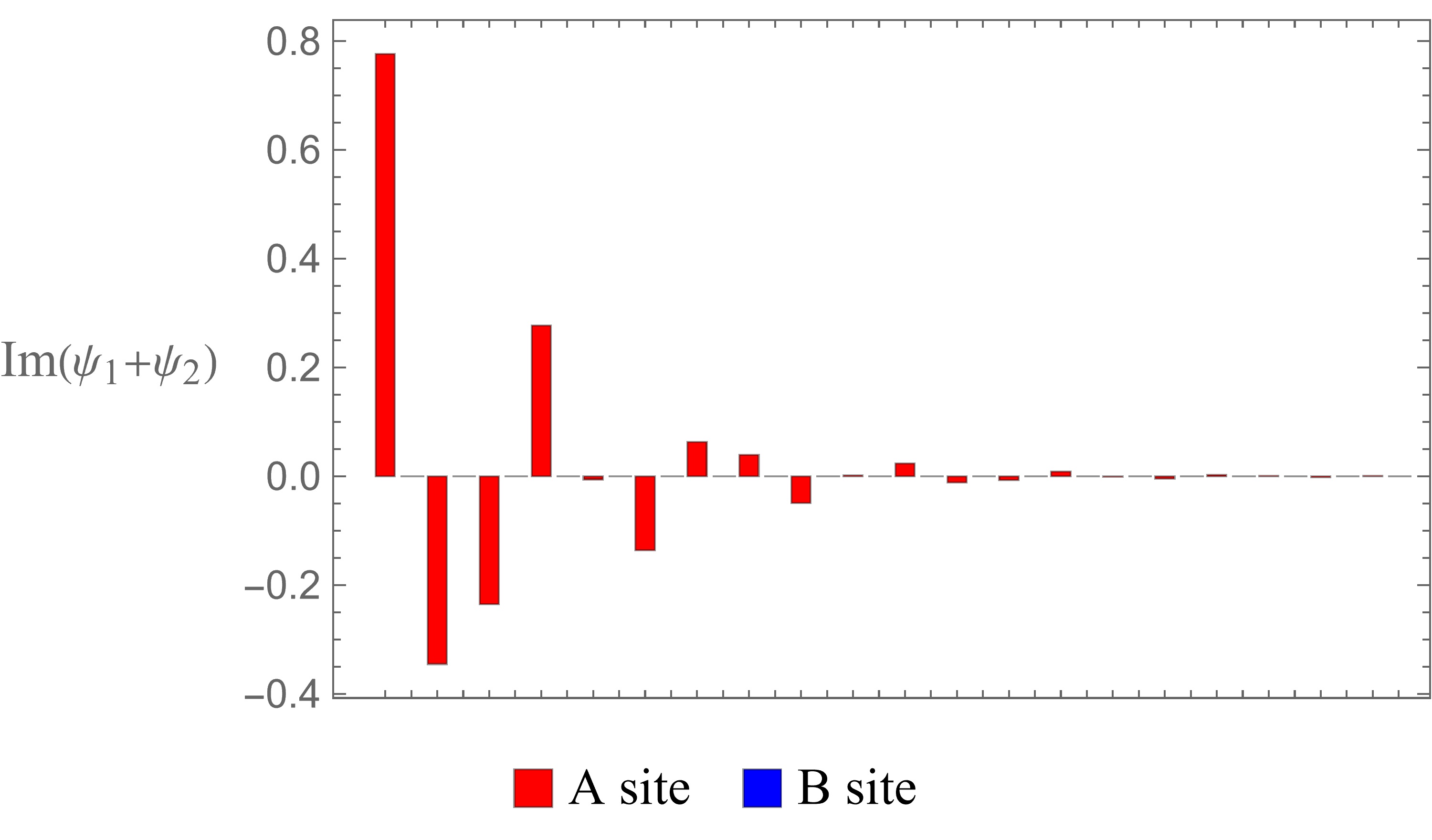}}\\
\caption{The energy spectrum of the NH type 2 extended SSH model with $40$ sites and the trajectory of $E(p)$ on the complex plane. Here, $\left( t_0^L, t_0^R, t_1^L, t_1^R, t_{-1}^L, t_{-1}^R \right)$
$= (2, 2, 9/2, 2, 1, 9/4)$. Thus, $(\tb_0, \tb_1, \tb_2) = (2, 3, 3/2)$ and it is in the topological phase with $\nb = 1$. Note that there are one left and one right edge states, which is consistent with the fact that $(\n_E^L, \n_E^R) = (1, -1)$. As mentioned, the energies of the eigenstates except for the edge states are encompassed by the trajectories of $E(p)$ on the complex plane.} \label{fig9 nHSSH-ext2}
\end{figure}

As an example, let's choose the parameters to be $\left( t_0^L, t_0^R, t_1^L, t_1^R, t_{-1}^L, t_{-1}^R \right) = (2, 2, 9/2, 2, 1, 9/4)$ so that $(\tb_0, \tb_1, \tb_{-1}) = (2, 3, 3/2)$, $\nb = 1$, $r=1$, and $(\n_E^L, \n_E^R) = (1, -1)$. As expected, there are two edge states with one on the left and the other on the right boundary. We show the energy spectrum along with the trajectory of $E(p)$ on the complex plane and the wave functions of the edge states in Fig.~\ref{fig9 nHSSH-ext2}. It is obvious that the NH type 2 extended SSH model we consider here is dual to the NH type 1 extended SSH model shown in Fig.~\ref{fig4 nHSSH-ext1} through the mapping given in eq.~(\ref{Type 1 and type 2 mapping}). In particular, Fig.~\ref{fig9 nHSSH-ext2}a and Fig.~\ref{fig4 nHSSH-ext1} are almost identical except for the existence of edge states.

\begin{figure}[hbt!]
\centering
\subfloat[]{\includegraphics[width=0.50\textwidth]{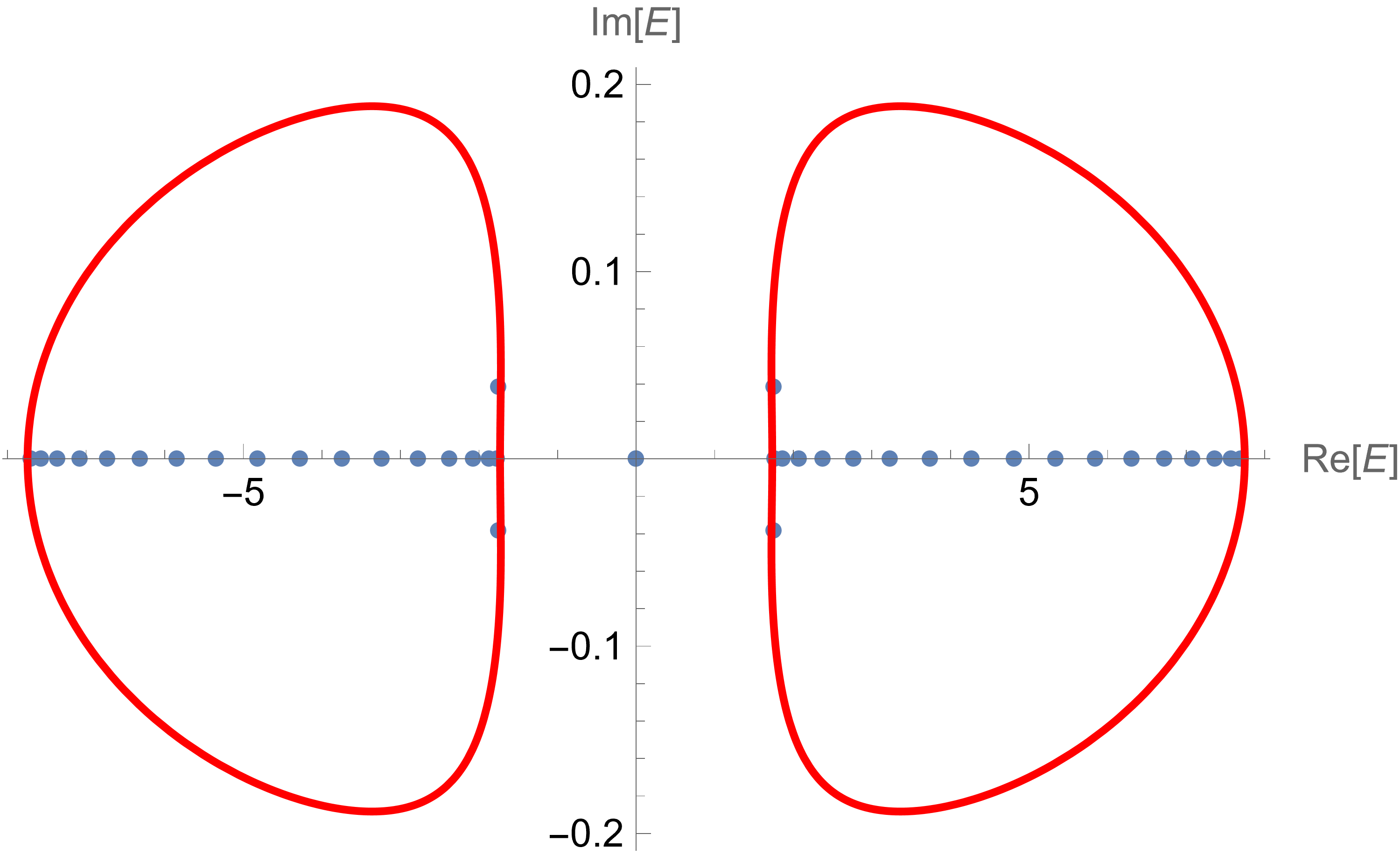}}\\
\subfloat[]{\includegraphics[width=0.30\textwidth]{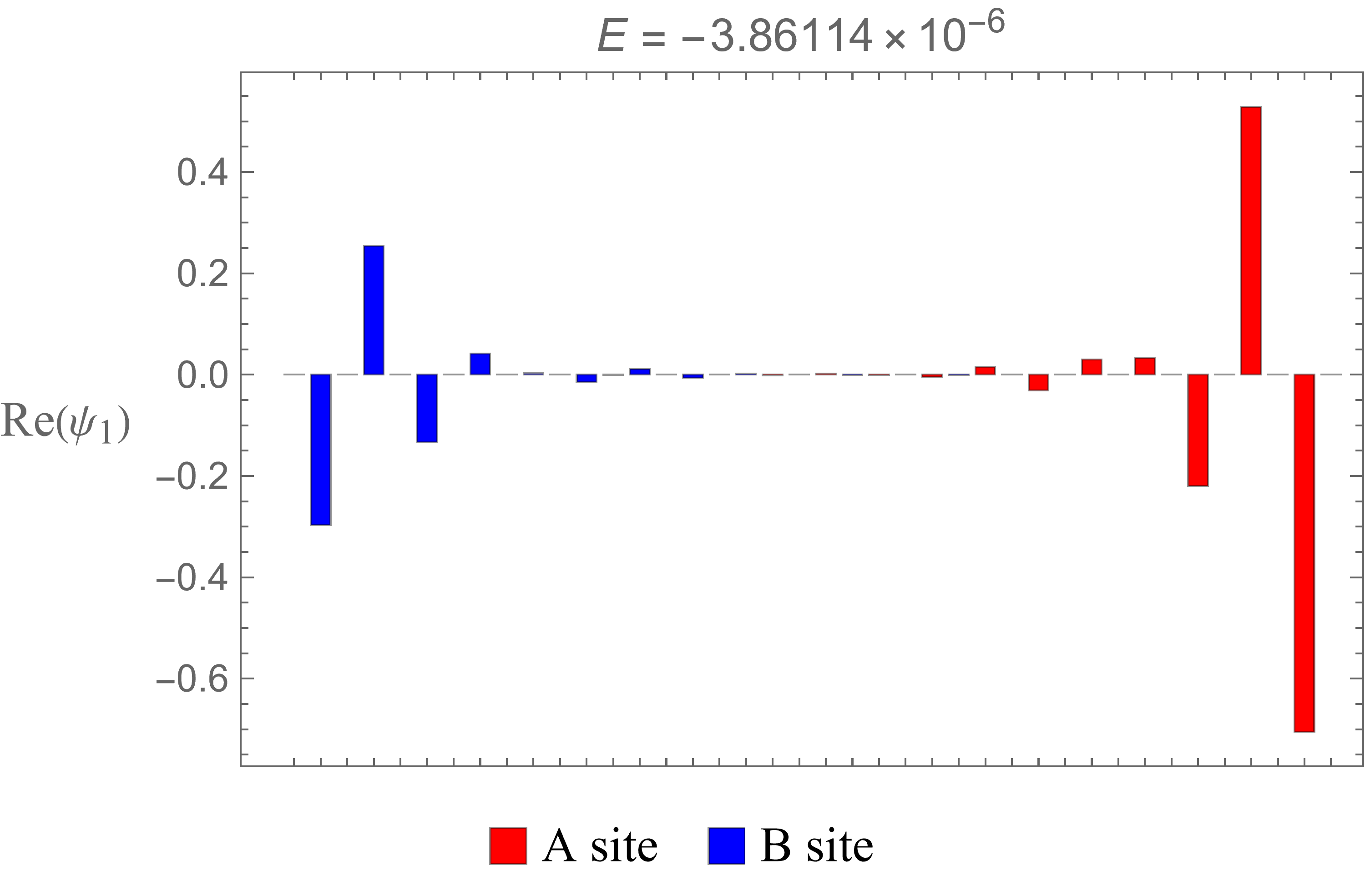}}\hskip 0.5cm
\subfloat[]{\includegraphics[width=0.30\textwidth]{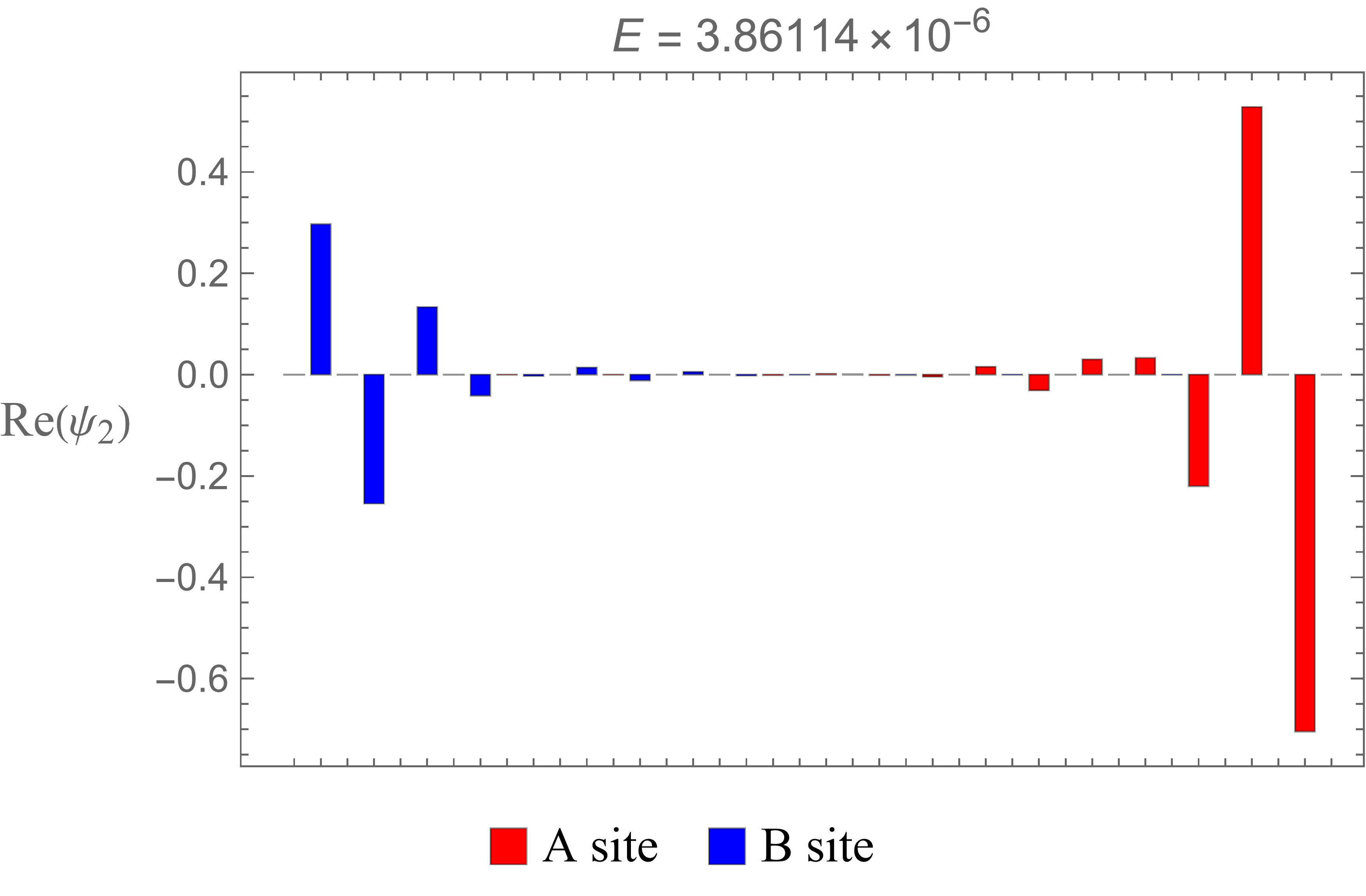}}\\
\subfloat[]{\includegraphics[width=0.30\textwidth]{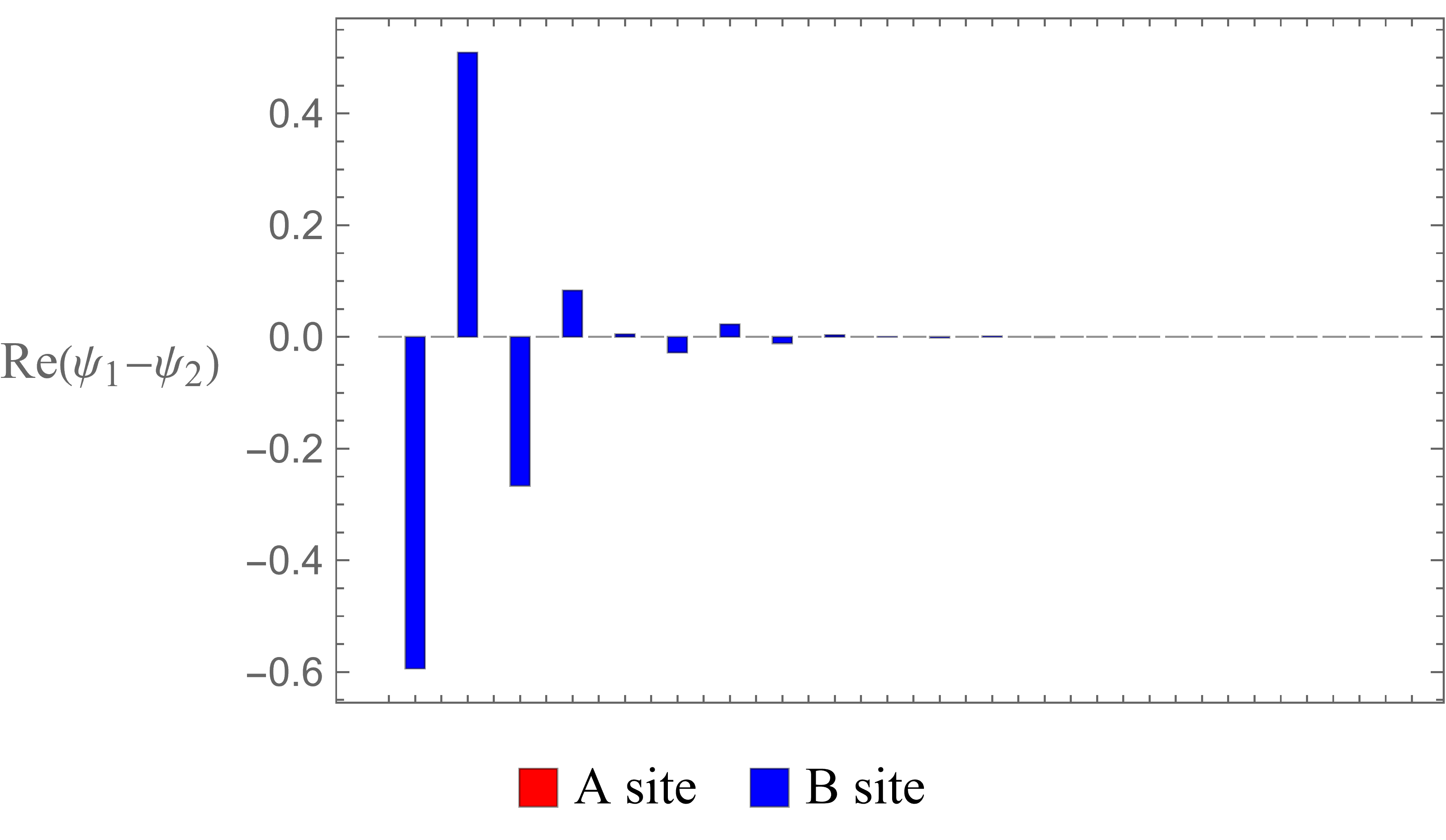}}\hskip 0.5cm
\subfloat[]{\includegraphics[width=0.33\textwidth]{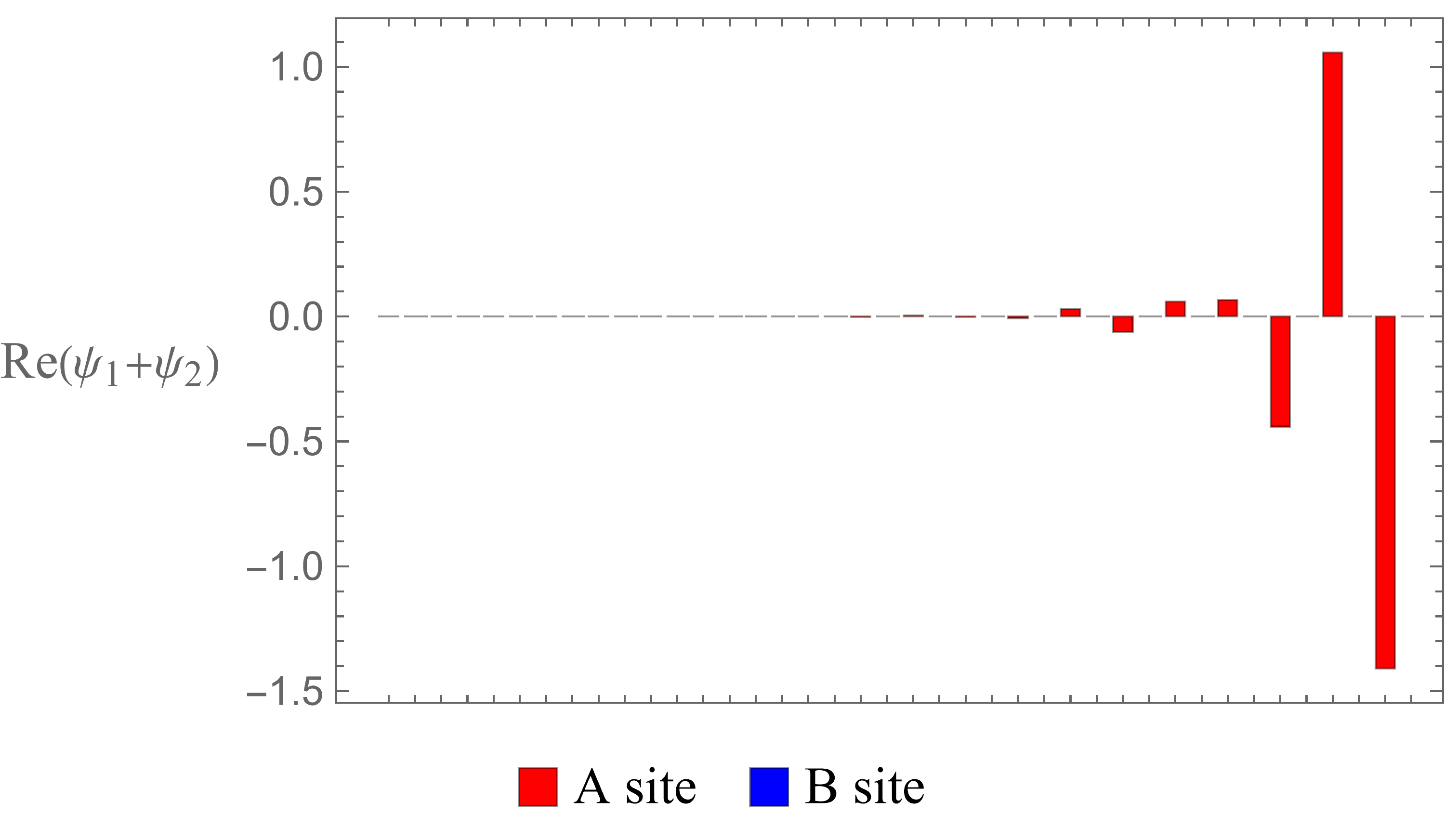}}\\
\caption{The energy spectrum of the NH type 2 extended SSH model with $40$ sites and the trajectory of $E(p)$ on the complex plane. Here, $\left( t_0^L, t_0^R, t_1^L, t_1^R, t_{-1}^L, t_{-1}^R \right)$
$=  (3, 3, 1, 1, 7/2, 4)$. Thus, $(\tb_0, \tb_1, \tb_2) =(3, 1, \sqrt{14})$ and it is in the topological phase with $\nb = -1$. Note that $(\n_E^L, \n_E^R) = (-1, 1)$ and the energies of the eigenstates except for the edge states are encompassed by the trajectories of $E(p)$ on the complex plane.} \label{fig10 nHSSH-ext2}
\end{figure}

Let's consider another example with $\left( t_0^L, t_0^R, t_1^L, t_1^R, t_{-1}^L, t_{-1}^R \right)$ $= (3, 3, 1, 1, 7/2, 4)$ so that $(\tb_0, \tb_1, \tb_{-1}) = (3, 1, \sqrt{14})$, $\nb = -1$, $r=(8/7)^{1/4}$, and $(\n_E^L, \n_E^R) = (-1, 1)$.  Again, there is one left and one right edge state. The energy spectrum along with the trajectory of $E(p)$ on the complex plane and the wave functions of the edge states are shown in Fig.~\ref{fig10 nHSSH-ext2}. The NH type 2 extended SSH model we consider here is dual to the NH type 1 extended SSH model, shown in Fig.~\ref{fig6 nHSSH-ext1}. Likewise, the energy spectra and the trajectories of $E(p)$ are both similar.

We also check the four cases considered in Ref.~\cite{GBZ4}, where $\left( t_0^L, t_0^R, t_1^L, t_1^R, t_{-1}^L, t_{-1}^R \right)$ $= (1.1+2/3, 1.1-2/3, 1, 1, 1/5, 1/5),$  $(-1.1+2/3, -1.1-2/3, 1, 1, 1/5, 1/5)$,  $(0.3, 0.3, 1.1+2/3, 1.1-2/3, 1/5, 1/5),$ and $(0.3, 0.3, 1.1-2/3, 1.1+2/3, 1/5, 1/5)$.  The corresponding $\nb$ are all 1, and $(\n_E^L, \n_E^R)  = (1,0), (0, -1), (1, -1)$ and $(1, -1),$ respectively.  Again, the total number of edge states and their locations are consistent with the criteria we obtain in the QH case.  

\vfill\eject


\begin{thebibliography}{99}
\bibitem{Review1} M. Z. Hasan and C. L. Kane, "Topological insulators," Rev. Mod. Phy., {\bf 82}, (2010).

\bibitem{Review2} X.-L. Qi and S.-C. Zhang, "Topological insulators and superconductors," Rev. Mod. Phy., {\bf 83}, (2011).

\bibitem{Periodic table1} A. P. Schnyder, S. Ryu, A. Furusaki, and A. W. W. Ludwig, "Classification of topological insulators and superconductors in three spatial dimensions," Phys. Rev. B {\bf 78}, 195125 (2008).

\bibitem{Periodic table2} A. Kitaev, "Periodic table for topological insulators and superconductors," AIP Conf. Proc. 1134, 22 (2009).

\bibitem{Periodic table3} A. P. Schnyder, S. Ryu, A. Furusaki, and A. W. W. Ludwig, "Classification of Topological Insulators and Superconductors," AIP Conf. Proc. 1134, 10 (2009).

\bibitem{SSH} W. P. Su, J. R. Schrieffer, and A. J. Heeger, "Solitons in Polyacetylene," Phys. Rev. Lett. {\bf 42}, 1698, (1979).

\bibitem{SSH-QD}  M. Kiczynski {\it et al}., ``Engineering topological states in atom-based semiconductor quantum dots,''  Nature 606, 694 (2022).

\bibitem{Zak} J. Zak, "Berry’s phase for energy bands in solids," Phys. Rev. Lett., {\bf 62}, 2747, (1989).

\bibitem{Berry} M. V. Berry,  "Quantal phase factors accompanying adiabatic changes," Proc. R. Soc. Lond. A. {\bf 392}, 45,  (1984).

\bibitem{Zak phase-OL}  M. Atala {\it et al}., ``Direct Measurement of the Zak phase in Topological Bloch Bands,''  Nature Phys. 9, 795–800 (2013).

\bibitem{Rice-Mele} J. K. Asboth, L. Oroszlany, and A. Palyi, "A Short Course on Topological Insulators," arXiv:1509.02295.

\bibitem{Thouless} D.J. Thouless, "Quantization of particle transport," Phys. Rev. B {\bf 27}, 6083–6087 (1983)

\bibitem{polarization1} R. D. King-Smith and D. Vanderbilt, "Theory of polarization of crystalline solids," Phys. Rev. B {\bf 47}, 1651(R) (1993).

\bibitem{polarization2} D. Vanderbilt and R. D. King-Smith, "Electric polarization as a bulk quantity and its relation to surface charge," Phys. Rev. B {\bf 48}, 4442 (1993).

\bibitem{Multi-band1} V. M. Martinez Alvarez and M. D. Coutinho-Filho, "Edge states in trimer lattices," Phys. Rev. A {\bf 99}, 013833 (2019).

\bibitem{Multi-band2} Y. Zhang, B. Ren, Y. Li, and F. Ye, "Topological states in the super-SSH model," Optics Express {\bf 29}, 42827, (2021).

\bibitem{Multi-band3} Chen-Shen Lee, Iao-Fai Io, and Hsien-chung Kao, “Winding number and Zak phase in multi-band SSH models,” Chinese Journal of Physics, {\bf 78}, 96 (2022).

\bibitem{Holonomy} R. Leone, "The geometry of (non)-Abelian adiabatic pumping," J. Phys. A: Math. Theor. {\bf 44}, 295301 (2011).

\bibitem{W-loop} F. Wilczek and A. Zee, "Appearance of Gauge Structure in Simple Dynamical Systems," Phys. Rev. Lett. {\bf 52}, 2111 (1984).

\bibitem{Kudin}  K. N. Kudin, R. Car, and R. Resta, "Berry phase approach to longitudinal dipole moments of infinite chains in electronic-structure methods with local basis sets," J. Chem. Phys. {\bf 126}, 234101 (2007).

\bibitem{Rhim} J.-W. Rhim, J. Behrends, and J. H. Bardarson, ``Bulk-boundary correspondence from the intercellular Zak phase,'' Phys. Rev. B {\bf 95}, 035421 (2017).

\bibitem{NH-Review1} E. J. Bergholtz, J. C. Budich, and F. K. Kunst,  ``Exceptional topology of non-Hermitian systems,'' Rev. Mod. Phys. {\bf 93}, 015005 (2021).

\bibitem{NH-Review2} N. Okuma and M. Sato, "Non-Hermitian Topological Phenomena: A Review," Annual Review of Condensed Matter Physics {\bf 14}, 83 (2022).

\bibitem{NH-Review3} Y. Ashida, Z. Gong, and M. Ueda, ``Non-Hermitian physics," Advances in Physics {\bf 69},  249 (2020)

\bibitem{NH-Review4} K. Ding, C. Fang, and G. Ma, "Non-Hermitian Topology and Exceptional-Point Geometries," Nature Reviews Physics {\bf 4}, 745 (2022).

\bibitem{NH-Exp1} C. Poli {\it et al}., “Selective enhancement of topologically induced interface states in a dielectric resonator chain,” Nat. Commun. {\bf }6, 6710 (2015).

\bibitem{NH-Exp2} J. M. Zeuner {\it et al}., “Observation of a Topological Transition in the Bulk of a Non-Hermitian System,” Phys. Rev. Lett. {\bf 115}, 040402 (2015).

\bibitem{NH-Exp3} W. Chen {\it et al}., “Exceptional points enhance sensing in an optical microcavity,” Nature (London) {\bf 548}, 192 (2017).

\bibitem{NH-Exp4} H. Hodaei  {\it et al}., “Enhanced sensitivity at higher-order exceptional points,” Nature
(London) {\bf 548}, 187 (2017).

\bibitem{NH-Exp5} S. Weimann  {\it et al}., “Topologically protected bound states in photonic parity-timesymmetric
crystals,” Nat. Mater. {\bf 16}, 433 (2017).

\bibitem{NH-Exp6} M. A. Bandres  {\it et al}., “Topological insulator laser: Experiments,” Science 359, eaar4005 (2018).

\bibitem{NH-Exp7} H. Zhou  {\it et al}., “Observation of bulk Fermi arc and polarization half charge from paired exceptional points,” Science {\bf 359}, 1009–1012 (2018).

\bibitem{NH-Exp8} A. Cerjan  {\it et al}., “Experimental realization of a Weyl exceptional ring,” Nat. Photonics {\bf 13}, 623–628 (2019).

\bibitem{NH-Exp9} T. Helbig {\it et al}., “Generalized bulk-boundary correspondence in non-Hermitian topolectrical circuits,” Nat. Phys. {\bf 16}, 747–750 (2020).

\bibitem{NH-Exp10}  L. Xiao{\it et al}.,  "Non-Hermitian bulk–boundary correspondence in quantum dynamics,"  Nature Physics {\bf 16}, 761 (2020).

\bibitem{Hatano-Nelson} Hatano, N., and D. R. Nelson, “Localization Transitions in Non-Hermitian Quantum Mechanics,” Phys. Rev. Lett. {\bf 77}, 570 (1996).

\bibitem{Spectral winding number1}  Z. Gong {\it et al}., “Topological Phases of Non-Hermitian Systems,” Phys. Rev. X {\bf 8}, 031079 (2018).

\bibitem{Spectral winding number2}  H. Shen, B. Zhen, and L. Fu, “Topological Band Theory for Non-Hermitian Hamiltonians,” Phys. Rev. Lett. {\bf 120}, 146402 (2018).

\bibitem{NH Periodic table} K. Kawabata, N. Okuma, and M. Sato, “Non-Bloch band theory of non-Hermitian Hamiltonians in the symplectic class,” Phys. Rev. B {\bf 101}, 195147 (2020).

\bibitem{Skin-effect1}  F. K. Kunst,  E. Edvardsson, J. C. Budich, and E. J. Bergholtz, “Biorthogonal Bulk-Boundary Correspondence in Non-Hermitian Systems,” Phys. Rev. Lett. {\bf 121}, 026808 (2018).

\bibitem{Skin-effect2}  Martinez Alvarez, V. M., J. E. Barrios Vargas, and L. E. F. Foa Torres, “Non-Hermitian robust edge states in one dimension: Anomalous localization and eigenspace condensation at exceptional points,” Phys. Rev. B {\bf 97}, 121401 (2018).

\bibitem{Skin-effect3}  Xiong, Y., “Why does bulk boundary correspondence fail in some non-Hermitian topological models,” J. Phys. Commun. {\bf 2}, 035043 (2018).

\bibitem{Skin-effect4}  S. Yao and Z. Wang, “Edge States and Topological Invariants of Non-Hermitian Systems,” Phys. Rev. Lett. {\bf 121}, 086803 (2018).

\bibitem{BBC} T. E. Lee, “Anomalous Edge State in a Non-Hermitian Lattice,” Phys. Rev. Lett. {\bf 116}, 133903 (2016).

\bibitem{GBZ1} S. Yao, F. Song, and Z.Wang, “Non-Hermitian Chern Bands,” Phys. Rev. Lett. {\bf 121}, 136802 (2018).

\bibitem{GBZ3} Deng, T. S., and W. Yi, “Non-Bloch topological invariants in a non-Hermitian domain wall system,” Phys. Rev. B {\bf 100}, 035102 (2019).

\bibitem{GBZ4} K. Yokomizo, and S. Murakami, “Non-Bloch Band Theory of Non-Hermitian Systems,” Phys. Rev. Lett. {\bf 123}, 066404 (2019).

\bibitem{GBZ5} Z. Yang, K. Zhang, C. Fang, and J. Hu, “Non-Hermitian Bulk-Boundary Correspondence and Auxiliary Generalized Brillouin Zone Theory,” Phys. Rev. Lett. {\bf 125}, 226402 (2020).

\bibitem{Biorthogonal1}  Edvardsson, E., F. K. Kunst, and E. J. Bergholtz, “Non-Hermitian extensions of higher-order topological phases and their biorthogonal bulk-boundary correspondence,” Phys. Rev. B {\bf 99}, 081302 (2019).

\bibitem{Biorthogonal2}  Edvardsson, E., F. K. Kunst, T. Yoshida, and E. J. Bergholtz,“Phase transitions and generalized biorthogonal polarization in non-Hermitian systems,” Phys. Rev. Research {\bf 2}, 043046 (2020).

\bibitem{Biorthogonal-polarization} S. Yao, F. Song, and Z. Wang, “Non-Hermitian Chern Bands,” Phys. Rev. Lett. {\bf 121}, 136802 (2018).

\bibitem{PH} A. Mostafazadeh, “Pseudo-Hermiticity versus PT symmetry: The necessary condition for the reality of the spectrum of a non-Hermitian Hamiltonian,” J. Math. Phys. (N.Y.) {\bf 43}, 205–214 (2002).

\bibitem{NH-SSH1} C. Yin {\it et al}., ``Geometrical meaning of winding number and its characterization of topological phases in one-dimensional chiral non-Hermitian systems,'' Phys. Rev. A {\bf 97}, 052115 (2018).

\bibitem{Chebyshev} R. Wakatsuki, M. Ezawa, and N. Nagaosa, Phys. Rev. B {\bf 89}, 174514 (2014).

\bibitem{MPBC} Ken-Ichiro Imura and Yositake Takane, “Generalized bulk-edge correspondence for non-Hermitian topological systems,” Phys.Rev. B {\bf 100}, 165430, (2019).

\bibitem{Ext-SSH}  Han-Ting Chen, Chia-Hsun Chang, and Hsien-chung Kao, "Connection between the winding number and the Chern number,'' Chinese Journal of Physics, {\bf 72}, 50-68, (2021).

\bibitem{CZM} H. C. Kao, "Chiral zero modes in superconducting nanowires with Dresselhaus spin-orbit coupling," Phys. Rev. B {\bf 90},  245435 (2014).
\end{thebibliography}
\end{document}